\documentclass[%
 reprint,
 amsmath,
 amssymb,
 aps,
prd,
showkeys,
floatfix,
]{revtex4-2}

\usepackage{breqn}
\usepackage{booktabs}
\usepackage[table]{xcolor}
\makeatletter
\let\cat@comma@active\@empty
\makeatother

\usepackage{graphicx}
\usepackage{dcolumn}
\usepackage{bm}
\usepackage{orcidlink}                                 
\usepackage{aas_macros} 
\usepackage{hyperref}
\hypersetup{
 colorlinks=true,
 linktoc=all,
 linkcolor=blue,
 citecolor=blue,
 urlcolor=blue
 }
 
\usepackage{listings}
\newcommand{\pysr}{\texttt{pySR}} 
\newcommand{\kt}{\tilde{k}_2} 
\newcommand{\rt}{\tilde{R}} 
\newcommand{\dr}{\Delta R}
\newcommand{\logl}{{\log_{10}\Lambda}} 

\begin{document}

   \title{Approximating neutron-star radii using gravitational-wave only measurements with symbolic regression}
   

\author{M. Bejger\orcidlink{0000-0002-4991-8213}}
\email{e-mail: bejger@camk.edu.pl}
\affiliation{INFN Sezione di Ferrara, Via Saragat 1, 44122 Ferrara, Italy} 
\affiliation{Nicolaus Copernicus Astronomical Center, Polish Academy of Sciences, Bartycka 18, 00-716, Warsaw, Poland} 


 
  \begin{abstract}
   {Gravitational waves emitted by binary neutron-star inspirals carry information on components' masses and tidal deformabilities, but not directly radii, which are measured by electromagnetic observations of neutron stars. To improve the multi-messenger astronomy studies of neutron stars, an expression for neutron-star radii as a function of gravitational-wave only data would be advantageous, as it would allow to compare information from two different channels.} 
   {In order to do so, a symbolic regression method, {\pysr}, is trained on TOV solutions to piecewise polytropic EOS input to discover an approximate symbolic expression for the neutron-star radius as a function of gravitational-wave measurements only.}
   {The approximation is tested on piecewise polytropic EOS NS data, as well as on NS sequences based on selected realistic (non-polytropic) dense-matter theory EOSs, achieving consistent agreement between the ground truth values and the symbolic approximation for a broad range of NS parameters covering current astrophysical observations, with average radii differences of few hundred meters. Additionally, the approximation is applied to the GW170817 gravitational-wave mass and tidal deformability posteriors, and compared to reported inferred radius distributions.}
   \end{abstract} 

   \maketitle

\section{Introduction}

Efficient use of information from electromagnetic (EM) observations and gravitational-wave (GW) signals, e.g. the GW170817 detection \citep{2017PhRvL.119p1101A} by Advanced LIGO \citep{2015CQGra..32g4001L} and Advanced Virgo \citep{2015CQGra..32b4001A} detectors, is the goal of multi-messenger astronomy \citep{2017ApJ...848L..12A}. Neutron stars (NSs) are among prime sources of both EM and GW emission; see e.g. \cite{Haensel:2007yy,2019gwa..book.....A} for textbook introductions. These compact, relativistic objects of a typical size of 20 km and a mass of $1-2\,M_\odot$, contain very dense (supra-nuclear) matter with as yet unknown composition and properties, like the equation of state (EOS, a relation describing the pressure as a function of density). EM observations of NSs provide measurements of their masses $M$ and radii $R$, through e.g. ray-tracing of light from NS surfaces in spacetime curved by their gravity \citep{2021ApJ...918L..28M}, as demonstrated by the NICER project \citep{NICER}. GWs detected so far are emitted during last moments of the binary compact systems; if at least one of the components is a NS, it may be tidally deformed by the external gravitational potential of its companion, introducing an extra energy loss which alters the detected GW waveform. In principle, GWs emitted by binary NSs (BNS) contain information on the masses of components $M_i$, and their tidal deformabilities $\Lambda_i$. Assuming that all NSs are composed of the same EOS, describing the ground state of cold, catalysed matter, mass $M$, radius $R$ and tidal deformability $\Lambda$ are related through this one EOS. So far there were many attempts to relate the NS parameters with each other, to either infer the unknown dense-matter EOS \citep{PhysRevLett.121.091102,2018ApJ...857L..23R,PhysRevLett.120.261103,2020GReGr..52..109C}, or provide an EOS-independent ''universal relation'' between the NS parameters, see, e.g. \citep{2013PhRvD..88b3009Y,2018PhRvD..97j4036C,2019PhRvD..99l3026K}. Machine learning (ML) is also applied in this respect, see \citep{PhysRevD.101.054016,2020A&A...642A..78M,PhysRevD.106.103023,2023JCAP...12..022F,2024ApJ...965...47G,PhysRevD.111.023035}, and recent reviews of ML applications to GW astrophysics \citep{2021MLS&T...2a1002C,2024arXiv240107406S,2025LRR....28....2C}. 

This work focuses on a symbolic regression (SR) method to find an approximate, but sufficiently accurate relation between the NS radius $R$ (an EM observable), and $M$ and $\Lambda$, obtained in GW measurements of BNS inspirals. SR allows for model discovery: a search for the optimal functional form of an expression that best describes the underlying data, see \cite{Angelis2023,Tenachi_2023} for recent reviews. In particular, it differs from the standard ''black box'' approaches, e.g. neural networks, because from the point of view of the requirement for eXplainable Artificial Intelligence (XAI) \citep{SAEED2023110273}, it addresses the problems of interpretability (symbolic expression is straightforward to understand) and explainability (result of an expression are easy to justify). SR was applied to GW and black hole-related problems\citep{2022mla..confE..25W,2025AAS...24545007N,2025arXiv250418270Y},  specifically recently in the context of mapping NS parameters with its EOS   \citep{2025PhLB..86539470P,2024PhRvD.110j3042I}. 

Here, I search for a general-purpose ''universal'' relation between global parameters describing the NS for any given EOS ($R$, $M$, $\Lambda$), to provide an ''EOS-independent'' $\rt(M, \Lambda)$ approximation, hopefully acceptable for a broad selection of EOSs. This text is composed as follows: Sec.~\ref{sec:method} introduces the general idea and the methods used, with details in the Appendix~\ref{sec:app_model_details}, Sec.~\ref{sec:data} describes the training and validation data, with details in the Appendix~\ref{sec:app_data_prep_details}, Sect.~\ref{sec:results} discusses the results and additional experiments with data outside the training dataset, and Sec.~\ref{sec:conclusions} contains brief summary and conclusions. 
\section{Methods} 
\label{sec:method} 

As demonstrated by the GW170817 event \citep{2017PhRvL.119p1101A}, NS tidal deformability induced by the external gravitational potential of its companion is detectable with the GW detectors during last stages of BNS inspirals. From the point of view of the NS structure in the simplest spherically-symmetric approximation, NS mass $M$ and radius $R$ are obtained by solving a set of ordinary differential equations, the Tolman-Oppenheimer-Volkoff (TOV) system \citep{1939PhRv...55..364T,1939PhRv...55..374O} with a given EOS as input. Calculation of the static tidal deformability $\Lambda$ in the lowest (quadrupole) approximation requires solving a similar ordinary differential equation; specifically, $M$, $R$ and $\Lambda$ are related to another NS functional, the $k_2$ tidal Love number \citep{1911spge.book.....L,2008PhRvD..77b1502F,2017PhRvD..95h3014V} as follows: 
\begin{equation} 
    k_2 = \frac{3\Lambda}{2}\left(\frac{GM}{Rc^2}\right)^5.
    \label{eq:k2} 
\end{equation} 
The aim of this work is to approximate (reconstruct, estimate) the NS radius $R$, restricting only to NS parameters obtained via GW detections during the BNS inspirals: $M$ and $\Lambda$. This requires to represent $k_2$ with sufficient accuracy as a function of $M$ and $\Lambda$ only, ${\kt}(M,\,\Lambda)$. Approximation for the radius $\rt$ is then 
\begin{equation} 
    \rt(M,\,\Lambda) = \left(\frac{3\Lambda}{2{\kt}(M,\,\Lambda)}\right)^{1/5}\frac{GM}{c^2}.
    \label{eq:rt} 
\end{equation} 
For the representation of $\kt$ to be practically useful, it has to connect in a simple way the NS global parameters spanning an astrophysically-relevant range of masses, as well as a range of EOS parameters underlying these parameters. 

To obtain interpretable solution to explain (or at least, describe) relationships in the data, I have resorted to SR as a tool for model discovery, specifically the {\pysr} \citep{2023arXiv230501582C} implementation of SR methods with the \texttt{python} and \texttt{julia} languages \citep{python,doi:10.1137/141000671}. The {\pysr} library combines the usability of \texttt{python}, which serves as a language wrapper for the high-performance symbolic capabilities of \texttt{julia}'s \texttt{SymbolicRegression.jl} package to carry out the computationally intensive tasks of finding optimal symbolic expressions. The method uses a combination of evolutionary algorithms (specifically genetic programming) and symbolic expression trees to discover equations that best fit given data, under an assumed set of available operations and functions (unary and binary operators: addition, multiplication, $\log()$, power etc.). The algorithm evolves towards the ''best'' expression by changing and recombining trial expressions based on how well they fit the training data, how well they are constrained by the training objective (the loss function), and by monitoring the Pareto front, i.e. accuracy vs complexity of the solution \citep{Smits2005}. 

Appendix~\ref{sec:app_model_details} contains summary of the technical details and input to the main SR function, \texttt{PySRRegressor()}, as well as description of additional hyperparameter optimisation used to obtain the results. 

\section{Data} 
\label{sec:data} 
The training and validation data is adopted from \cite{2020A&A...642A..78M}, see their Sec.~3, and Appendix~\ref{sec:app_data_prep_details} for details. The datasets consist of standard sequences of solutions to the TOV equations for the astrophysically-relevant range of NS masses (from $0.5\,M_\odot$ to $M_{max}$, with $M_{max}$ depending on the EOS), augmented with solutions for the Love number $k_2$ and the tidal deformability $\Lambda$. 

\begin{figure}
\includegraphics[width=\hsize]{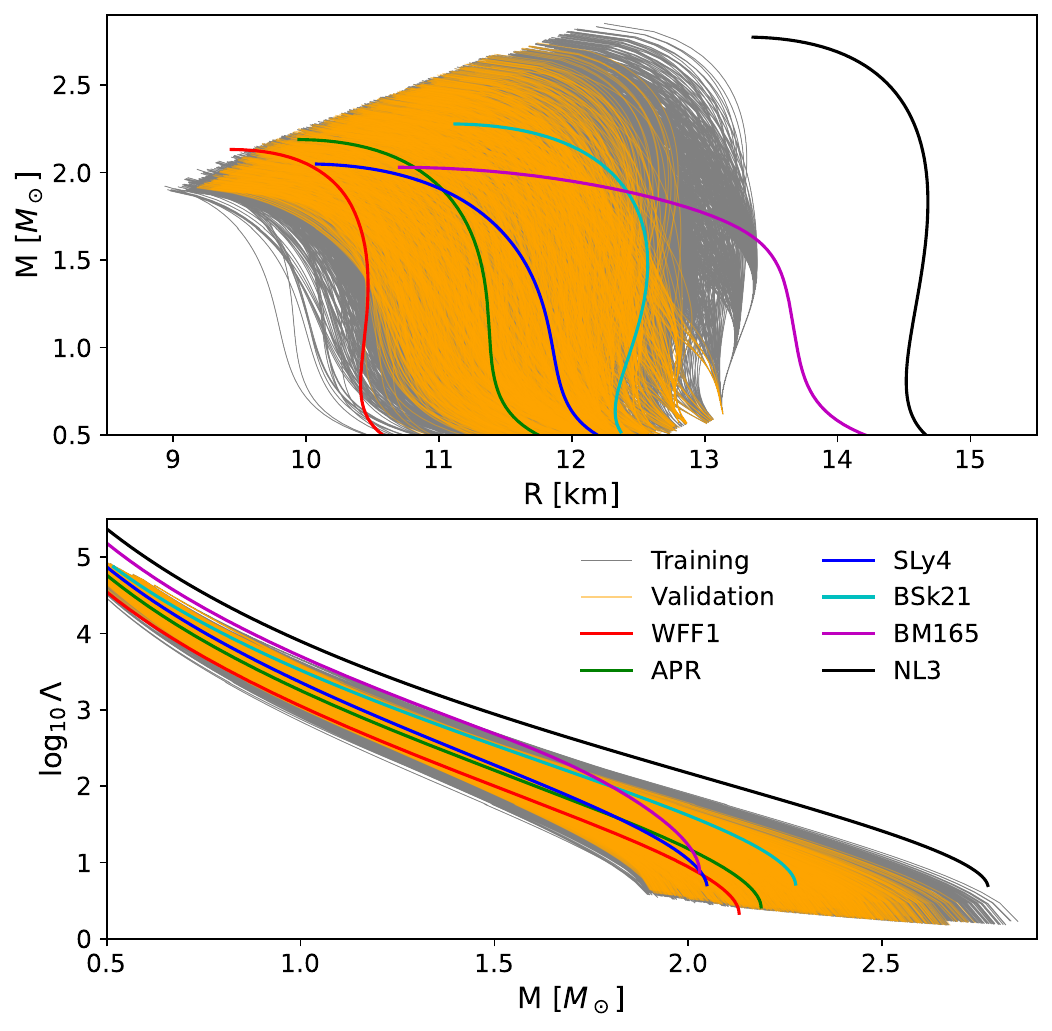}
    \caption{Training and validation datasets, and NS sequences based on selected EOSs using realistic theories of dense-matter interaction. Top panel: mass-radius $M(R)$ with training dataset (grey lines) and validation dataset (orange lines), and selected realistic dense-matter NS sequences for post-training validation purposes. Bottom panel: logarithm of $\Lambda$ as a function of mass $M$. See the legend and text for details on the realistic dense-matter EOSs.}
    \label{fig:mr_dataset}
\end{figure}

The input to the TOV equations is in form of piecewise relativistic polytropic EOS \citep{1965ApJ...142.1541T} with adiabatic indices $\gamma$ in range between 2 and 4 in three polytropic segments, for densities above 0.1 fm$^{-3}$ (approximately the nuclear saturation density). For smaller densities, the crust EOS is assumed to be known. It is represented by a relevant low-density part of the SLy4 EOS \citep{DouchinH2001,1994A&A...283..313H}. The parametric EOS model proposed in \cite{2020A&A...642A..78M} (and used here) allows to cover an astrophysically-interesting part of the $M(R)$ and $\logl(M)$ plane, as shown in Fig.~\ref{fig:mr_dataset}. 

For comparison, a set of non-polytropic EOSs, based on realistic theories of dense-matter interactions, spanning various regions of the $M(R)$ plane and the training data set is displayed as well: WFF1 \citep{1988PhRvC..38.1010W}, APR \citep{1998PhRvC..58.1804A}, SLy4 \citep{DouchinH2001}, BSk21 \citep{2010PhRvC..82c5804G},  BM165 \citep{2012A&A...543A.157B} and NL3 \citep{PhysRevC.55.540,2012NuPhA.895...44S} EOSs. 

\section{Results} 
\label{sec:results} 

It turns out it is in fact possible to approximately represent $k_2$ as a relatively simple function of $M$ and $\Lambda$ only, i.e. $\kt(M, \Lambda)\approx k_2$. SR search for a well-behaving expression for $\kt$ was agnostic in the sense that, in addition to set of operators and other assumptions described in Appendix~\ref{sec:app_model_details}, no prior form of the expression was assumed. Among expressions yielding the lowest, but similar mean square error (MSE): $2.75{\times}10^{-6}$ on the validation set, $2.68{\times}10^{-6}$ on the training set, I've chosen the following: 
\begin{eqnarray} 
    \kt(M, \Lambda) &=& 0.0351\,\left(\logl\right)^2 \times \nonumber \\ 
    &\times& \left(\frac{M}{M + 2.05} - 0.109\right) + 0.0137, 
    \label{eq:the_expression} 
\end{eqnarray} 
where mass $M$ is expressed in $M_\odot$ (see Appendix~\ref{sec:app_model_details} for description explaining the choice of this expression, and Table~\ref{tab:other_expressions} for examples of other expressions of varying accuracy and complexity). Since $k_2$ is positive, this approximation is limited to $M \gtrapprox 0.25\,M_\odot$, i.e. well within the current applications for astrophysical NSs.

\subsection{Robustness tests with polytropic and realistic EOSs}
\label{sec:robustness_tests}

\begin{figure}
\centering
\includegraphics[width=1.02\hsize]{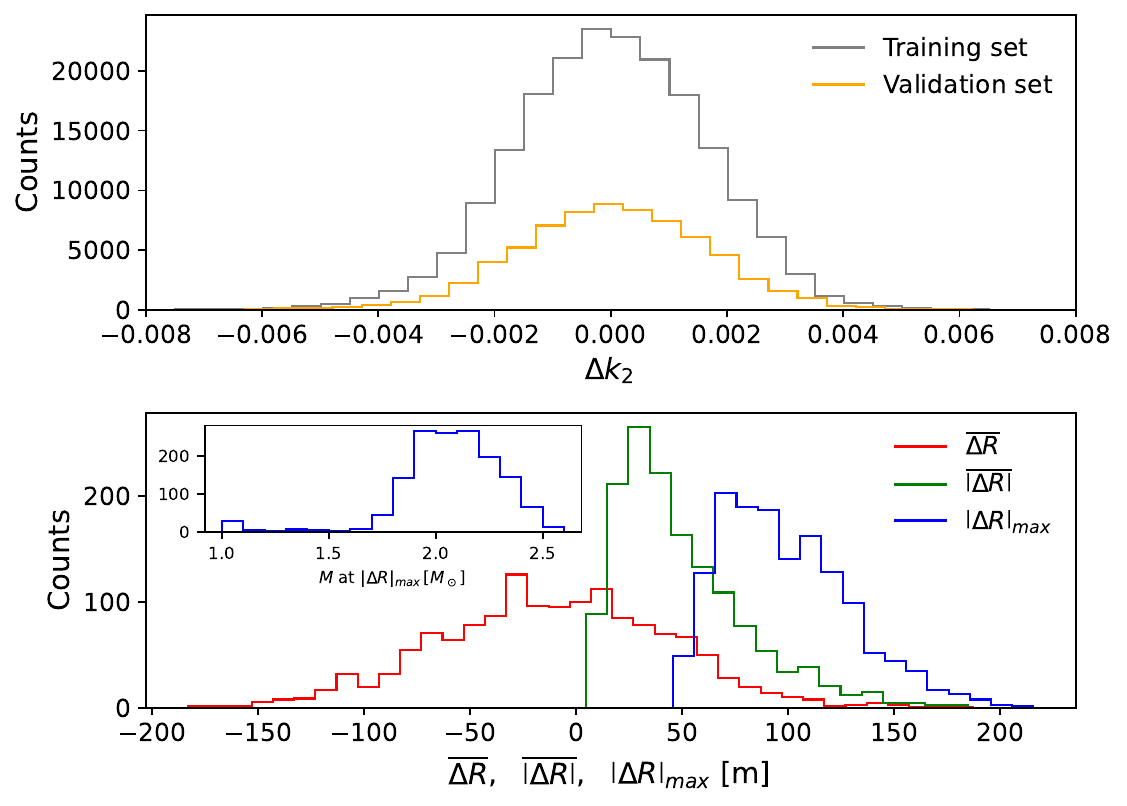}
    \caption{Tests of Eqs.~\ref{eq:rt} and \ref{eq:the_expression} on the training and validation datasets. Top panel: distributions of $\Delta k_2 = k_2 - \kt$ for the whole training and validation datasets; see Appendix~\ref{sec:app_data_prep_details} for details on how they were obtained. Bottom panel: distributions of three metrics related to $\Delta R = R - \rt$, evaluated for each validation dataset EOS along its corresponding NS sequence, for a range of masses $M\in (1\,M_\odot, M_{max})$, with $M_{max}$ value depending on the EOS used for a given NS sequence. $\overline{\dr}$ denotes mean value of $\dr$, $\overline{|\dr|}$ is mean absolute value of $\dr$, $|\dr|_{max}$ is maximum absolute value of $\dr$. The inset shows the distribution of mass at $|\dr|_{max}$.}
    \label{fig:pred_hist}
\end{figure}

Figure~\ref{fig:pred_hist} shows the distribution of $\Delta k_2 = k_2 - \kt$ (top panel) for the training and validation datasets, and $\dr = R - \rt$ (bottom panel) for the validation dataset only, i.e. 4-tuples of $(M, R, k_2, \Lambda)$, and resulting $\kt(M, \Lambda)$ from Eq.~\ref{eq:the_expression} and $\rt(M, \Lambda)$ from Eq.~\ref{eq:rt}. 

Top panel presents distribution of $\Delta k_2$ for the two datasets in the full range of masses $M\in (0.5\,M_\odot, M_{max})$ - the distributions are centred around zero difference, which is expected if the approximation does not have systematic biases. The bottom panel contains distribution of metrics related to $\dr$, restricted to 4-tuples with NS masses $M\in(1\,M_\odot, M_{max})$, with three radius-related metrics shown for this mass range, defined as follows. 

The $\overline{\dr}$ represents the mean $\dr$ along each NS sequence belonging to the validation dataset. On average the $\overline{\dr}$ values are distributed around the zero, i.e. on average $\rt$ is near the ground truth $R$ in the mass range studied. Second metric, the average absolute $\dr$, $\overline{|\dr|}$, peaks around 50 m, with the tail of the distribution reaching 200 m. Third metric is the distribution of the maximum absolute value (along each NS sequence) of $\dr$, $|\dr|_{max}$, with average values around 100 m. 

The inset displays the distribution for mass values for $|\dr|_{max}$. Concentration of $|\dr|_{max}$ near $M_{max}$ is caused by relatively small change of $M$ and $\Lambda$, small values of $\Lambda$ itself, and relatively large variation of $R$ values near $M_{max}$, which together leads to larger than the average approximation errors.

\begin{figure}
\centering
\includegraphics[width=\hsize]{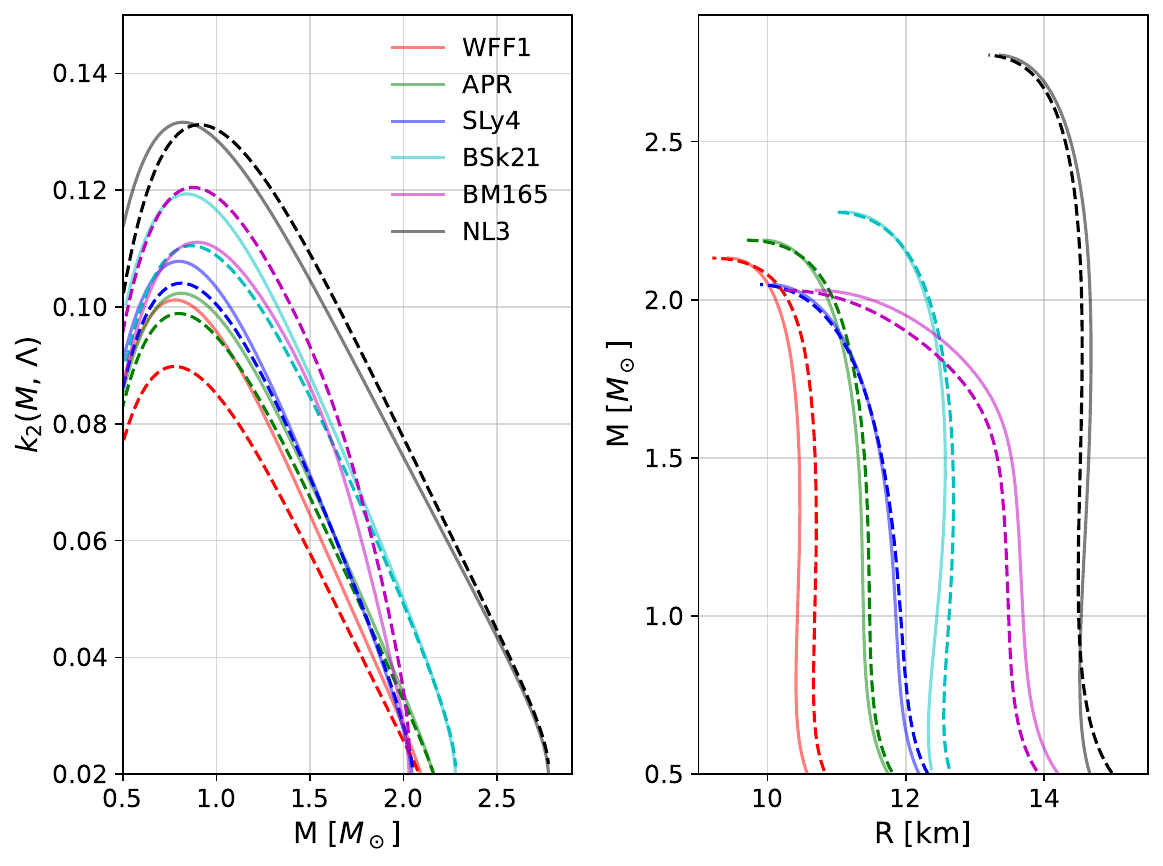}
    \caption{Ground truth $k_2(M,\Lambda)$ and $M(R)$, and approximations $\kt(M,\Lambda)$ and $M(\rt)$ for NS sequences based on selected EOSs of dense matter. Left panel: comparison of $k_2$ (solid curves) and $\kt$ (dashed curves). Right panel: comparison for $M(R)$ (solid curves) and $M(\rt)$ (dashed curves). See Tab.~\ref{tab:real} for a summary on approximation errors.}
    \label{fig:real}
\end{figure}

Equation~\ref{eq:the_expression} was also applied to the NS TOV sequences based on realistic dense-matter EOSs. Figure~\ref{fig:real} shows pairs of curves: solid for ground truth TOV solutions, and dashed curves for approximations obtained by means of Eqs.~\ref{eq:rt} and \ref{eq:the_expression}. Note that these realistic EOSs are non-polytropic relations between pressure and density, and some of them result in NS sequences partially or fully outside of the training or validation dataset, e.g. the WFF1, BM165 and NL3 EOSs. Despite this, errors in $R$ approximation based on Eqs.~\ref{eq:rt} and \ref{eq:the_expression} are within reasonable range, as summarized in Tab.~\ref{tab:real}. In addition, the proposed $\kt$ approximation is able to recover the $M(R)$ sequence for instances very different from the training dataset.  

\begin{table} 
{\small
\setlength{\tabcolsep}{12pt}
\begin{tabular}{@{}lcccc@{}}
\toprule\toprule
EOS &  $|\dr|_{max}$ & $\overline{\dr}$  
    & $\overline{|\dr|}$ & $M_{|\dr|_{max}}$ \\ 
  ~ & [m] & [m] & [m] & $[M_\odot]$ \\ 
\midrule 
WFF1   & 251.2 & -153.9 & 186.2 & 1.011 \\ 
APR    & 247.9 & -44.1 & 87.2 & 2.188 \\ 
SLy4   & 186.5 & 18.3 & 49.5 & 2.048 \\ 
BSk21  & 174.1 & -49.0 & 80.0 & 1.016 \\ 
BM165  & 452.6 & 299.8 & 299.8 & 2.030 \\ 
NL3    & 167.6 & 100.6 & 100.6 & 2.773  \\
\bottomrule\bottomrule
\end{tabular}
}
\caption{Approximation errors of Eqs.~\ref{eq:rt} and \ref{eq:the_expression} for NS sequences using realistic EOSs of dense matter for masses $M\in (1\,M_\odot, M_{max})$. $|\dr|_{max}$: maximum absolute value of $\dr$; $\overline{\dr}$: mean value of $\dr$; $\overline{|\dr|}$: mean absolute value of $\dr$; $M_{|\dr|_{max}}$ denotes mass at which $|\dr|$ is the largest.}
\label{tab:real} 
\end{table} 

\subsection{Radius recovery for intersecting NS sequences and NS with a strong phase transition}
\label{sec:same_compactness} 

\begin{figure}
\centering
\includegraphics[width=\hsize]{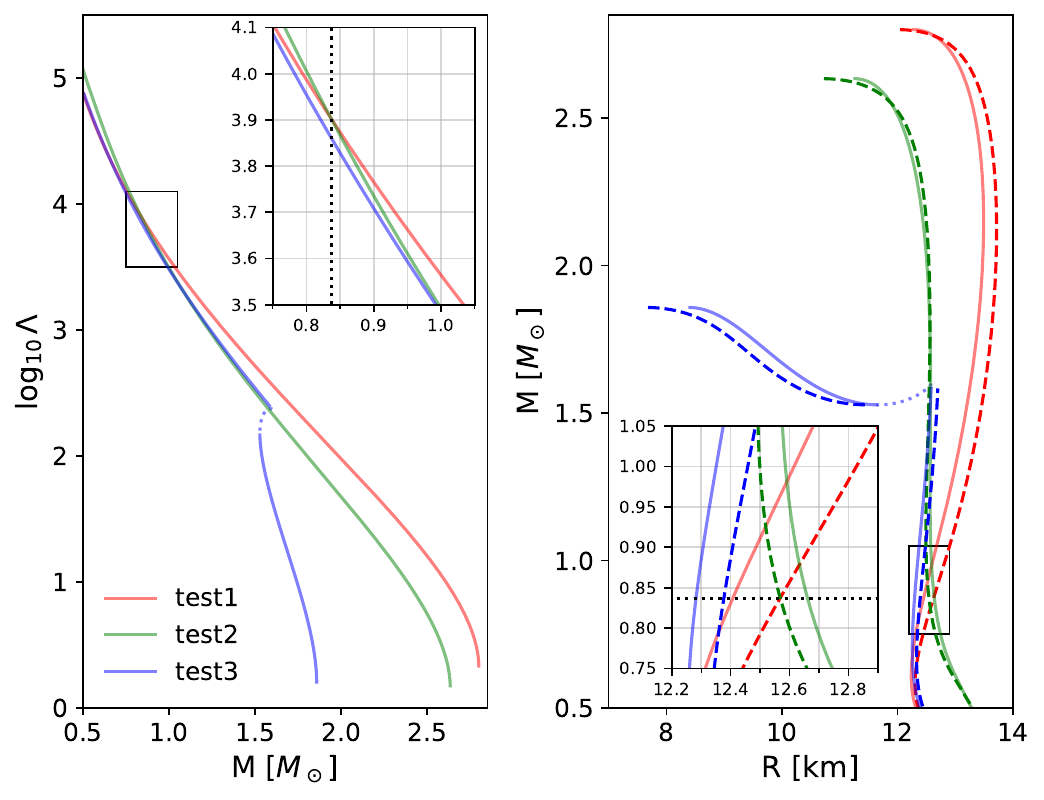}
    \caption{Two intersecting NS sequences with different EOSs (red and green curves), and one NS sequence (blue curve) featuring a large density jump leading to an unstable branch (dotted blue curve). Left panel: $\logl$ as a function of $M$; the inset plot shows a close up with a vertical dotted black line marking the intersection mass, $M \approx 0.837\,M_\odot$. Right panel: $M(R)$ (solid curves) and $M(\rt)$ (dashed curves). The approximated radius $\rt$ value at the intersection is close to the ground truth radius $R$, but it is recovered at different mass; horizontal dotted black line marks the mass value from the left plot.}
    \label{fig:test}
\end{figure}

For more tests, additional three polytrope-based EOSs with parameters outside the training dataset were manufactured. First two EOSs are meant to check how the approximation handles intersecting $M(R)$ sequences, i.e. two NSs with the same compactness $M/R$, but different EOSs. The setup is presented in the right panel of Fig.~\ref{fig:test}, with ground truth $M(R)$ plotted as solid curves, and approximations $M(\rt)$ as dashed curves. Left panel of Fig.~\ref{fig:test} shows the NS sequences in $\logl(M)$ plane. The parameters of the test EOSs are, for the test1 EOS: SLy4 crust-core EOS connected at $n_0=0.25\,\mathrm{fm}^{-3}$ to $\gamma=3.5$ relativistic polytrope EOS; for the test2 EOS: SLy4 crust EOS connected at $n_0=0.15\,\mathrm{fm}^{-3}$ to  $\gamma_1=2$ EOS, which changes to $\gamma_2=4$ at $n_0=0.275\,\mathrm{fm}^{-3}$. 

The $\rt$ value at the intersection is recovered with accuracy similar to previous tests, and is very close to the ground truth $R$ value, as can be seen from the inset plot. However, it occurs at a mass value close to the mass at the intersection in the $\logl(M)$. This suggests that point values of $M$ and $\Lambda$ may not be sufficient to recover the $R$ value precisely, and some additional input information should be provided, e.g. in the form of $\mathrm{d}\Lambda/\mathrm{d}M$. This is outside of the scope of this work, which aims at recovering $R$ from point estimates of $M$ and $\Lambda$ only (as they come from the GW waveforms), but shall be addressed in a future project. 

In case of the test3 EOS, the parameters are: SLy4 crust EOS connected at $n_0=0.2\,\mathrm{fm}^{-3}$ to $\gamma_1=3.75$ EOS, which ends at $n_1=0.425\,\mathrm{fm}^{-3}$; next polytropic segment with $\gamma_2=4.5$ starts at $n_2=0.74375\,\mathrm{fm}^{-3}$. This setup is approximating a first-order phase transition with a density jump, here $n_2/n_1 = 1.75$, large enough to destabilize the NS \citep{1971SvA....15..347S}. The unstable region in the NS sequence is marked with dotted curves in both panels in Fig.~\ref{fig:test}. The $\rt$ approximation is able to recover the $M(R)$ values, despite the fact of being in a large part outside of the training dataset range. However, a larger discrepancy in $\dr$ near the $M_{max}$ is due to decreasing values of $k_2$ and $\Lambda$, and a small range of $M$ near the $M_{max}$. $\rt(\Lambda)$ approximated by Eqs. \ref{eq:rt} and \ref{eq:the_expression} is proportional to $\left(\Lambda/\log^2_{10}(\Lambda)\right)^{1/5}$, which grows with decreasing $\Lambda$ and has a singularity at $\Lambda=1$. 


\begin{figure*}
\includegraphics[width=0.3375\textwidth]{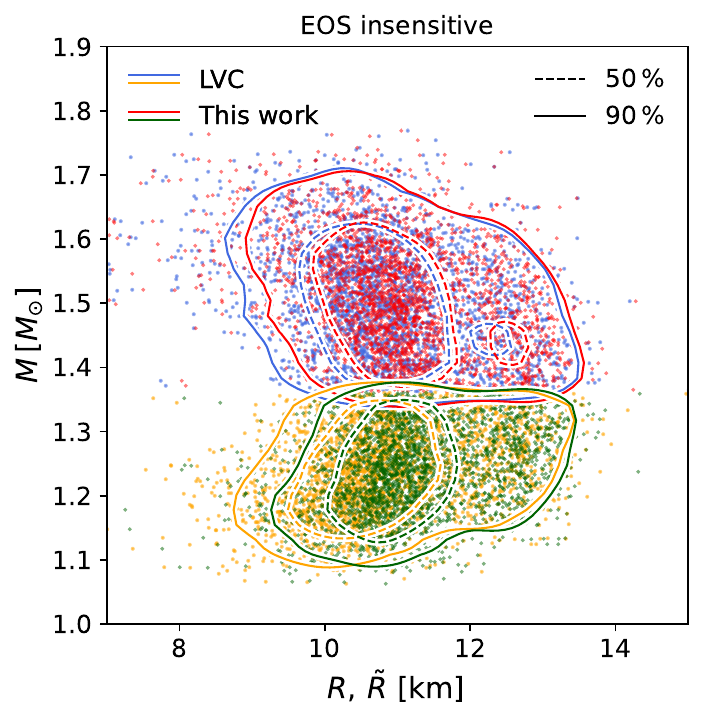} 
\includegraphics[width=0.325\textwidth]{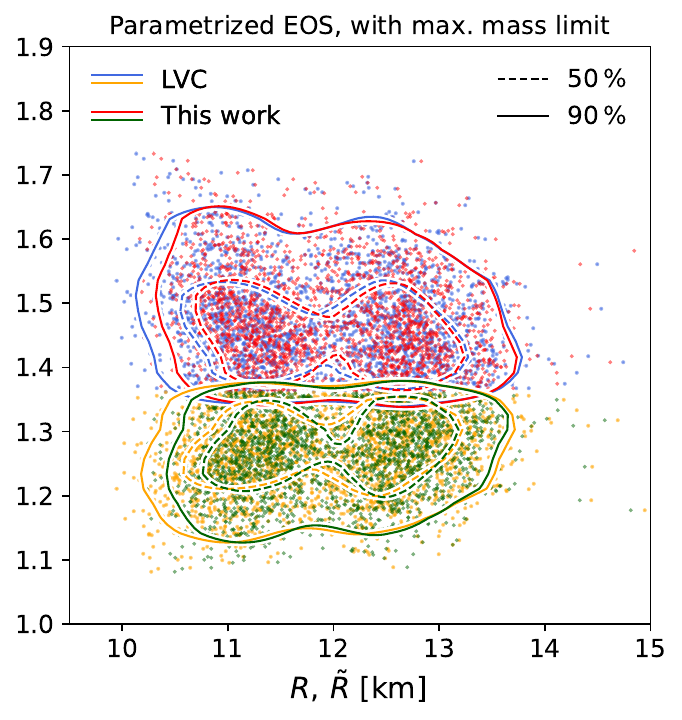}
\includegraphics[width=0.325\textwidth]{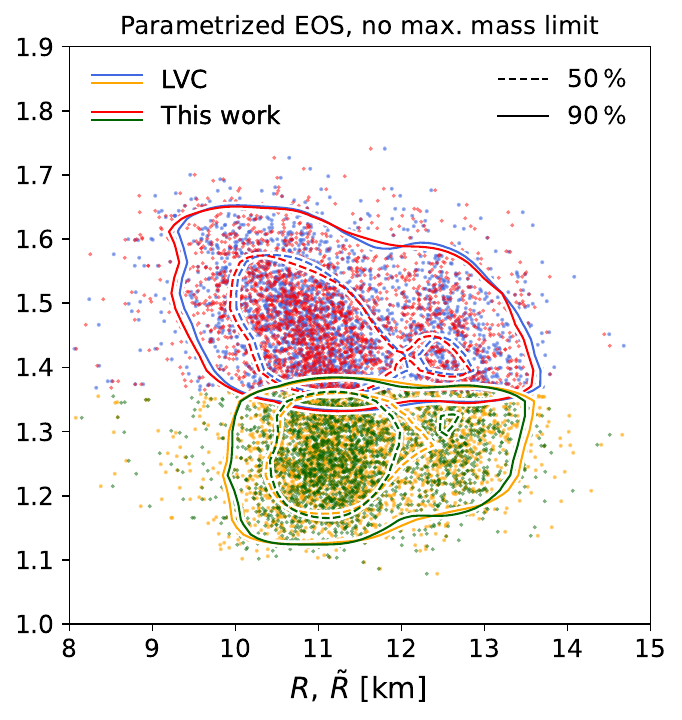}
  \caption{Comparison of the results from Fig. 3 of the LIGO-Virgo Collaboration \citep{2018PhRvL.121p1101A} and this work, presenting posteriors for the masses $M$, and inferred radii $R$ and $\rt$ for both BNS components of the GW170817 event, using EOS-insensitive relations (left panel), and parametrized EOS where a lower limit on the maximum mass was imposed (middle panel), or wasn't imposed (right panel); see \cite{2018PhRvL.121p1101A} for details. Dashed and solid curve regions represent 50\% and 90\% cumulative density regions.}
  \label{fig:gw_posteriors} 
\end{figure*}

\subsection{Radius recovery for GW170817 BNS components}
\label{sec:gw170817} 

The $\rt$ approximation was also applied to the data inferred from the GW170817 detection \cite{2017PhRvL.119p1101A}. \citet{2018PhRvL.121p1101A} in their Fig. 3 presented the recovered radii for both BNS components, using various assumptions about the EOS; see \citeauthor{P1800115} for the data behind the figure. Here, in Fig.~\ref{fig:gw_posteriors} I have compared the radii estimates in \cite{2018PhRvL.121p1101A} using EOS-insensitive relation, and relations derived from the analysis of parametrized EOS with and without imposing a lower limit on the maximum NS mass, with the $\rt(M, \Lambda)$ approximation using Eqs.~\ref{eq:rt} and \ref{eq:the_expression}, using the values of $M$ and $\Lambda$ from \citeauthor{P1800115}. 

In addition to individual points, $M(R)$ and $M(\rt)$, 50\% and 90\% cumulative density regions are computed using the kernel density estimation (KDE) method, with the \texttt{gaussian\_kde()} implementation from \texttt{scipy} \cite{2020SciPy-NMeth}.

Results in Fig.~\ref{fig:gw_posteriors} show that this approximation is comparable in precision with previously proposed relations. An added value is that an explicit (computationally-straightforward) form of the expression, directly returning values of radii from pairs of mass-tidal deformability values, is the simplest inference method possible, whereas other methods involve evaluating multiple relations obtained and tuned with various EOS datasets: to infer radius, one needs either to combine relations based on parameters of the binary, $\Lambda_s = (\Lambda_2+\Lambda_1)/2$, $\Lambda_a=(\Lambda_2-\Lambda_1)/2$, mass ratio $q=M_2/M_1$, and compactness $M/R$ as a function of $\Lambda$, or integrate the TOV equations to compute the radius for each sample in the spectral EOS parametrization (for the parametrized EOS case), see \cite{2018PhRvL.121p1101A} for details.

\section{Conclusions} 
\label{sec:conclusions}

This SR model discovery exploration provides an interpretable, explainable and computationally-inexpensive ML solution connecting two different information channels of multi-messenger astronomy: NS radius inference from GW-only measurements. The method takes as input individual $(M, \Lambda)$ points to infer corresponding radius approximation value $\rt$, i.e. it does not depend on the global structure of the $(M, \Lambda)$ or $(M, R)$ sequences. Although the model was trained on a set of piecewise polytropic EOSs, it shows substantial capabilities to extrapolate outside the training dataset parameter space. Additionally, it is largely EOS-independent: was tested on a selection of piecewise polytropic EOSs outside the training dataset, as well as  various non-polytropic EOSs, showing a reasonable robustness (predictability) of the output. 

Apart from stand alone usage (independent inference of radius values for individual $(M, \Lambda)$ measurements), it may be a pre-processing step before more refined parameter estimation methods, to e.g. constrain possible parameter space for NS sequences and underlying EOSs. Several improvements are possible and shall be addressed in the future. Among them is constraining the space of solutions towards manifestly physically-inspired results, and introducing extra information supplied at the training stage, but hidden (not available) during validation stage, e.g. on differential aspects of NS sequences for input features and the target features, like inclusion, in addition to point values of $M$, $R$ and $\Lambda$, also $\mathrm{d}\Lambda/\mathrm{d}M$ and $\mathrm{d}R/\mathrm{d}M$ behaviour in the loss function, which in general should improve the accuracy of the prediction. 

The data that support the findings of this article are openly available at \url{https://github.com/mbejger/pysr_r-as-mlambda}. 

\begin{acknowledgments}
This research was partially supported by the European Union-Next Generation EU, Mission 4 Component 1 CUP J53D23001550006 with the PRIN Project No. 202275HT58 and the Polish National Science Center OPUS grant no. 2021/43/B/ST9/01714. H. Du Bois is acknowledged for providing initial inspiration regarding research on the optimal expression. 
\end{acknowledgments}

\appendix

\section{Symbolic regression model details} 
\label{sec:app_model_details}
 
The \texttt{PySRRegressor} function is used to search for an optimal expression given the data and constraints. 
In this work, I have adopted addition \texttt{''+''}, subtraction \texttt{''-''}, multiplication \texttt{''*''}, division \texttt{''/''}, and a power function $\mathtt{base}^{\mathtt{exponent}}$ 
as binary operators. To constrain the complexity of the power function, the \texttt{base} is allowed to have any complexity, whereas the \texttt{exponent} has complexity of 1: a constant, or one of the independent input variables (features). In order to work with features of similar magnitude, input variables are $M$ and $\logl$. 

The objective of the search is to minimize the \texttt{elementwise\_loss} function, chosen here to be the MSE (\texttt{L2DistLoss()} in {\pysr}). Additionally, the search was encouraged to constrain all the constants to be positive.\footnote{How to constrain all the constants to be positive: \url{https://github.com/MilesCranmer/PySR/discussions/449}} The hyperparameters of the search related to how the data is divided into batches during the training, how many populations of which size are used, how deep the graphs may be, and how to punish complexity (with a multiplicative \texttt{parsimony} parameter decreasing the overall score for complex expressions) were selected with the \texttt{optuna} optimization framework \citep{akiba2019optuna} by demanding the smallest MSE on the validation dataset. They are (see the {\pysr} Application Programming Interface (API) \cite{pysr2025} for detailed description):
\begin{lstlisting}
    batching=True,              
    batch_size=4096,            
    niterations=100,            
    populations=150,            
    population_size=64,         
    ncycles_per_iteration=100,  
    model_selection=''best'',
    parsimony=1e-5,             
    maxdepth=7,
    turbo=True
\end{lstlisting}
Individual data instances - ($M, R, \kt, \Lambda$) 4-tuples described in Appendix~\ref{sec:app_data_prep_details} - are chosen at random from the dataset during the training, and provided to the \texttt{PySRRegressor} function in batches, so that no knowledge about NS sequences related to a given EOS is directly trained, to make sure that the expression works on average equally well for all NS sequences. From the point of view of practical usefulness, it is desirable that the expression is sufficiently accurate, but also as simple as possible. This means that one has to balance (optimize) the overall accuracy (loss) and complexity (the number of nodes in an expression tree, meaning the number of operators included in the expression) \cite{2023arXiv230501582C}; more complex expressions are often more accurate. The {\pysr} implementation defines the ''best'' expression as one with the highest score (negated derivative of the log-loss with respect to complexity) among expressions with a loss better than at least 1.5 times the value of the most accurate model, see the {\pysr} API \cite{pysr2025} for details. The \texttt{turbo=True} setting allows to use \texttt{julia} loop vectorisation \texttt{LoopVectorization.jl} for speed up. In terms of wall clock time, one regression run on a dataset described below takes approx. 20 minutes on an \texttt{AMD Ryzen 9 5950X} 16-core processor CPU. Table~\ref{tab:other_expressions} contains the ''hall of fame'' summary of symbolic expressions found in the search presented in this work, with the ''best'' expression, Eq.~\ref{eq:the_expression}, emphasized with top and bottom horizontal lines; expressions more complex than the chosen ''best'' one are not providing substantially better accuracy (loss).

\renewcommand{\arraystretch}{1.75}
\renewcommand{\thetable}{A}
\begin{table*}[t]
\resizebox{\textwidth}{!}{
\setlength{\tabcolsep}{2pt}
\begin{tabular}{cccc}
\toprule\toprule
Equation & Complexity & Loss & Score \\
\midrule
$y = 0.0718$ & $1$ & $0.000803$ & $0.0$ \\
$y = x_{1} \cdot 0.0273$ & $3$ & $0.000142$ & $0.866$ \\
$y = x_{1}^{0.0477} - 0.968$ & $5$ & $8.61 \cdot 10^{-5}$ & $0.250$ \\
$y = x_{0} x_{1}^{1.80} \cdot 0.0110$ & $7$ & $1.53 \cdot 10^{-5}$ & $0.863$ \\
$y = x_{0} x_{1}^{1.92} \cdot 0.00867 + 0.00864$ & $9$ & $1.04 \cdot 10^{-5}$ & $0.195$ \\
\begin{minipage}{0.8\linewidth} \vspace{-1em} \begin{dmath*} y = x_{1} \left(0.0327 - \left(- 0.00568 x_{1} + \frac{0.0219}{x_{0}}\right)\right) \end{dmath*} \end{minipage} & $11$ & $4.29 \cdot 10^{-6}$ & $0.442$ \\
\begin{minipage}{0.8\linewidth} \vspace{-1em} \begin{dmath*} y = \left(x_{1} + 0.110\right) \left(x_{1} \cdot 0.00629 - \left(\left(-1\right) 0.0300 + \frac{0.0222}{x_{0}}\right)\right) \end{dmath*} \end{minipage} & $13$ & $4.11 \cdot 10^{-6}$ & $0.0214$ \\ 
\toprule
\begin{minipage}{0.8\linewidth} \vspace{-1em} \begin{dmath*} y = x_{1} x_{1} \cdot 0.0351 \left(\frac{x_{0}}{x_{0} + 2.05} - 0.109\right) + 0.0137 \end{dmath*} \end{minipage} & $15$ & $2.68 \cdot 10^{-6}$ & $0.213$ \\ \bottomrule 
\begin{minipage}{0.8\linewidth} \vspace{-1em} \begin{dmath*} y = \left(\frac{x_{0}}{x_{0} + 1.80} - 0.123\right) \left(x_{1} + 0.142\right) x_{1} \cdot 0.0318 + 0.0126 \end{dmath*} \end{minipage} & $17$ & $2.63 \cdot 10^{-6}$ & $0.00946$ \\
\begin{minipage}{0.8\linewidth} \vspace{-1em} \begin{dmath*} y = x_{1} \left(x_{1} x_{0} \cdot 0.0262 \frac{1}{x_{0} + 2.48} - \left(\left(-1\right) 0.00906 + \frac{0.00766}{x_{0}}\right)\right) + 0.0104 \end{dmath*} \end{minipage} & $19$ & $2.52 \cdot 10^{-6}$ & $0.0220$ \\
\begin{minipage}{0.8\linewidth} \vspace{-1em} \begin{dmath*} y = x_{1} \left(x_{1} x_{0} \cdot 0.0270 \frac{1}{x_{0} + 2.48} - \left(\left(-1\right) 0.00929 + \frac{0.00948}{x_{0} + 0.0938}\right)\right) + 0.0104 \end{dmath*} \end{minipage} & $21$ & $2.51 \cdot 10^{-6}$ & $0.00113$ \\
\begin{minipage}{0.8\linewidth} \vspace{-1em} \begin{dmath*} y = x_{1} \left(\frac{0.0287 x_{0} x_{1}}{x_{0} + 2.74} - \left(\left(-1\right) 0.00744 + \frac{0.00724}{x_{0}}\right)\right) + \frac{0.0946}{x_{0} + 6.25} \end{dmath*} \end{minipage} & $23$ & $2.50 \cdot 10^{-6}$ & $0.00290$ \\
\begin{minipage}{0.8\linewidth} \vspace{-1em} \begin{dmath*} y = x_{1} \left(- (\left(-1\right) 0.00778 + \frac{0.00666}{x_{0}}) + \frac{0.0298 \left(x_{0} x_{1} - 0.261\right)}{x_{0} + 2.77}\right) + \frac{0.0817}{x_{0} + 4.76} \end{dmath*} \end{minipage} & $25$ & $2.49 \cdot 10^{-6}$ & $0.00152$ \\
\begin{minipage}{0.8\linewidth} \vspace{-1em} \begin{dmath*} y = x_{1} \left(\frac{0.0296 x_{0} x_{1}}{x_{0} + 2.74} - \left(\left(-1\right) 0.00592 + \frac{0.00702}{x_{0}}\right)\right) + \frac{0.0936}{{0.106}^{x_{1}} + x_{0} + 5.32} \end{dmath*} \end{minipage} & $27$ & $2.48 \cdot 10^{-6}$ & $0.000874$ \\
\begin{minipage}{0.8\linewidth} \vspace{-1em} \begin{dmath*} y = x_{1} \left(\frac{0.0304 x_{0} x_{1}}{x_{0} + 2.74} - \left(\left(-1\right) 0.00458 + \frac{0.00677}{x_{0}}\right)\right) + \frac{0.101}{\frac{{0.106}^{x_{1}}}{x_{1}} + x_{0} + 5.32} \end{dmath*} \end{minipage} & $29$ & $2.47 \cdot 10^{-6}$ & $0.00197$ \\
\bottomrule\bottomrule
\label{tab:other_expressions}
\end{tabular}
}
\caption{The ''hall of fame'' of symbolic expressions found using the dataset described in Sec.~\ref{sec:data} and the Appendix~\ref{sec:app_data_prep_details}. The table is a verbatim output of \texttt{model.latex_table()} from the {\pysr} API \cite{pysr2025}: $y$ denotes $\kt$ ($k_2$ approximation), $x_0$ is $M/M_\odot$ and $x_1$ is $\log_{10}\Lambda$. The complexity column describes the complexity of the final expression (1 for a constant or a single variable, 3 for expression like $a+b$, $a\cdot b$ etc.) Loss is the \texttt{L2DistLoss()} (MSE loss). Score is defined as the negated derivative of the log-loss with respect to complexity. The ''best'' equation with complexity 15 (Eq.~\ref{eq:the_expression}) is featured between horizontal lines. See \url{https://github.com/mbejger/pysr_r-as-mlambda} for the computed model, and the {\pysr} API \cite{pysr2025} for details of the implementation.} 
\end{table*}
 
\section{Data preparation details} 
\label{sec:app_data_prep_details} 

\begin{figure}[ht]
\centering
\includegraphics[width=\hsize]{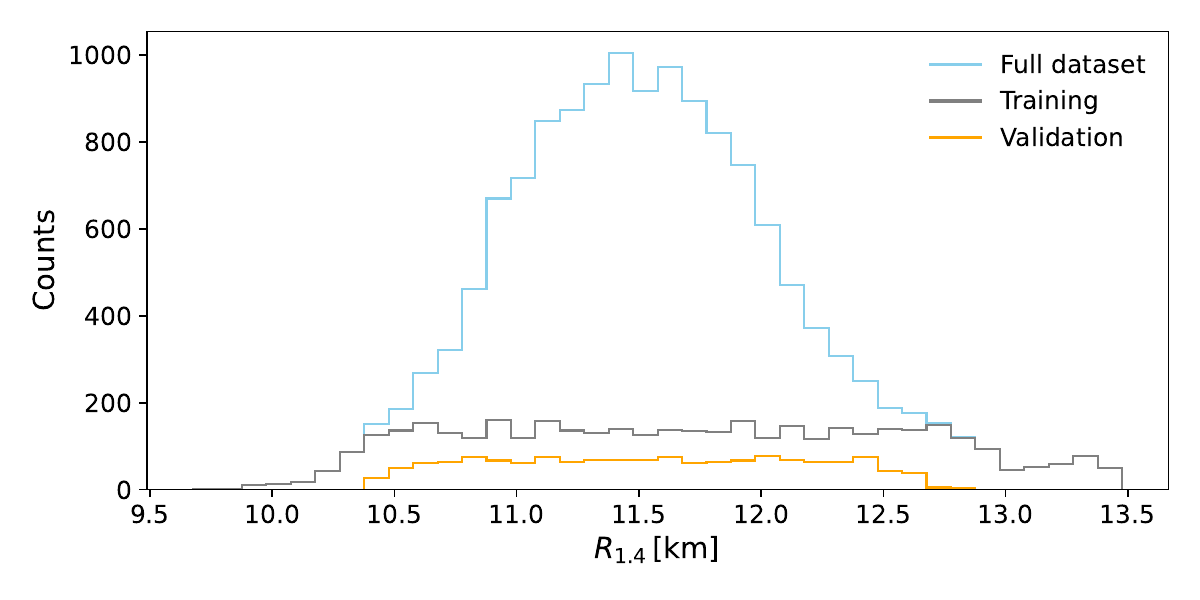}
    \caption{Histograms of $R_{1.4}$, NS radius at $M=1.4\,M_\odot$, values for the training and validation dataset compared with the original dataset \cite{2020A&A...642A..78M}.}
    \label{fig:data_sampling}
\end{figure}

The data from \cite{2020A&A...642A..78M} is originally uniformly distributed from the point of view of the piecewise-polytropic parameters describing the EOSs, which results in non-uniform distribution of the TOV sequences of NS global parameters. A distribution of $R_{1.4}$, NS radius at $M=1.4\,M_\odot$, is presented in Fig.~\ref{fig:data_sampling}. 
In order to make the training and validation datasets more uniform in terms of distribution of their TOV sequences in the $M(R)$ plane, I have divided the range of $R_{1.4} \in (9.5, 13.5)$ km into $N=20$ bins, and selectively sampled the original distribution of TOV ($M, R, k_2, \Lambda$) sequences. 

For the training dataset, each bin contains 250 randomly-selected TOV sequences. In case of very small and very large radii, bins didn't contain 250 instances, therefore all instances for those particular bins were selected. Similarly, for the validation dataset, 100 randomly-selected TOV sequences were chosen from the remaining data. The training and validation datasets are composed of 3938 and 1423 TOV sequences. Each TOV sequence consists of 50 4-tuples of ($M, R, k_2, \Lambda$), obtained for a range of masses between $0.2\,M_\odot$ and the respective EOS $M_{max}$. For the training and validation datasets, only ($M, R, k_2, \Lambda$) 4-tuples with $M>0.5\,M_\odot$ and $M_{max} > 1.9\,M_\odot$ were selected from the \cite{2020A&A...642A..78M} dataset. They are presented in a form of $M(R)$ and $\logl(M)$ sequences in Fig.~\ref{fig:mr_dataset}. The training and validation datasets are composed of 192205 and 71091 ($M, R, k_2, \Lambda$) 4-tuples, respectively. 

\bibliography{references} 

\begin{thebibliography}{57}%
\makeatletter
\providecommand \@ifxundefined [1]{%
 \@ifx{#1\undefined}
}%
\providecommand \@ifnum [1]{%
 \ifnum #1\expandafter \@firstoftwo
 \else \expandafter \@secondoftwo
 \fi
}%
\providecommand \@ifx [1]{%
 \ifx #1\expandafter \@firstoftwo
 \else \expandafter \@secondoftwo
 \fi
}%
\providecommand \natexlab [1]{#1}%
\providecommand \enquote  [1]{``#1''}%
\providecommand \bibnamefont  [1]{#1}%
\providecommand \bibfnamefont [1]{#1}%
\providecommand \citenamefont [1]{#1}%
\providecommand \href@noop [0]{\@secondoftwo}%
\providecommand \href [0]{\begingroup \@sanitize@url \@href}%
\providecommand \@href[1]{\@@startlink{#1}\@@href}%
\providecommand \@@href[1]{\endgroup#1\@@endlink}%
\providecommand \@sanitize@url [0]{\catcode `\\12\catcode `\$12\catcode
  `\&12\catcode `\#12\catcode `\^12\catcode `\_12\catcode `\%12\relax}%
\providecommand \@@startlink[1]{}%
\providecommand \@@endlink[0]{}%
\providecommand \url  [0]{\begingroup\@sanitize@url \@url }%
\providecommand \@url [1]{\endgroup\@href {#1}{\urlprefix }}%
\providecommand \urlprefix  [0]{URL }%
\providecommand \Eprint [0]{\href }%
\providecommand \doibase [0]{https://doi.org/}%
\providecommand \selectlanguage [0]{\@gobble}%
\providecommand \bibinfo  [0]{\@secondoftwo}%
\providecommand \bibfield  [0]{\@secondoftwo}%
\providecommand \translation [1]{[#1]}%
\providecommand \BibitemOpen [0]{}%
\providecommand \bibitemStop [0]{}%
\providecommand \bibitemNoStop [0]{.\EOS\space}%
\providecommand \EOS [0]{\spacefactor3000\relax}%
\providecommand \BibitemShut  [1]{\csname bibitem#1\endcsname}%
\let\auto@bib@innerbib\@empty
\bibitem [{\citenamefont {{Abbott}}\ \emph
  {et~al.}(2017{\natexlab{a}})\citenamefont {{Abbott}}, \citenamefont
  {{Abbott}}, \citenamefont {{Abbott}}, \citenamefont {{Acernese}},
  \citenamefont {{Ackley}}, \citenamefont {{Adams}}, \citenamefont {{Adams}},
  \citenamefont {{Addesso}}, \citenamefont {{Adhikari}}, \citenamefont
  {{Adya}}, \citenamefont {{Affeldt}}, \citenamefont {{Afrough}}, \citenamefont
  {{Agarwal}}, \citenamefont {{Agathos}}, \citenamefont {{Agatsuma}},
  \citenamefont {{Aggarwal}}, \citenamefont {{Aguiar}}, \citenamefont
  {{Aiello}}, \citenamefont {{Ain}}, \citenamefont {{Ajith}}, \citenamefont
  {{Allen}}, \citenamefont {{Allen}}, \citenamefont {{Allocca}}, \citenamefont
  {{Altin}}, \citenamefont {{Amato}}, \citenamefont {{Ananyeva}}, \citenamefont
  {{Anderson}}, \citenamefont {{Anderson}}, \citenamefont {{Angelova}},
  \citenamefont {{Antier}}, \citenamefont {{Appert}}, \citenamefont {{Arai}},
  \citenamefont {{Araya}}, \citenamefont {{Areeda}}, \citenamefont {{Arnaud}},
  \citenamefont {{Arun}}, \citenamefont {{Ascenzi}}, \citenamefont {{Ashton}},
  \citenamefont {{Ast}}, \citenamefont {{Aston}}, \citenamefont {{Astone}},
  \citenamefont {{Atallah}}, \citenamefont {{Aufmuth}}, \citenamefont
  {{Aulbert}}, \citenamefont {{AultONeal}}, \citenamefont {{Austin}},
  \citenamefont {{Avila-Alvarez}}, \citenamefont {{Babak}}, \citenamefont
  {{Bacon}}, \citenamefont {{Bader}}, \citenamefont {{Bae}}, \citenamefont
  {{Bailes}}, \citenamefont {{Baker}}, \citenamefont {{Baldaccini}},
  \citenamefont {{Ballardin}}, \citenamefont {{Ballmer}}, \citenamefont
  {{Banagiri}}, \citenamefont {{Barayoga}}, \citenamefont {{Barclay}},
  \citenamefont {{Barish}}, \citenamefont {{Barker}}, \citenamefont
  {{Barkett}}, \citenamefont {{Barone}}, \citenamefont {{Barr}}, \citenamefont
  {{Barsotti}}, \citenamefont {{Barsuglia}}, \citenamefont {{Barta}},
  \citenamefont {{Barthelmy}}, \citenamefont {{Bartlett}}, \citenamefont
  {{Bartos}}, \citenamefont {{Bassiri}}, \citenamefont {{Basti}}, \citenamefont
  {{Batch}}, \citenamefont {{Bawaj}}, \citenamefont {{Bayley}}, \citenamefont
  {{Bazzan}}, \citenamefont {{B{\'e}csy}}, \citenamefont {{Beer}},
  \citenamefont {{Bejger}}, \citenamefont {{Belahcene}}, \citenamefont
  {{Bell}}, \citenamefont {{Berger}}, \citenamefont {{Bergmann}}, \citenamefont
  {{Bernuzzi}}, \citenamefont {{Bero}}, \citenamefont {{Berry}}, \citenamefont
  {{Bersanetti}}, \citenamefont {{Bertolini}}, \citenamefont {{Betzwieser}},
  \citenamefont {{Bhagwat}}, \citenamefont {{Bhandare}}, \citenamefont
  {{Bilenko}}, \citenamefont {{Billingsley}}, \citenamefont {{Billman}},
  \citenamefont {{Birch}}, \citenamefont {{Birney}}, \citenamefont
  {{Birnholtz}}, \citenamefont {{Biscans}}, \citenamefont {{Biscoveanu}},
  \citenamefont {{Bisht}}, \citenamefont {{Bitossi}}, \citenamefont {{Biwer}},
  \citenamefont {{Bizouard}}, \citenamefont {{Blackburn}}, \citenamefont
  {{Blackman}}, \citenamefont {{Blair}}, \citenamefont {{Blair}}, \citenamefont
  {{Blair}}, \citenamefont {{Bloemen}}, \citenamefont {{Bock}}, \citenamefont
  {{Bode}}, \citenamefont {{Boer}}, \citenamefont {{Bogaert}}, \citenamefont
  {{Bohe}}, \citenamefont {{Bondu}}, \citenamefont {{Bonilla}}, \citenamefont
  {{Bonnand}}, \citenamefont {{Boom}}, \citenamefont {{Bork}}, \citenamefont
  {{Boschi}}, \citenamefont {{Bose}}, \citenamefont {{Bossie}}, \citenamefont
  {{Bouffanais}}, \citenamefont {{Bozzi}}, \citenamefont {{Bradaschia}},
  \citenamefont {{Brady}}, \citenamefont {{Branchesi}}, \citenamefont {{Brau}},
  \citenamefont {{Briant}}, \citenamefont {{Brillet}}, \citenamefont
  {{Brinkmann}}, \citenamefont {{Brisson}}, \citenamefont {{Brockill}},
  \citenamefont {{Broida}}, \citenamefont {{Brooks}}, \citenamefont {{Brown}},
  \citenamefont {{Brown}}, \citenamefont {{Brunett}}, \citenamefont
  {{Buchanan}}, \citenamefont {{Buikema}}, \citenamefont {{Bulik}},
  \citenamefont {{Bulten}}, \citenamefont {{Buonanno}}, \citenamefont
  {{Buskulic}}, \citenamefont {{Buy}}, \citenamefont {{Byer}}, \citenamefont
  {{Cabero}}, \citenamefont {{Cadonati}}, \citenamefont {{Cagnoli}},
  \citenamefont {{Cahillane}}, \citenamefont {{Calder{\'o}n Bustillo}},
  \citenamefont {{Callister}}, \citenamefont {{Calloni}}, \citenamefont
  {{Camp}}, \citenamefont {{Canepa}}, \citenamefont {{Canizares}},
  \citenamefont {{Cannon}}, \citenamefont {{Cao}}, \citenamefont {{Cao}},
  \citenamefont {{Capano}}, \citenamefont {{Capocasa}}, \citenamefont
  {{Carbognani}}, \citenamefont {{Caride}}, \citenamefont {{Carney}},
  \citenamefont {{Carullo}}, \citenamefont {{Casanueva Diaz}}, \citenamefont
  {{Casentini}}, \citenamefont {{Caudill}}, \citenamefont {{Cavagli{\`a}}},
  \citenamefont {{Cavalier}}, \citenamefont {{Cavalieri}}, \citenamefont
  {{Cella}}, \citenamefont {{Cepeda}}, \citenamefont {{Cerd{\'a}-Dur{\'a}n}},
  \citenamefont {{Cerretani}}, \citenamefont {{Cesarini}}, \citenamefont
  {{Chamberlin}}, \citenamefont {{Chan}}, \citenamefont {{Chao}}, \citenamefont
  {{Charlton}}, \citenamefont {{Chase}}, \citenamefont {{Chassande-Mottin}},
  \citenamefont {{Chatterjee}}, \citenamefont {{Chatziioannou}}, \citenamefont
  {{Cheeseboro}}, \citenamefont {{Chen}}, \citenamefont {{Chen}}, \citenamefont
  {{Chen}}, \citenamefont {{Cheng}}, \citenamefont {{Chia}}, \citenamefont
  {{Chincarini}}, \citenamefont {{Chiummo}}, \citenamefont {{Chmiel}},
  \citenamefont {{Cho}}, \citenamefont {{Cho}}, \citenamefont {{Chow}},
  \citenamefont {{Christensen}}, \citenamefont {{Chu}}, \citenamefont
  {{Chua}},\ and\ \citenamefont {{Chua}}}]{2017PhRvL.119p1101A}%
  \BibitemOpen
  \bibfield  {author} {\bibinfo {author} {\bibfnamefont {B.~P.}\ \bibnamefont
  {{Abbott}}}, \bibinfo {author} {\bibfnamefont {R.}~\bibnamefont {{Abbott}}},
  \bibinfo {author} {\bibfnamefont {T.~D.}\ \bibnamefont {{Abbott}}}, \bibinfo
  {author} {\bibfnamefont {F.}~\bibnamefont {{Acernese}}}, \bibinfo {author}
  {\bibfnamefont {K.}~\bibnamefont {{Ackley}}}, \bibinfo {author}
  {\bibfnamefont {C.}~\bibnamefont {{Adams}}}, \bibinfo {author} {\bibfnamefont
  {T.}~\bibnamefont {{Adams}}}, \bibinfo {author} {\bibfnamefont
  {P.}~\bibnamefont {{Addesso}}}, \bibinfo {author} {\bibfnamefont {R.~X.}\
  \bibnamefont {{Adhikari}}}, \bibinfo {author} {\bibfnamefont {V.~B.}\
  \bibnamefont {{Adya}}}, \bibinfo {author} {\bibfnamefont {C.}~\bibnamefont
  {{Affeldt}}}, \bibinfo {author} {\bibfnamefont {M.}~\bibnamefont
  {{Afrough}}}, \bibinfo {author} {\bibfnamefont {B.}~\bibnamefont
  {{Agarwal}}}, \bibinfo {author} {\bibfnamefont {M.}~\bibnamefont
  {{Agathos}}}, \bibinfo {author} {\bibfnamefont {K.}~\bibnamefont
  {{Agatsuma}}}, \bibinfo {author} {\bibfnamefont {N.}~\bibnamefont
  {{Aggarwal}}}, \bibinfo {author} {\bibfnamefont {O.~D.}\ \bibnamefont
  {{Aguiar}}}, \bibinfo {author} {\bibfnamefont {L.}~\bibnamefont {{Aiello}}},
  \bibinfo {author} {\bibfnamefont {A.}~\bibnamefont {{Ain}}}, \bibinfo
  {author} {\bibfnamefont {P.}~\bibnamefont {{Ajith}}}, \bibinfo {author}
  {\bibfnamefont {B.}~\bibnamefont {{Allen}}}, \bibinfo {author} {\bibfnamefont
  {G.}~\bibnamefont {{Allen}}}, \bibinfo {author} {\bibfnamefont
  {A.}~\bibnamefont {{Allocca}}}, \bibinfo {author} {\bibfnamefont {P.~A.}\
  \bibnamefont {{Altin}}}, \bibinfo {author} {\bibfnamefont {A.}~\bibnamefont
  {{Amato}}}, \bibinfo {author} {\bibfnamefont {A.}~\bibnamefont {{Ananyeva}}},
  \bibinfo {author} {\bibfnamefont {S.~B.}\ \bibnamefont {{Anderson}}},
  \bibinfo {author} {\bibfnamefont {W.~G.}\ \bibnamefont {{Anderson}}},
  \bibinfo {author} {\bibfnamefont {S.~V.}\ \bibnamefont {{Angelova}}},
  \bibinfo {author} {\bibfnamefont {S.}~\bibnamefont {{Antier}}}, \bibinfo
  {author} {\bibfnamefont {S.}~\bibnamefont {{Appert}}}, \bibinfo {author}
  {\bibfnamefont {K.}~\bibnamefont {{Arai}}}, \bibinfo {author} {\bibfnamefont
  {M.~C.}\ \bibnamefont {{Araya}}}, \bibinfo {author} {\bibfnamefont {J.~S.}\
  \bibnamefont {{Areeda}}}, \bibinfo {author} {\bibfnamefont {N.}~\bibnamefont
  {{Arnaud}}}, \bibinfo {author} {\bibfnamefont {K.~G.}\ \bibnamefont
  {{Arun}}}, \bibinfo {author} {\bibfnamefont {S.}~\bibnamefont {{Ascenzi}}},
  \bibinfo {author} {\bibfnamefont {G.}~\bibnamefont {{Ashton}}}, \bibinfo
  {author} {\bibfnamefont {M.}~\bibnamefont {{Ast}}}, \bibinfo {author}
  {\bibfnamefont {S.~M.}\ \bibnamefont {{Aston}}}, \bibinfo {author}
  {\bibfnamefont {P.}~\bibnamefont {{Astone}}}, \bibinfo {author}
  {\bibfnamefont {D.~V.}\ \bibnamefont {{Atallah}}}, \bibinfo {author}
  {\bibfnamefont {P.}~\bibnamefont {{Aufmuth}}}, \bibinfo {author}
  {\bibfnamefont {C.}~\bibnamefont {{Aulbert}}}, \bibinfo {author}
  {\bibfnamefont {K.}~\bibnamefont {{AultONeal}}}, \bibinfo {author}
  {\bibfnamefont {C.}~\bibnamefont {{Austin}}}, \bibinfo {author}
  {\bibfnamefont {A.}~\bibnamefont {{Avila-Alvarez}}}, \bibinfo {author}
  {\bibfnamefont {S.}~\bibnamefont {{Babak}}}, \bibinfo {author} {\bibfnamefont
  {P.}~\bibnamefont {{Bacon}}}, \bibinfo {author} {\bibfnamefont {M.~K.~M.}\
  \bibnamefont {{Bader}}}, \bibinfo {author} {\bibfnamefont {S.}~\bibnamefont
  {{Bae}}}, \bibinfo {author} {\bibfnamefont {M.}~\bibnamefont {{Bailes}}},
  \bibinfo {author} {\bibfnamefont {P.~T.}\ \bibnamefont {{Baker}}}, \bibinfo
  {author} {\bibfnamefont {F.}~\bibnamefont {{Baldaccini}}}, \bibinfo {author}
  {\bibfnamefont {G.}~\bibnamefont {{Ballardin}}}, \bibinfo {author}
  {\bibfnamefont {S.~W.}\ \bibnamefont {{Ballmer}}}, \bibinfo {author}
  {\bibfnamefont {S.}~\bibnamefont {{Banagiri}}}, \bibinfo {author}
  {\bibfnamefont {J.~C.}\ \bibnamefont {{Barayoga}}}, \bibinfo {author}
  {\bibfnamefont {S.~E.}\ \bibnamefont {{Barclay}}}, \bibinfo {author}
  {\bibfnamefont {B.~C.}\ \bibnamefont {{Barish}}}, \bibinfo {author}
  {\bibfnamefont {D.}~\bibnamefont {{Barker}}}, \bibinfo {author}
  {\bibfnamefont {K.}~\bibnamefont {{Barkett}}}, \bibinfo {author}
  {\bibfnamefont {F.}~\bibnamefont {{Barone}}}, \bibinfo {author}
  {\bibfnamefont {B.}~\bibnamefont {{Barr}}}, \bibinfo {author} {\bibfnamefont
  {L.}~\bibnamefont {{Barsotti}}}, \bibinfo {author} {\bibfnamefont
  {M.}~\bibnamefont {{Barsuglia}}}, \bibinfo {author} {\bibfnamefont
  {D.}~\bibnamefont {{Barta}}}, \bibinfo {author} {\bibfnamefont {S.~D.}\
  \bibnamefont {{Barthelmy}}}, \bibinfo {author} {\bibfnamefont
  {J.}~\bibnamefont {{Bartlett}}}, \bibinfo {author} {\bibfnamefont
  {I.}~\bibnamefont {{Bartos}}}, \bibinfo {author} {\bibfnamefont
  {R.}~\bibnamefont {{Bassiri}}}, \bibinfo {author} {\bibfnamefont
  {A.}~\bibnamefont {{Basti}}}, \bibinfo {author} {\bibfnamefont {J.~C.}\
  \bibnamefont {{Batch}}}, \bibinfo {author} {\bibfnamefont {M.}~\bibnamefont
  {{Bawaj}}}, \bibinfo {author} {\bibfnamefont {J.~C.}\ \bibnamefont
  {{Bayley}}}, \bibinfo {author} {\bibfnamefont {M.}~\bibnamefont {{Bazzan}}},
  \bibinfo {author} {\bibfnamefont {B.}~\bibnamefont {{B{\'e}csy}}}, \bibinfo
  {author} {\bibfnamefont {C.}~\bibnamefont {{Beer}}}, \bibinfo {author}
  {\bibfnamefont {M.}~\bibnamefont {{Bejger}}}, \bibinfo {author}
  {\bibfnamefont {I.}~\bibnamefont {{Belahcene}}}, \bibinfo {author}
  {\bibfnamefont {A.~S.}\ \bibnamefont {{Bell}}}, \bibinfo {author}
  {\bibfnamefont {B.~K.}\ \bibnamefont {{Berger}}}, \bibinfo {author}
  {\bibfnamefont {G.}~\bibnamefont {{Bergmann}}}, \bibinfo {author}
  {\bibfnamefont {S.}~\bibnamefont {{Bernuzzi}}}, \bibinfo {author}
  {\bibfnamefont {J.~J.}\ \bibnamefont {{Bero}}}, \bibinfo {author}
  {\bibfnamefont {C.~P.~L.}\ \bibnamefont {{Berry}}}, \bibinfo {author}
  {\bibfnamefont {D.}~\bibnamefont {{Bersanetti}}}, \bibinfo {author}
  {\bibfnamefont {A.}~\bibnamefont {{Bertolini}}}, \bibinfo {author}
  {\bibfnamefont {J.}~\bibnamefont {{Betzwieser}}}, \bibinfo {author}
  {\bibfnamefont {S.}~\bibnamefont {{Bhagwat}}}, \bibinfo {author}
  {\bibfnamefont {R.}~\bibnamefont {{Bhandare}}}, \bibinfo {author}
  {\bibfnamefont {I.~A.}\ \bibnamefont {{Bilenko}}}, \bibinfo {author}
  {\bibfnamefont {G.}~\bibnamefont {{Billingsley}}}, \bibinfo {author}
  {\bibfnamefont {C.~R.}\ \bibnamefont {{Billman}}}, \bibinfo {author}
  {\bibfnamefont {J.}~\bibnamefont {{Birch}}}, \bibinfo {author} {\bibfnamefont
  {R.}~\bibnamefont {{Birney}}}, \bibinfo {author} {\bibfnamefont
  {O.}~\bibnamefont {{Birnholtz}}}, \bibinfo {author} {\bibfnamefont
  {S.}~\bibnamefont {{Biscans}}}, \bibinfo {author} {\bibfnamefont
  {S.}~\bibnamefont {{Biscoveanu}}}, \bibinfo {author} {\bibfnamefont
  {A.}~\bibnamefont {{Bisht}}}, \bibinfo {author} {\bibfnamefont
  {M.}~\bibnamefont {{Bitossi}}}, \bibinfo {author} {\bibfnamefont
  {C.}~\bibnamefont {{Biwer}}}, \bibinfo {author} {\bibfnamefont {M.~A.}\
  \bibnamefont {{Bizouard}}}, \bibinfo {author} {\bibfnamefont {J.~K.}\
  \bibnamefont {{Blackburn}}}, \bibinfo {author} {\bibfnamefont
  {J.}~\bibnamefont {{Blackman}}}, \bibinfo {author} {\bibfnamefont {C.~D.}\
  \bibnamefont {{Blair}}}, \bibinfo {author} {\bibfnamefont {D.~G.}\
  \bibnamefont {{Blair}}}, \bibinfo {author} {\bibfnamefont {R.~M.}\
  \bibnamefont {{Blair}}}, \bibinfo {author} {\bibfnamefont {S.}~\bibnamefont
  {{Bloemen}}}, \bibinfo {author} {\bibfnamefont {O.}~\bibnamefont {{Bock}}},
  \bibinfo {author} {\bibfnamefont {N.}~\bibnamefont {{Bode}}}, \bibinfo
  {author} {\bibfnamefont {M.}~\bibnamefont {{Boer}}}, \bibinfo {author}
  {\bibfnamefont {G.}~\bibnamefont {{Bogaert}}}, \bibinfo {author}
  {\bibfnamefont {A.}~\bibnamefont {{Bohe}}}, \bibinfo {author} {\bibfnamefont
  {F.}~\bibnamefont {{Bondu}}}, \bibinfo {author} {\bibfnamefont
  {E.}~\bibnamefont {{Bonilla}}}, \bibinfo {author} {\bibfnamefont
  {R.}~\bibnamefont {{Bonnand}}}, \bibinfo {author} {\bibfnamefont {B.~A.}\
  \bibnamefont {{Boom}}}, \bibinfo {author} {\bibfnamefont {R.}~\bibnamefont
  {{Bork}}}, \bibinfo {author} {\bibfnamefont {V.}~\bibnamefont {{Boschi}}},
  \bibinfo {author} {\bibfnamefont {S.}~\bibnamefont {{Bose}}}, \bibinfo
  {author} {\bibfnamefont {K.}~\bibnamefont {{Bossie}}}, \bibinfo {author}
  {\bibfnamefont {Y.}~\bibnamefont {{Bouffanais}}}, \bibinfo {author}
  {\bibfnamefont {A.}~\bibnamefont {{Bozzi}}}, \bibinfo {author} {\bibfnamefont
  {C.}~\bibnamefont {{Bradaschia}}}, \bibinfo {author} {\bibfnamefont {P.~R.}\
  \bibnamefont {{Brady}}}, \bibinfo {author} {\bibfnamefont {M.}~\bibnamefont
  {{Branchesi}}}, \bibinfo {author} {\bibfnamefont {J.~E.}\ \bibnamefont
  {{Brau}}}, \bibinfo {author} {\bibfnamefont {T.}~\bibnamefont {{Briant}}},
  \bibinfo {author} {\bibfnamefont {A.}~\bibnamefont {{Brillet}}}, \bibinfo
  {author} {\bibfnamefont {M.}~\bibnamefont {{Brinkmann}}}, \bibinfo {author}
  {\bibfnamefont {V.}~\bibnamefont {{Brisson}}}, \bibinfo {author}
  {\bibfnamefont {P.}~\bibnamefont {{Brockill}}}, \bibinfo {author}
  {\bibfnamefont {J.~E.}\ \bibnamefont {{Broida}}}, \bibinfo {author}
  {\bibfnamefont {A.~F.}\ \bibnamefont {{Brooks}}}, \bibinfo {author}
  {\bibfnamefont {D.~A.}\ \bibnamefont {{Brown}}}, \bibinfo {author}
  {\bibfnamefont {D.~D.}\ \bibnamefont {{Brown}}}, \bibinfo {author}
  {\bibfnamefont {S.}~\bibnamefont {{Brunett}}}, \bibinfo {author}
  {\bibfnamefont {C.~C.}\ \bibnamefont {{Buchanan}}}, \bibinfo {author}
  {\bibfnamefont {A.}~\bibnamefont {{Buikema}}}, \bibinfo {author}
  {\bibfnamefont {T.}~\bibnamefont {{Bulik}}}, \bibinfo {author} {\bibfnamefont
  {H.~J.}\ \bibnamefont {{Bulten}}}, \bibinfo {author} {\bibfnamefont
  {A.}~\bibnamefont {{Buonanno}}}, \bibinfo {author} {\bibfnamefont
  {D.}~\bibnamefont {{Buskulic}}}, \bibinfo {author} {\bibfnamefont
  {C.}~\bibnamefont {{Buy}}}, \bibinfo {author} {\bibfnamefont {R.~L.}\
  \bibnamefont {{Byer}}}, \bibinfo {author} {\bibfnamefont {M.}~\bibnamefont
  {{Cabero}}}, \bibinfo {author} {\bibfnamefont {L.}~\bibnamefont
  {{Cadonati}}}, \bibinfo {author} {\bibfnamefont {G.}~\bibnamefont
  {{Cagnoli}}}, \bibinfo {author} {\bibfnamefont {C.}~\bibnamefont
  {{Cahillane}}}, \bibinfo {author} {\bibfnamefont {J.}~\bibnamefont
  {{Calder{\'o}n Bustillo}}}, \bibinfo {author} {\bibfnamefont {T.~A.}\
  \bibnamefont {{Callister}}}, \bibinfo {author} {\bibfnamefont
  {E.}~\bibnamefont {{Calloni}}}, \bibinfo {author} {\bibfnamefont {J.~B.}\
  \bibnamefont {{Camp}}}, \bibinfo {author} {\bibfnamefont {M.}~\bibnamefont
  {{Canepa}}}, \bibinfo {author} {\bibfnamefont {P.}~\bibnamefont
  {{Canizares}}}, \bibinfo {author} {\bibfnamefont {K.~C.}\ \bibnamefont
  {{Cannon}}}, \bibinfo {author} {\bibfnamefont {H.}~\bibnamefont {{Cao}}},
  \bibinfo {author} {\bibfnamefont {J.}~\bibnamefont {{Cao}}}, \bibinfo
  {author} {\bibfnamefont {C.~D.}\ \bibnamefont {{Capano}}}, \bibinfo {author}
  {\bibfnamefont {E.}~\bibnamefont {{Capocasa}}}, \bibinfo {author}
  {\bibfnamefont {F.}~\bibnamefont {{Carbognani}}}, \bibinfo {author}
  {\bibfnamefont {S.}~\bibnamefont {{Caride}}}, \bibinfo {author}
  {\bibfnamefont {M.~F.}\ \bibnamefont {{Carney}}}, \bibinfo {author}
  {\bibfnamefont {G.}~\bibnamefont {{Carullo}}}, \bibinfo {author}
  {\bibfnamefont {J.}~\bibnamefont {{Casanueva Diaz}}}, \bibinfo {author}
  {\bibfnamefont {C.}~\bibnamefont {{Casentini}}}, \bibinfo {author}
  {\bibfnamefont {S.}~\bibnamefont {{Caudill}}}, \bibinfo {author}
  {\bibfnamefont {M.}~\bibnamefont {{Cavagli{\`a}}}}, \bibinfo {author}
  {\bibfnamefont {F.}~\bibnamefont {{Cavalier}}}, \bibinfo {author}
  {\bibfnamefont {R.}~\bibnamefont {{Cavalieri}}}, \bibinfo {author}
  {\bibfnamefont {G.}~\bibnamefont {{Cella}}}, \bibinfo {author} {\bibfnamefont
  {C.~B.}\ \bibnamefont {{Cepeda}}}, \bibinfo {author} {\bibfnamefont
  {P.}~\bibnamefont {{Cerd{\'a}-Dur{\'a}n}}}, \bibinfo {author} {\bibfnamefont
  {G.}~\bibnamefont {{Cerretani}}}, \bibinfo {author} {\bibfnamefont
  {E.}~\bibnamefont {{Cesarini}}}, \bibinfo {author} {\bibfnamefont {S.~J.}\
  \bibnamefont {{Chamberlin}}}, \bibinfo {author} {\bibfnamefont
  {M.}~\bibnamefont {{Chan}}}, \bibinfo {author} {\bibfnamefont
  {S.}~\bibnamefont {{Chao}}}, \bibinfo {author} {\bibfnamefont
  {P.}~\bibnamefont {{Charlton}}}, \bibinfo {author} {\bibfnamefont
  {E.}~\bibnamefont {{Chase}}}, \bibinfo {author} {\bibfnamefont
  {E.}~\bibnamefont {{Chassande-Mottin}}}, \bibinfo {author} {\bibfnamefont
  {D.}~\bibnamefont {{Chatterjee}}}, \bibinfo {author} {\bibfnamefont
  {K.}~\bibnamefont {{Chatziioannou}}}, \bibinfo {author} {\bibfnamefont
  {B.~D.}\ \bibnamefont {{Cheeseboro}}}, \bibinfo {author} {\bibfnamefont
  {H.~Y.}\ \bibnamefont {{Chen}}}, \bibinfo {author} {\bibfnamefont
  {X.}~\bibnamefont {{Chen}}}, \bibinfo {author} {\bibfnamefont
  {Y.}~\bibnamefont {{Chen}}}, \bibinfo {author} {\bibfnamefont {H.~P.}\
  \bibnamefont {{Cheng}}}, \bibinfo {author} {\bibfnamefont {H.}~\bibnamefont
  {{Chia}}}, \bibinfo {author} {\bibfnamefont {A.}~\bibnamefont
  {{Chincarini}}}, \bibinfo {author} {\bibfnamefont {A.}~\bibnamefont
  {{Chiummo}}}, \bibinfo {author} {\bibfnamefont {T.}~\bibnamefont {{Chmiel}}},
  \bibinfo {author} {\bibfnamefont {H.~S.}\ \bibnamefont {{Cho}}}, \bibinfo
  {author} {\bibfnamefont {M.}~\bibnamefont {{Cho}}}, \bibinfo {author}
  {\bibfnamefont {J.~H.}\ \bibnamefont {{Chow}}}, \bibinfo {author}
  {\bibfnamefont {N.}~\bibnamefont {{Christensen}}}, \bibinfo {author}
  {\bibfnamefont {Q.}~\bibnamefont {{Chu}}}, \bibinfo {author} {\bibfnamefont
  {A.~J.~K.}\ \bibnamefont {{Chua}}},\ and\ \bibinfo {author} {\bibfnamefont
  {S.}~\bibnamefont {{Chua}}},\ }\bibfield  {title} {\bibinfo {title}
  {{GW170817: Observation of Gravitational Waves from a Binary Neutron Star
  Inspiral}},\ }\href {https://doi.org/10.1103/PhysRevLett.119.161101}
  {\bibfield  {journal} {\bibinfo  {journal} {\prl}\ }\textbf {\bibinfo
  {volume} {119}},\ \bibinfo {eid} {161101} (\bibinfo {year}
  {2017}{\natexlab{a}})},\ \Eprint {https://arxiv.org/abs/1710.05832}
  {arXiv:1710.05832 [gr-qc]} \BibitemShut {NoStop}%
\bibitem [{\citenamefont {{Aasi}}\ \emph {et~al.}(2015)\citenamefont {{Aasi}}
  \emph {et~al.}}]{2015CQGra..32g4001L}%
  \BibitemOpen
  \bibfield  {author} {\bibinfo {author} {\bibfnamefont {J.}~\bibnamefont
  {{Aasi}}} \emph {et~al.},\ }\bibfield  {title} {\bibinfo {title} {{Advanced
  LIGO}},\ }\href {https://doi.org/10.1088/0264-9381/32/7/074001} {\bibfield
  {journal} {\bibinfo  {journal} {Classical and Quantum Gravity}\ }\textbf
  {\bibinfo {volume} {32}},\ \bibinfo {eid} {074001} (\bibinfo {year}
  {2015})},\ \Eprint {https://arxiv.org/abs/1411.4547} {arXiv:1411.4547
  [gr-qc]} \BibitemShut {NoStop}%
\bibitem [{\citenamefont {{Acernese}}\ \emph {et~al.}(2015)\citenamefont
  {{Acernese}} \emph {et~al.}}]{2015CQGra..32b4001A}%
  \BibitemOpen
  \bibfield  {author} {\bibinfo {author} {\bibfnamefont {F.}~\bibnamefont
  {{Acernese}}} \emph {et~al.},\ }\bibfield  {title} {\bibinfo {title}
  {{Advanced Virgo: a second-generation interferometric gravitational wave
  detector}},\ }\href {https://doi.org/10.1088/0264-9381/32/2/024001}
  {\bibfield  {journal} {\bibinfo  {journal} {Classical and Quantum Gravity}\
  }\textbf {\bibinfo {volume} {32}},\ \bibinfo {eid} {024001} (\bibinfo {year}
  {2015})},\ \Eprint {https://arxiv.org/abs/1408.3978} {arXiv:1408.3978
  [gr-qc]} \BibitemShut {NoStop}%
\bibitem [{\citenamefont {{Abbott}}\ \emph
  {et~al.}(2017{\natexlab{b}})\citenamefont {{Abbott}}, \citenamefont
  {{Abbott}}, \citenamefont {{Abbott}}, \citenamefont {{Acernese}},
  \citenamefont {{Ackley}}, \citenamefont {{Adams}}, \citenamefont {{Adams}},
  \citenamefont {{Addesso}}, \citenamefont {{Adhikari}}, \citenamefont
  {{Adya}}, \citenamefont {{Affeldt}}, \citenamefont {{Afrough}}, \citenamefont
  {{Agarwal}}, \citenamefont {{Agathos}}, \citenamefont {{Agatsuma}},
  \citenamefont {{Aggarwal}}, \citenamefont {{Aguiar}}, \citenamefont
  {{Aiello}}, \citenamefont {{Ain}}, \citenamefont {{Ajith}}, \citenamefont
  {{Allen}}, \citenamefont {{Allen}}, \citenamefont {{Allocca}}, \citenamefont
  {{Altin}}, \citenamefont {{Amato}}, \citenamefont {{Ananyeva}}, \citenamefont
  {{Anderson}}, \citenamefont {{Anderson}}, \citenamefont {{Angelova}},
  \citenamefont {{Antier}}, \citenamefont {{Appert}}, \citenamefont {{Arai}},
  \citenamefont {{Araya}}, \citenamefont {{Areeda}}, \citenamefont {{Arnaud}},
  \citenamefont {{Arun}}, \citenamefont {{Ascenzi}}, \citenamefont {{Ashton}},
  \citenamefont {{Ast}}, \citenamefont {{Aston}}, \citenamefont {{Astone}},
  \citenamefont {{Atallah}}, \citenamefont {{Aufmuth}}, \citenamefont
  {{Aulbert}}, \citenamefont {{AultONeal}}, \citenamefont {{Austin}},
  \citenamefont {{Avila-Alvarez}}, \citenamefont {{Babak}}, \citenamefont
  {{Bacon}}, \citenamefont {{Bader}}, \citenamefont {{Bae}}, \citenamefont
  {{Baker}}, \citenamefont {{Baldaccini}}, \citenamefont {{Ballardin}},
  \citenamefont {{Ballmer}}, \citenamefont {{Banagiri}}, \citenamefont
  {{Barayoga}}, \citenamefont {{Barclay}}, \citenamefont {{Barish}},
  \citenamefont {{Barker}}, \citenamefont {{Barkett}}, \citenamefont
  {{Barone}}, \citenamefont {{Barr}}, \citenamefont {{Barsotti}}, \citenamefont
  {{Barsuglia}}, \citenamefont {{Barta}}, \citenamefont {{Barthelmy}},
  \citenamefont {{Bartlett}}, \citenamefont {{Bartos}}, \citenamefont
  {{Bassiri}}, \citenamefont {{Basti}}, \citenamefont {{Batch}}, \citenamefont
  {{Bawaj}}, \citenamefont {{Bayley}}, \citenamefont {{Bazzan}}, \citenamefont
  {{B{\'e}csy}}, \citenamefont {{Beer}}, \citenamefont {{Bejger}},
  \citenamefont {{Belahcene}}, \citenamefont {{Bell}}, \citenamefont
  {{Berger}}, \citenamefont {{Bergmann}}, \citenamefont {{Bero}}, \citenamefont
  {{Berry}}, \citenamefont {{Bersanetti}}, \citenamefont {{Bertolini}},
  \citenamefont {{Betzwieser}}, \citenamefont {{Bhagwat}}, \citenamefont
  {{Bhandare}}, \citenamefont {{Bilenko}}, \citenamefont {{Billingsley}},
  \citenamefont {{Billman}}, \citenamefont {{Birch}}, \citenamefont {{Birney}},
  \citenamefont {{Birnholtz}}, \citenamefont {{Biscans}}, \citenamefont
  {{Biscoveanu}}, \citenamefont {{Bisht}}, \citenamefont {{Bitossi}},
  \citenamefont {{Biwer}}, \citenamefont {{Bizouard}}, \citenamefont
  {{Blackburn}}, \citenamefont {{Blackman}}, \citenamefont {{Blair}},
  \citenamefont {{Blair}}, \citenamefont {{Blair}}, \citenamefont {{Bloemen}},
  \citenamefont {{Bock}}, \citenamefont {{Bode}}, \citenamefont {{Boer}},
  \citenamefont {{Bogaert}}, \citenamefont {{Bohe}}, \citenamefont {{Bondu}},
  \citenamefont {{Bonilla}}, \citenamefont {{Bonnand}}, \citenamefont {{Boom}},
  \citenamefont {{Bork}}, \citenamefont {{Boschi}}, \citenamefont {{Bose}},
  \citenamefont {{Bossie}}, \citenamefont {{Bouffanais}}, \citenamefont
  {{Bozzi}}, \citenamefont {{Bradaschia}}, \citenamefont {{Brady}},
  \citenamefont {{Branchesi}}, \citenamefont {{Brau}}, \citenamefont
  {{Briant}}, \citenamefont {{Brillet}}, \citenamefont {{Brinkmann}},
  \citenamefont {{Brisson}}, \citenamefont {{Brockill}}, \citenamefont
  {{Broida}}, \citenamefont {{Brooks}}, \citenamefont {{Brown}}, \citenamefont
  {{Brown}}, \citenamefont {{Brunett}}, \citenamefont {{Buchanan}},
  \citenamefont {{Buikema}}, \citenamefont {{Bulik}}, \citenamefont {{Bulten}},
  \citenamefont {{Buonanno}}, \citenamefont {{Buskulic}}, \citenamefont
  {{Buy}}, \citenamefont {{Byer}}, \citenamefont {{Cabero}}, \citenamefont
  {{Cadonati}}, \citenamefont {{Cagnoli}}, \citenamefont {{Cahillane}},
  \citenamefont {{Calder{\'o}n Bustillo}}, \citenamefont {{Callister}},
  \citenamefont {{Calloni}}, \citenamefont {{Camp}}, \citenamefont {{Canepa}},
  \citenamefont {{Canizares}}, \citenamefont {{Cannon}}, \citenamefont {{Cao}},
  \citenamefont {{Cao}}, \citenamefont {{Capano}}, \citenamefont {{Capocasa}},
  \citenamefont {{Carbognani}}, \citenamefont {{Caride}}, \citenamefont
  {{Carney}}, \citenamefont {{Casanueva Diaz}}, \citenamefont {{Casentini}},
  \citenamefont {{Caudill}}, \citenamefont {{Cavagli{\`a}}}, \citenamefont
  {{Cavalier}}, \citenamefont {{Cavalieri}}, \citenamefont {{Cella}},
  \citenamefont {{Cepeda}}, \citenamefont {{Cerd{\'a}-Dur{\'a}n}},
  \citenamefont {{Cerretani}}, \citenamefont {{Cesarini}}, \citenamefont
  {{Chamberlin}}, \citenamefont {{Chan}}, \citenamefont {{Chao}}, \citenamefont
  {{Charlton}}, \citenamefont {{Chase}}, \citenamefont {{Chassande-Mottin}},
  \citenamefont {{Chatterjee}}, \citenamefont {{Chatziioannou}}, \citenamefont
  {{Cheeseboro}}, \citenamefont {{Chen}}, \citenamefont {{Chen}}, \citenamefont
  {{Chen}}, \citenamefont {{Cheng}}, \citenamefont {{Chia}}, \citenamefont
  {{Chincarini}}, \citenamefont {{Chiummo}}, \citenamefont {{Chmiel}},
  \citenamefont {{Cho}}, \citenamefont {{Cho}}, \citenamefont {{Chow}},
  \citenamefont {{Christensen}}, \citenamefont {{Chu}}, \citenamefont {{Chua}},
  \citenamefont {{Chua}}, \citenamefont {{Chung}}, \citenamefont {{Chung}},\
  and\ \citenamefont {{Ciani}}}]{2017ApJ...848L..12A}%
  \BibitemOpen
  \bibfield  {author} {\bibinfo {author} {\bibfnamefont {B.~P.}\ \bibnamefont
  {{Abbott}}}, \bibinfo {author} {\bibfnamefont {R.}~\bibnamefont {{Abbott}}},
  \bibinfo {author} {\bibfnamefont {T.~D.}\ \bibnamefont {{Abbott}}}, \bibinfo
  {author} {\bibfnamefont {F.}~\bibnamefont {{Acernese}}}, \bibinfo {author}
  {\bibfnamefont {K.}~\bibnamefont {{Ackley}}}, \bibinfo {author}
  {\bibfnamefont {C.}~\bibnamefont {{Adams}}}, \bibinfo {author} {\bibfnamefont
  {T.}~\bibnamefont {{Adams}}}, \bibinfo {author} {\bibfnamefont
  {P.}~\bibnamefont {{Addesso}}}, \bibinfo {author} {\bibfnamefont {R.~X.}\
  \bibnamefont {{Adhikari}}}, \bibinfo {author} {\bibfnamefont {V.~B.}\
  \bibnamefont {{Adya}}}, \bibinfo {author} {\bibfnamefont {C.}~\bibnamefont
  {{Affeldt}}}, \bibinfo {author} {\bibfnamefont {M.}~\bibnamefont
  {{Afrough}}}, \bibinfo {author} {\bibfnamefont {B.}~\bibnamefont
  {{Agarwal}}}, \bibinfo {author} {\bibfnamefont {M.}~\bibnamefont
  {{Agathos}}}, \bibinfo {author} {\bibfnamefont {K.}~\bibnamefont
  {{Agatsuma}}}, \bibinfo {author} {\bibfnamefont {N.}~\bibnamefont
  {{Aggarwal}}}, \bibinfo {author} {\bibfnamefont {O.~D.}\ \bibnamefont
  {{Aguiar}}}, \bibinfo {author} {\bibfnamefont {L.}~\bibnamefont {{Aiello}}},
  \bibinfo {author} {\bibfnamefont {A.}~\bibnamefont {{Ain}}}, \bibinfo
  {author} {\bibfnamefont {P.}~\bibnamefont {{Ajith}}}, \bibinfo {author}
  {\bibfnamefont {B.}~\bibnamefont {{Allen}}}, \bibinfo {author} {\bibfnamefont
  {G.}~\bibnamefont {{Allen}}}, \bibinfo {author} {\bibfnamefont
  {A.}~\bibnamefont {{Allocca}}}, \bibinfo {author} {\bibfnamefont {P.~A.}\
  \bibnamefont {{Altin}}}, \bibinfo {author} {\bibfnamefont {A.}~\bibnamefont
  {{Amato}}}, \bibinfo {author} {\bibfnamefont {A.}~\bibnamefont {{Ananyeva}}},
  \bibinfo {author} {\bibfnamefont {S.~B.}\ \bibnamefont {{Anderson}}},
  \bibinfo {author} {\bibfnamefont {W.~G.}\ \bibnamefont {{Anderson}}},
  \bibinfo {author} {\bibfnamefont {S.~V.}\ \bibnamefont {{Angelova}}},
  \bibinfo {author} {\bibfnamefont {S.}~\bibnamefont {{Antier}}}, \bibinfo
  {author} {\bibfnamefont {S.}~\bibnamefont {{Appert}}}, \bibinfo {author}
  {\bibfnamefont {K.}~\bibnamefont {{Arai}}}, \bibinfo {author} {\bibfnamefont
  {M.~C.}\ \bibnamefont {{Araya}}}, \bibinfo {author} {\bibfnamefont {J.~S.}\
  \bibnamefont {{Areeda}}}, \bibinfo {author} {\bibfnamefont {N.}~\bibnamefont
  {{Arnaud}}}, \bibinfo {author} {\bibfnamefont {K.~G.}\ \bibnamefont
  {{Arun}}}, \bibinfo {author} {\bibfnamefont {S.}~\bibnamefont {{Ascenzi}}},
  \bibinfo {author} {\bibfnamefont {G.}~\bibnamefont {{Ashton}}}, \bibinfo
  {author} {\bibfnamefont {M.}~\bibnamefont {{Ast}}}, \bibinfo {author}
  {\bibfnamefont {S.~M.}\ \bibnamefont {{Aston}}}, \bibinfo {author}
  {\bibfnamefont {P.}~\bibnamefont {{Astone}}}, \bibinfo {author}
  {\bibfnamefont {D.~V.}\ \bibnamefont {{Atallah}}}, \bibinfo {author}
  {\bibfnamefont {P.}~\bibnamefont {{Aufmuth}}}, \bibinfo {author}
  {\bibfnamefont {C.}~\bibnamefont {{Aulbert}}}, \bibinfo {author}
  {\bibfnamefont {K.}~\bibnamefont {{AultONeal}}}, \bibinfo {author}
  {\bibfnamefont {C.}~\bibnamefont {{Austin}}}, \bibinfo {author}
  {\bibfnamefont {A.}~\bibnamefont {{Avila-Alvarez}}}, \bibinfo {author}
  {\bibfnamefont {S.}~\bibnamefont {{Babak}}}, \bibinfo {author} {\bibfnamefont
  {P.}~\bibnamefont {{Bacon}}}, \bibinfo {author} {\bibfnamefont {M.~K.~M.}\
  \bibnamefont {{Bader}}}, \bibinfo {author} {\bibfnamefont {S.}~\bibnamefont
  {{Bae}}}, \bibinfo {author} {\bibfnamefont {P.~T.}\ \bibnamefont {{Baker}}},
  \bibinfo {author} {\bibfnamefont {F.}~\bibnamefont {{Baldaccini}}}, \bibinfo
  {author} {\bibfnamefont {G.}~\bibnamefont {{Ballardin}}}, \bibinfo {author}
  {\bibfnamefont {S.~W.}\ \bibnamefont {{Ballmer}}}, \bibinfo {author}
  {\bibfnamefont {S.}~\bibnamefont {{Banagiri}}}, \bibinfo {author}
  {\bibfnamefont {J.~C.}\ \bibnamefont {{Barayoga}}}, \bibinfo {author}
  {\bibfnamefont {S.~E.}\ \bibnamefont {{Barclay}}}, \bibinfo {author}
  {\bibfnamefont {B.~C.}\ \bibnamefont {{Barish}}}, \bibinfo {author}
  {\bibfnamefont {D.}~\bibnamefont {{Barker}}}, \bibinfo {author}
  {\bibfnamefont {K.}~\bibnamefont {{Barkett}}}, \bibinfo {author}
  {\bibfnamefont {F.}~\bibnamefont {{Barone}}}, \bibinfo {author}
  {\bibfnamefont {B.}~\bibnamefont {{Barr}}}, \bibinfo {author} {\bibfnamefont
  {L.}~\bibnamefont {{Barsotti}}}, \bibinfo {author} {\bibfnamefont
  {M.}~\bibnamefont {{Barsuglia}}}, \bibinfo {author} {\bibfnamefont
  {D.}~\bibnamefont {{Barta}}}, \bibinfo {author} {\bibfnamefont {S.~D.}\
  \bibnamefont {{Barthelmy}}}, \bibinfo {author} {\bibfnamefont
  {J.}~\bibnamefont {{Bartlett}}}, \bibinfo {author} {\bibfnamefont
  {I.}~\bibnamefont {{Bartos}}}, \bibinfo {author} {\bibfnamefont
  {R.}~\bibnamefont {{Bassiri}}}, \bibinfo {author} {\bibfnamefont
  {A.}~\bibnamefont {{Basti}}}, \bibinfo {author} {\bibfnamefont {J.~C.}\
  \bibnamefont {{Batch}}}, \bibinfo {author} {\bibfnamefont {M.}~\bibnamefont
  {{Bawaj}}}, \bibinfo {author} {\bibfnamefont {J.~C.}\ \bibnamefont
  {{Bayley}}}, \bibinfo {author} {\bibfnamefont {M.}~\bibnamefont {{Bazzan}}},
  \bibinfo {author} {\bibfnamefont {B.}~\bibnamefont {{B{\'e}csy}}}, \bibinfo
  {author} {\bibfnamefont {C.}~\bibnamefont {{Beer}}}, \bibinfo {author}
  {\bibfnamefont {M.}~\bibnamefont {{Bejger}}}, \bibinfo {author}
  {\bibfnamefont {I.}~\bibnamefont {{Belahcene}}}, \bibinfo {author}
  {\bibfnamefont {A.~S.}\ \bibnamefont {{Bell}}}, \bibinfo {author}
  {\bibfnamefont {B.~K.}\ \bibnamefont {{Berger}}}, \bibinfo {author}
  {\bibfnamefont {G.}~\bibnamefont {{Bergmann}}}, \bibinfo {author}
  {\bibfnamefont {J.~J.}\ \bibnamefont {{Bero}}}, \bibinfo {author}
  {\bibfnamefont {C.~P.~L.}\ \bibnamefont {{Berry}}}, \bibinfo {author}
  {\bibfnamefont {D.}~\bibnamefont {{Bersanetti}}}, \bibinfo {author}
  {\bibfnamefont {A.}~\bibnamefont {{Bertolini}}}, \bibinfo {author}
  {\bibfnamefont {J.}~\bibnamefont {{Betzwieser}}}, \bibinfo {author}
  {\bibfnamefont {S.}~\bibnamefont {{Bhagwat}}}, \bibinfo {author}
  {\bibfnamefont {R.}~\bibnamefont {{Bhandare}}}, \bibinfo {author}
  {\bibfnamefont {I.~A.}\ \bibnamefont {{Bilenko}}}, \bibinfo {author}
  {\bibfnamefont {G.}~\bibnamefont {{Billingsley}}}, \bibinfo {author}
  {\bibfnamefont {C.~R.}\ \bibnamefont {{Billman}}}, \bibinfo {author}
  {\bibfnamefont {J.}~\bibnamefont {{Birch}}}, \bibinfo {author} {\bibfnamefont
  {R.}~\bibnamefont {{Birney}}}, \bibinfo {author} {\bibfnamefont
  {O.}~\bibnamefont {{Birnholtz}}}, \bibinfo {author} {\bibfnamefont
  {S.}~\bibnamefont {{Biscans}}}, \bibinfo {author} {\bibfnamefont
  {S.}~\bibnamefont {{Biscoveanu}}}, \bibinfo {author} {\bibfnamefont
  {A.}~\bibnamefont {{Bisht}}}, \bibinfo {author} {\bibfnamefont
  {M.}~\bibnamefont {{Bitossi}}}, \bibinfo {author} {\bibfnamefont
  {C.}~\bibnamefont {{Biwer}}}, \bibinfo {author} {\bibfnamefont {M.~A.}\
  \bibnamefont {{Bizouard}}}, \bibinfo {author} {\bibfnamefont {J.~K.}\
  \bibnamefont {{Blackburn}}}, \bibinfo {author} {\bibfnamefont
  {J.}~\bibnamefont {{Blackman}}}, \bibinfo {author} {\bibfnamefont {C.~D.}\
  \bibnamefont {{Blair}}}, \bibinfo {author} {\bibfnamefont {D.~G.}\
  \bibnamefont {{Blair}}}, \bibinfo {author} {\bibfnamefont {R.~M.}\
  \bibnamefont {{Blair}}}, \bibinfo {author} {\bibfnamefont {S.}~\bibnamefont
  {{Bloemen}}}, \bibinfo {author} {\bibfnamefont {O.}~\bibnamefont {{Bock}}},
  \bibinfo {author} {\bibfnamefont {N.}~\bibnamefont {{Bode}}}, \bibinfo
  {author} {\bibfnamefont {M.}~\bibnamefont {{Boer}}}, \bibinfo {author}
  {\bibfnamefont {G.}~\bibnamefont {{Bogaert}}}, \bibinfo {author}
  {\bibfnamefont {A.}~\bibnamefont {{Bohe}}}, \bibinfo {author} {\bibfnamefont
  {F.}~\bibnamefont {{Bondu}}}, \bibinfo {author} {\bibfnamefont
  {E.}~\bibnamefont {{Bonilla}}}, \bibinfo {author} {\bibfnamefont
  {R.}~\bibnamefont {{Bonnand}}}, \bibinfo {author} {\bibfnamefont {B.~A.}\
  \bibnamefont {{Boom}}}, \bibinfo {author} {\bibfnamefont {R.}~\bibnamefont
  {{Bork}}}, \bibinfo {author} {\bibfnamefont {V.}~\bibnamefont {{Boschi}}},
  \bibinfo {author} {\bibfnamefont {S.}~\bibnamefont {{Bose}}}, \bibinfo
  {author} {\bibfnamefont {K.}~\bibnamefont {{Bossie}}}, \bibinfo {author}
  {\bibfnamefont {Y.}~\bibnamefont {{Bouffanais}}}, \bibinfo {author}
  {\bibfnamefont {A.}~\bibnamefont {{Bozzi}}}, \bibinfo {author} {\bibfnamefont
  {C.}~\bibnamefont {{Bradaschia}}}, \bibinfo {author} {\bibfnamefont {P.~R.}\
  \bibnamefont {{Brady}}}, \bibinfo {author} {\bibfnamefont {M.}~\bibnamefont
  {{Branchesi}}}, \bibinfo {author} {\bibfnamefont {J.~E.}\ \bibnamefont
  {{Brau}}}, \bibinfo {author} {\bibfnamefont {T.}~\bibnamefont {{Briant}}},
  \bibinfo {author} {\bibfnamefont {A.}~\bibnamefont {{Brillet}}}, \bibinfo
  {author} {\bibfnamefont {M.}~\bibnamefont {{Brinkmann}}}, \bibinfo {author}
  {\bibfnamefont {V.}~\bibnamefont {{Brisson}}}, \bibinfo {author}
  {\bibfnamefont {P.}~\bibnamefont {{Brockill}}}, \bibinfo {author}
  {\bibfnamefont {J.~E.}\ \bibnamefont {{Broida}}}, \bibinfo {author}
  {\bibfnamefont {A.~F.}\ \bibnamefont {{Brooks}}}, \bibinfo {author}
  {\bibfnamefont {D.~A.}\ \bibnamefont {{Brown}}}, \bibinfo {author}
  {\bibfnamefont {D.~D.}\ \bibnamefont {{Brown}}}, \bibinfo {author}
  {\bibfnamefont {S.}~\bibnamefont {{Brunett}}}, \bibinfo {author}
  {\bibfnamefont {C.~C.}\ \bibnamefont {{Buchanan}}}, \bibinfo {author}
  {\bibfnamefont {A.}~\bibnamefont {{Buikema}}}, \bibinfo {author}
  {\bibfnamefont {T.}~\bibnamefont {{Bulik}}}, \bibinfo {author} {\bibfnamefont
  {H.~J.}\ \bibnamefont {{Bulten}}}, \bibinfo {author} {\bibfnamefont
  {A.}~\bibnamefont {{Buonanno}}}, \bibinfo {author} {\bibfnamefont
  {D.}~\bibnamefont {{Buskulic}}}, \bibinfo {author} {\bibfnamefont
  {C.}~\bibnamefont {{Buy}}}, \bibinfo {author} {\bibfnamefont {R.~L.}\
  \bibnamefont {{Byer}}}, \bibinfo {author} {\bibfnamefont {M.}~\bibnamefont
  {{Cabero}}}, \bibinfo {author} {\bibfnamefont {L.}~\bibnamefont
  {{Cadonati}}}, \bibinfo {author} {\bibfnamefont {G.}~\bibnamefont
  {{Cagnoli}}}, \bibinfo {author} {\bibfnamefont {C.}~\bibnamefont
  {{Cahillane}}}, \bibinfo {author} {\bibfnamefont {J.}~\bibnamefont
  {{Calder{\'o}n Bustillo}}}, \bibinfo {author} {\bibfnamefont {T.~A.}\
  \bibnamefont {{Callister}}}, \bibinfo {author} {\bibfnamefont
  {E.}~\bibnamefont {{Calloni}}}, \bibinfo {author} {\bibfnamefont {J.~B.}\
  \bibnamefont {{Camp}}}, \bibinfo {author} {\bibfnamefont {M.}~\bibnamefont
  {{Canepa}}}, \bibinfo {author} {\bibfnamefont {P.}~\bibnamefont
  {{Canizares}}}, \bibinfo {author} {\bibfnamefont {K.~C.}\ \bibnamefont
  {{Cannon}}}, \bibinfo {author} {\bibfnamefont {H.}~\bibnamefont {{Cao}}},
  \bibinfo {author} {\bibfnamefont {J.}~\bibnamefont {{Cao}}}, \bibinfo
  {author} {\bibfnamefont {C.~D.}\ \bibnamefont {{Capano}}}, \bibinfo {author}
  {\bibfnamefont {E.}~\bibnamefont {{Capocasa}}}, \bibinfo {author}
  {\bibfnamefont {F.}~\bibnamefont {{Carbognani}}}, \bibinfo {author}
  {\bibfnamefont {S.}~\bibnamefont {{Caride}}}, \bibinfo {author}
  {\bibfnamefont {M.~F.}\ \bibnamefont {{Carney}}}, \bibinfo {author}
  {\bibfnamefont {J.}~\bibnamefont {{Casanueva Diaz}}}, \bibinfo {author}
  {\bibfnamefont {C.}~\bibnamefont {{Casentini}}}, \bibinfo {author}
  {\bibfnamefont {S.}~\bibnamefont {{Caudill}}}, \bibinfo {author}
  {\bibfnamefont {M.}~\bibnamefont {{Cavagli{\`a}}}}, \bibinfo {author}
  {\bibfnamefont {F.}~\bibnamefont {{Cavalier}}}, \bibinfo {author}
  {\bibfnamefont {R.}~\bibnamefont {{Cavalieri}}}, \bibinfo {author}
  {\bibfnamefont {G.}~\bibnamefont {{Cella}}}, \bibinfo {author} {\bibfnamefont
  {C.~B.}\ \bibnamefont {{Cepeda}}}, \bibinfo {author} {\bibfnamefont
  {P.}~\bibnamefont {{Cerd{\'a}-Dur{\'a}n}}}, \bibinfo {author} {\bibfnamefont
  {G.}~\bibnamefont {{Cerretani}}}, \bibinfo {author} {\bibfnamefont
  {E.}~\bibnamefont {{Cesarini}}}, \bibinfo {author} {\bibfnamefont {S.~J.}\
  \bibnamefont {{Chamberlin}}}, \bibinfo {author} {\bibfnamefont
  {M.}~\bibnamefont {{Chan}}}, \bibinfo {author} {\bibfnamefont
  {S.}~\bibnamefont {{Chao}}}, \bibinfo {author} {\bibfnamefont
  {P.}~\bibnamefont {{Charlton}}}, \bibinfo {author} {\bibfnamefont
  {E.}~\bibnamefont {{Chase}}}, \bibinfo {author} {\bibfnamefont
  {E.}~\bibnamefont {{Chassande-Mottin}}}, \bibinfo {author} {\bibfnamefont
  {D.}~\bibnamefont {{Chatterjee}}}, \bibinfo {author} {\bibfnamefont
  {K.}~\bibnamefont {{Chatziioannou}}}, \bibinfo {author} {\bibfnamefont
  {B.~D.}\ \bibnamefont {{Cheeseboro}}}, \bibinfo {author} {\bibfnamefont
  {H.~Y.}\ \bibnamefont {{Chen}}}, \bibinfo {author} {\bibfnamefont
  {X.}~\bibnamefont {{Chen}}}, \bibinfo {author} {\bibfnamefont
  {Y.}~\bibnamefont {{Chen}}}, \bibinfo {author} {\bibfnamefont {H.~P.}\
  \bibnamefont {{Cheng}}}, \bibinfo {author} {\bibfnamefont {H.}~\bibnamefont
  {{Chia}}}, \bibinfo {author} {\bibfnamefont {A.}~\bibnamefont
  {{Chincarini}}}, \bibinfo {author} {\bibfnamefont {A.}~\bibnamefont
  {{Chiummo}}}, \bibinfo {author} {\bibfnamefont {T.}~\bibnamefont {{Chmiel}}},
  \bibinfo {author} {\bibfnamefont {H.~S.}\ \bibnamefont {{Cho}}}, \bibinfo
  {author} {\bibfnamefont {M.}~\bibnamefont {{Cho}}}, \bibinfo {author}
  {\bibfnamefont {J.~H.}\ \bibnamefont {{Chow}}}, \bibinfo {author}
  {\bibfnamefont {N.}~\bibnamefont {{Christensen}}}, \bibinfo {author}
  {\bibfnamefont {Q.}~\bibnamefont {{Chu}}}, \bibinfo {author} {\bibfnamefont
  {A.~J.~K.}\ \bibnamefont {{Chua}}}, \bibinfo {author} {\bibfnamefont
  {S.}~\bibnamefont {{Chua}}}, \bibinfo {author} {\bibfnamefont {A.~K.~W.}\
  \bibnamefont {{Chung}}}, \bibinfo {author} {\bibfnamefont {S.}~\bibnamefont
  {{Chung}}},\ and\ \bibinfo {author} {\bibfnamefont {G.}~\bibnamefont
  {{Ciani}}},\ }\bibfield  {title} {\bibinfo {title} {{Multi-messenger
  Observations of a Binary Neutron Star Merger}},\ }\href
  {https://doi.org/10.3847/2041-8213/aa91c9} {\bibfield  {journal} {\bibinfo
  {journal} {\apjl}\ }\textbf {\bibinfo {volume} {848}},\ \bibinfo {eid} {L12}
  (\bibinfo {year} {2017}{\natexlab{b}})},\ \Eprint
  {https://arxiv.org/abs/1710.05833} {arXiv:1710.05833 [astro-ph.HE]}
  \BibitemShut {NoStop}%
\bibitem [{\citenamefont {Haensel}\ \emph {et~al.}(2007)\citenamefont
  {Haensel}, \citenamefont {Potekhin},\ and\ \citenamefont
  {Yakovlev}}]{Haensel:2007yy}%
  \BibitemOpen
  \bibfield  {author} {\bibinfo {author} {\bibfnamefont {P.}~\bibnamefont
  {Haensel}}, \bibinfo {author} {\bibfnamefont {A.~Y.}\ \bibnamefont
  {Potekhin}},\ and\ \bibinfo {author} {\bibfnamefont {D.~G.}\ \bibnamefont
  {Yakovlev}},\ }\href {https://doi.org/10.1007/978-0-387-47301-7} {\emph
  {\bibinfo {title} {{Neutron stars 1: Equation of state and structure}}}},\
  Vol.\ \bibinfo {volume} {326}\ (\bibinfo  {publisher} {Springer},\ \bibinfo
  {address} {New York, USA},\ \bibinfo {year} {2007})\BibitemShut {NoStop}%
\bibitem [{\citenamefont {{Andersson}}(2019)}]{2019gwa..book.....A}%
  \BibitemOpen
  \bibfield  {author} {\bibinfo {author} {\bibfnamefont {N.}~\bibnamefont
  {{Andersson}}},\ }\href
  {https://doi.org/10.1093/oso/9780198568032.001.0001/oso-9780198568032} {\emph
  {\bibinfo {title} {{Gravitational-Wave Astronomy: Exploring the Dark Side of
  the Universe}}}}\ (\bibinfo  {publisher} {Oxford University Press},\ \bibinfo
  {year} {2019})\BibitemShut {NoStop}%
\bibitem [{\citenamefont {{Miller}}\ \emph {et~al.}(2021)\citenamefont
  {{Miller}}, \citenamefont {{Lamb}}, \citenamefont {{Dittmann}}, \citenamefont
  {{Bogdanov}}, \citenamefont {{Arzoumanian}}, \citenamefont {{Gendreau}},
  \citenamefont {{Guillot}}, \citenamefont {{Ho}}, \citenamefont {{Lattimer}},
  \citenamefont {{Loewenstein}}, \citenamefont {{Morsink}}, \citenamefont
  {{Ray}}, \citenamefont {{Wolff}}, \citenamefont {{Baker}}, \citenamefont
  {{Cazeau}}, \citenamefont {{Manthripragada}}, \citenamefont {{Markwardt}},
  \citenamefont {{Okajima}}, \citenamefont {{Pollard}}, \citenamefont
  {{Cognard}}, \citenamefont {{Cromartie}}, \citenamefont {{Fonseca}},
  \citenamefont {{Guillemot}}, \citenamefont {{Kerr}}, \citenamefont
  {{Parthasarathy}}, \citenamefont {{Pennucci}}, \citenamefont {{Ransom}},\
  and\ \citenamefont {{Stairs}}}]{2021ApJ...918L..28M}%
  \BibitemOpen
  \bibfield  {author} {\bibinfo {author} {\bibfnamefont {M.~C.}\ \bibnamefont
  {{Miller}}}, \bibinfo {author} {\bibfnamefont {F.~K.}\ \bibnamefont
  {{Lamb}}}, \bibinfo {author} {\bibfnamefont {A.~J.}\ \bibnamefont
  {{Dittmann}}}, \bibinfo {author} {\bibfnamefont {S.}~\bibnamefont
  {{Bogdanov}}}, \bibinfo {author} {\bibfnamefont {Z.}~\bibnamefont
  {{Arzoumanian}}}, \bibinfo {author} {\bibfnamefont {K.~C.}\ \bibnamefont
  {{Gendreau}}}, \bibinfo {author} {\bibfnamefont {S.}~\bibnamefont
  {{Guillot}}}, \bibinfo {author} {\bibfnamefont {W.~C.~G.}\ \bibnamefont
  {{Ho}}}, \bibinfo {author} {\bibfnamefont {J.~M.}\ \bibnamefont
  {{Lattimer}}}, \bibinfo {author} {\bibfnamefont {M.}~\bibnamefont
  {{Loewenstein}}}, \bibinfo {author} {\bibfnamefont {S.~M.}\ \bibnamefont
  {{Morsink}}}, \bibinfo {author} {\bibfnamefont {P.~S.}\ \bibnamefont
  {{Ray}}}, \bibinfo {author} {\bibfnamefont {M.~T.}\ \bibnamefont {{Wolff}}},
  \bibinfo {author} {\bibfnamefont {C.~L.}\ \bibnamefont {{Baker}}}, \bibinfo
  {author} {\bibfnamefont {T.}~\bibnamefont {{Cazeau}}}, \bibinfo {author}
  {\bibfnamefont {S.}~\bibnamefont {{Manthripragada}}}, \bibinfo {author}
  {\bibfnamefont {C.~B.}\ \bibnamefont {{Markwardt}}}, \bibinfo {author}
  {\bibfnamefont {T.}~\bibnamefont {{Okajima}}}, \bibinfo {author}
  {\bibfnamefont {S.}~\bibnamefont {{Pollard}}}, \bibinfo {author}
  {\bibfnamefont {I.}~\bibnamefont {{Cognard}}}, \bibinfo {author}
  {\bibfnamefont {H.~T.}\ \bibnamefont {{Cromartie}}}, \bibinfo {author}
  {\bibfnamefont {E.}~\bibnamefont {{Fonseca}}}, \bibinfo {author}
  {\bibfnamefont {L.}~\bibnamefont {{Guillemot}}}, \bibinfo {author}
  {\bibfnamefont {M.}~\bibnamefont {{Kerr}}}, \bibinfo {author} {\bibfnamefont
  {A.}~\bibnamefont {{Parthasarathy}}}, \bibinfo {author} {\bibfnamefont
  {T.~T.}\ \bibnamefont {{Pennucci}}}, \bibinfo {author} {\bibfnamefont
  {S.}~\bibnamefont {{Ransom}}},\ and\ \bibinfo {author} {\bibfnamefont
  {I.}~\bibnamefont {{Stairs}}},\ }\bibfield  {title} {\bibinfo {title} {{The
  Radius of PSR J0740+6620 from NICER and XMM-Newton Data}},\ }\href
  {https://doi.org/10.3847/2041-8213/ac089b} {\bibfield  {journal} {\bibinfo
  {journal} {\apjl}\ }\textbf {\bibinfo {volume} {918}},\ \bibinfo {eid} {L28}
  (\bibinfo {year} {2021})},\ \Eprint {https://arxiv.org/abs/2105.06979}
  {arXiv:2105.06979 [astro-ph.HE]} \BibitemShut {NoStop}%
\bibitem [{\citenamefont {Gendreau}\ \emph {et~al.}(2012)\citenamefont
  {Gendreau}, \citenamefont {Arzoumanian},\ and\ \citenamefont
  {Okajima}}]{NICER}%
  \BibitemOpen
  \bibfield  {author} {\bibinfo {author} {\bibfnamefont {K.~C.}\ \bibnamefont
  {Gendreau}}, \bibinfo {author} {\bibfnamefont {Z.}~\bibnamefont
  {Arzoumanian}},\ and\ \bibinfo {author} {\bibfnamefont {T.}~\bibnamefont
  {Okajima}},\ }\bibfield  {title} {\bibinfo {title} {{The Neutron star
  Interior Composition ExploreR (NICER): an Explorer mission of opportunity for
  soft x-ray timing spectroscopy}},\ }in\ \href
  {https://doi.org/10.1117/12.926396} {\emph {\bibinfo {booktitle} {Space
  Telescopes and Instrumentation 2012: Ultraviolet to Gamma Ray}}},\ Vol.\
  \bibinfo {volume} {8443},\ \bibinfo {editor} {edited by\ \bibinfo {editor}
  {\bibfnamefont {T.}~\bibnamefont {Takahashi}}, \bibinfo {editor}
  {\bibfnamefont {S.~S.}\ \bibnamefont {Murray}},\ and\ \bibinfo {editor}
  {\bibfnamefont {J.-W.~A.}\ \bibnamefont {den Herder}}},\ \bibinfo
  {organization} {International Society for Optics and Photonics}\ (\bibinfo
  {publisher} {SPIE},\ \bibinfo {year} {2012})\ pp.\ \bibinfo {pages} {322 --
  329}\BibitemShut {NoStop}%
\bibitem [{\citenamefont {De}\ \emph {et~al.}(2018)\citenamefont {De},
  \citenamefont {Finstad}, \citenamefont {Lattimer}, \citenamefont {Brown},
  \citenamefont {Berger},\ and\ \citenamefont
  {Biwer}}]{PhysRevLett.121.091102}%
  \BibitemOpen
  \bibfield  {author} {\bibinfo {author} {\bibfnamefont {S.}~\bibnamefont
  {De}}, \bibinfo {author} {\bibfnamefont {D.}~\bibnamefont {Finstad}},
  \bibinfo {author} {\bibfnamefont {J.~M.}\ \bibnamefont {Lattimer}}, \bibinfo
  {author} {\bibfnamefont {D.~A.}\ \bibnamefont {Brown}}, \bibinfo {author}
  {\bibfnamefont {E.}~\bibnamefont {Berger}},\ and\ \bibinfo {author}
  {\bibfnamefont {C.~M.}\ \bibnamefont {Biwer}},\ }\bibfield  {title} {\bibinfo
  {title} {Tidal deformabilities and radii of neutron stars from the
  observation of gw170817},\ }\href
  {https://doi.org/10.1103/PhysRevLett.121.091102} {\bibfield  {journal}
  {\bibinfo  {journal} {Phys. Rev. Lett.}\ }\textbf {\bibinfo {volume} {121}},\
  \bibinfo {pages} {091102} (\bibinfo {year} {2018})}\BibitemShut {NoStop}%
\bibitem [{\citenamefont {{Raithel}}\ \emph {et~al.}(2018)\citenamefont
  {{Raithel}}, \citenamefont {{{\"O}zel}},\ and\ \citenamefont
  {{Psaltis}}}]{2018ApJ...857L..23R}%
  \BibitemOpen
  \bibfield  {author} {\bibinfo {author} {\bibfnamefont {C.~A.}\ \bibnamefont
  {{Raithel}}}, \bibinfo {author} {\bibfnamefont {F.}~\bibnamefont
  {{{\"O}zel}}},\ and\ \bibinfo {author} {\bibfnamefont {D.}~\bibnamefont
  {{Psaltis}}},\ }\bibfield  {title} {\bibinfo {title} {{Tidal Deformability
  from GW170817 as a Direct Probe of the Neutron Star Radius}},\ }\href
  {https://doi.org/10.3847/2041-8213/aabcbf} {\bibfield  {journal} {\bibinfo
  {journal} {\apjl}\ }\textbf {\bibinfo {volume} {857}},\ \bibinfo {eid} {L23}
  (\bibinfo {year} {2018})},\ \Eprint {https://arxiv.org/abs/1803.07687}
  {arXiv:1803.07687 [astro-ph.HE]} \BibitemShut {NoStop}%
\bibitem [{\citenamefont {Most}\ \emph {et~al.}(2018)\citenamefont {Most},
  \citenamefont {Weih}, \citenamefont {Rezzolla},\ and\ \citenamefont
  {Schaffner-Bielich}}]{PhysRevLett.120.261103}%
  \BibitemOpen
  \bibfield  {author} {\bibinfo {author} {\bibfnamefont {E.~R.}\ \bibnamefont
  {Most}}, \bibinfo {author} {\bibfnamefont {L.~R.}\ \bibnamefont {Weih}},
  \bibinfo {author} {\bibfnamefont {L.}~\bibnamefont {Rezzolla}},\ and\
  \bibinfo {author} {\bibfnamefont {J.}~\bibnamefont {Schaffner-Bielich}},\
  }\bibfield  {title} {\bibinfo {title} {New constraints on radii and tidal
  deformabilities of neutron stars from gw170817},\ }\href
  {https://doi.org/10.1103/PhysRevLett.120.261103} {\bibfield  {journal}
  {\bibinfo  {journal} {Phys. Rev. Lett.}\ }\textbf {\bibinfo {volume} {120}},\
  \bibinfo {pages} {261103} (\bibinfo {year} {2018})}\BibitemShut {NoStop}%
\bibitem [{\citenamefont {{Chatziioannou}}(2020)}]{2020GReGr..52..109C}%
  \BibitemOpen
  \bibfield  {author} {\bibinfo {author} {\bibfnamefont {K.}~\bibnamefont
  {{Chatziioannou}}},\ }\bibfield  {title} {\bibinfo {title} {{Neutron-star
  tidal deformability and equation-of-state constraints}},\ }\href
  {https://doi.org/10.1007/s10714-020-02754-3} {\bibfield  {journal} {\bibinfo
  {journal} {General Relativity and Gravitation}\ }\textbf {\bibinfo {volume}
  {52}},\ \bibinfo {eid} {109} (\bibinfo {year} {2020})},\ \Eprint
  {https://arxiv.org/abs/2006.03168} {arXiv:2006.03168 [gr-qc]} \BibitemShut
  {NoStop}%
\bibitem [{\citenamefont {{Yagi}}\ and\ \citenamefont
  {{Yunes}}(2013)}]{2013PhRvD..88b3009Y}%
  \BibitemOpen
  \bibfield  {author} {\bibinfo {author} {\bibfnamefont {K.}~\bibnamefont
  {{Yagi}}}\ and\ \bibinfo {author} {\bibfnamefont {N.}~\bibnamefont
  {{Yunes}}},\ }\bibfield  {title} {\bibinfo {title} {{I-Love-Q relations in
  neutron stars and their applications to astrophysics, gravitational waves,
  and fundamental physics}},\ }\href
  {https://doi.org/10.1103/PhysRevD.88.023009} {\bibfield  {journal} {\bibinfo
  {journal} {\prd}\ }\textbf {\bibinfo {volume} {88}},\ \bibinfo {eid} {023009}
  (\bibinfo {year} {2013})},\ \Eprint {https://arxiv.org/abs/1303.1528}
  {arXiv:1303.1528 [gr-qc]} \BibitemShut {NoStop}%
\bibitem [{\citenamefont {{Chatziioannou}}\ \emph {et~al.}(2018)\citenamefont
  {{Chatziioannou}}, \citenamefont {{Haster}},\ and\ \citenamefont
  {{Zimmerman}}}]{2018PhRvD..97j4036C}%
  \BibitemOpen
  \bibfield  {author} {\bibinfo {author} {\bibfnamefont {K.}~\bibnamefont
  {{Chatziioannou}}}, \bibinfo {author} {\bibfnamefont {C.-J.}\ \bibnamefont
  {{Haster}}},\ and\ \bibinfo {author} {\bibfnamefont {A.}~\bibnamefont
  {{Zimmerman}}},\ }\bibfield  {title} {\bibinfo {title} {{Measuring the
  neutron star tidal deformability with equation-of-state-independent relations
  and gravitational waves}},\ }\href
  {https://doi.org/10.1103/PhysRevD.97.104036} {\bibfield  {journal} {\bibinfo
  {journal} {\prd}\ }\textbf {\bibinfo {volume} {97}},\ \bibinfo {eid} {104036}
  (\bibinfo {year} {2018})},\ \Eprint {https://arxiv.org/abs/1804.03221}
  {arXiv:1804.03221 [gr-qc]} \BibitemShut {NoStop}%
\bibitem [{\citenamefont {{Kumar}}\ and\ \citenamefont
  {{Landry}}(2019)}]{2019PhRvD..99l3026K}%
  \BibitemOpen
  \bibfield  {author} {\bibinfo {author} {\bibfnamefont {B.}~\bibnamefont
  {{Kumar}}}\ and\ \bibinfo {author} {\bibfnamefont {P.}~\bibnamefont
  {{Landry}}},\ }\bibfield  {title} {\bibinfo {title} {{Inferring neutron star
  properties from GW170817 with universal relations}},\ }\href
  {https://doi.org/10.1103/PhysRevD.99.123026} {\bibfield  {journal} {\bibinfo
  {journal} {\prd}\ }\textbf {\bibinfo {volume} {99}},\ \bibinfo {eid} {123026}
  (\bibinfo {year} {2019})},\ \Eprint {https://arxiv.org/abs/1902.04557}
  {arXiv:1902.04557 [gr-qc]} \BibitemShut {NoStop}%
\bibitem [{\citenamefont {Fujimoto}\ \emph {et~al.}(2020)\citenamefont
  {Fujimoto}, \citenamefont {Fukushima},\ and\ \citenamefont
  {Murase}}]{PhysRevD.101.054016}%
  \BibitemOpen
  \bibfield  {author} {\bibinfo {author} {\bibfnamefont {Y.}~\bibnamefont
  {Fujimoto}}, \bibinfo {author} {\bibfnamefont {K.}~\bibnamefont
  {Fukushima}},\ and\ \bibinfo {author} {\bibfnamefont {K.}~\bibnamefont
  {Murase}},\ }\bibfield  {title} {\bibinfo {title} {Mapping neutron star data
  to the equation of state using the deep neural network},\ }\href
  {https://doi.org/10.1103/PhysRevD.101.054016} {\bibfield  {journal} {\bibinfo
   {journal} {Phys. Rev. D}\ }\textbf {\bibinfo {volume} {101}},\ \bibinfo
  {pages} {054016} (\bibinfo {year} {2020})}\BibitemShut {NoStop}%
\bibitem [{\citenamefont {{Morawski}}\ and\ \citenamefont
  {{Bejger}}(2020)}]{2020A&A...642A..78M}%
  \BibitemOpen
  \bibfield  {author} {\bibinfo {author} {\bibfnamefont {F.}~\bibnamefont
  {{Morawski}}}\ and\ \bibinfo {author} {\bibfnamefont {M.}~\bibnamefont
  {{Bejger}}},\ }\bibfield  {title} {\bibinfo {title} {{Neural network
  reconstruction of the dense matter equation of state derived from the
  parameters of neutron stars}},\ }\href
  {https://doi.org/10.1051/0004-6361/202038130} {\bibfield  {journal} {\bibinfo
   {journal} {\aap}\ }\textbf {\bibinfo {volume} {642}},\ \bibinfo {eid} {A78}
  (\bibinfo {year} {2020})},\ \Eprint {https://arxiv.org/abs/2006.07194}
  {arXiv:2006.07194 [astro-ph.HE]} \BibitemShut {NoStop}%
\bibitem [{\citenamefont {Ferreira}\ \emph {et~al.}(2022)\citenamefont
  {Ferreira}, \citenamefont {Carvalho},\ and\ \citenamefont
  {Provid\^encia}}]{PhysRevD.106.103023}%
  \BibitemOpen
  \bibfield  {author} {\bibinfo {author} {\bibfnamefont {M.}~\bibnamefont
  {Ferreira}}, \bibinfo {author} {\bibfnamefont {V.}~\bibnamefont {Carvalho}},\
  and\ \bibinfo {author} {\bibfnamefont {C.~m.~c.}\ \bibnamefont
  {Provid\^encia}},\ }\bibfield  {title} {\bibinfo {title} {Extracting nuclear
  matter properties from the neutron star matter equation of state using deep
  neural networks},\ }\href {https://doi.org/10.1103/PhysRevD.106.103023}
  {\bibfield  {journal} {\bibinfo  {journal} {Phys. Rev. D}\ }\textbf {\bibinfo
  {volume} {106}},\ \bibinfo {pages} {103023} (\bibinfo {year}
  {2022})}\BibitemShut {NoStop}%
\bibitem [{\citenamefont {{Farrell}}\ \emph {et~al.}(2023)\citenamefont
  {{Farrell}}, \citenamefont {{Baldi}}, \citenamefont {{Ott}}, \citenamefont
  {{Ghosh}}, \citenamefont {{Steiner}}, \citenamefont {{Kavitkar}},
  \citenamefont {{Lindblom}}, \citenamefont {{Whiteson}},\ and\ \citenamefont
  {{Weber}}}]{2023JCAP...12..022F}%
  \BibitemOpen
  \bibfield  {author} {\bibinfo {author} {\bibfnamefont {D.}~\bibnamefont
  {{Farrell}}}, \bibinfo {author} {\bibfnamefont {P.}~\bibnamefont {{Baldi}}},
  \bibinfo {author} {\bibfnamefont {J.}~\bibnamefont {{Ott}}}, \bibinfo
  {author} {\bibfnamefont {A.}~\bibnamefont {{Ghosh}}}, \bibinfo {author}
  {\bibfnamefont {A.~W.}\ \bibnamefont {{Steiner}}}, \bibinfo {author}
  {\bibfnamefont {A.}~\bibnamefont {{Kavitkar}}}, \bibinfo {author}
  {\bibfnamefont {L.}~\bibnamefont {{Lindblom}}}, \bibinfo {author}
  {\bibfnamefont {D.}~\bibnamefont {{Whiteson}}},\ and\ \bibinfo {author}
  {\bibfnamefont {F.}~\bibnamefont {{Weber}}},\ }\bibfield  {title} {\bibinfo
  {title} {{Deducing neutron star equation of state from telescope spectra with
  machine-learning-derived likelihoods}},\ }\href
  {https://doi.org/10.1088/1475-7516/2023/12/022} {\bibfield  {journal}
  {\bibinfo  {journal} {\jcap}\ }\textbf {\bibinfo {volume} {2023}},\ \bibinfo
  {eid} {022} (\bibinfo {year} {2023})},\ \Eprint
  {https://arxiv.org/abs/2305.07442} {arXiv:2305.07442 [astro-ph.HE]}
  \BibitemShut {NoStop}%
\bibitem [{\citenamefont {{Guo}}\ \emph {et~al.}(2024)\citenamefont {{Guo}},
  \citenamefont {{Xiong}}, \citenamefont {{Ma}},\ and\ \citenamefont
  {{Ma}}}]{2024ApJ...965...47G}%
  \BibitemOpen
  \bibfield  {author} {\bibinfo {author} {\bibfnamefont {L.-J.}\ \bibnamefont
  {{Guo}}}, \bibinfo {author} {\bibfnamefont {J.-Y.}\ \bibnamefont {{Xiong}}},
  \bibinfo {author} {\bibfnamefont {Y.}~\bibnamefont {{Ma}}},\ and\ \bibinfo
  {author} {\bibfnamefont {Y.-L.}\ \bibnamefont {{Ma}}},\ }\bibfield  {title}
  {\bibinfo {title} {{Insights into Neutron Star Equation of State by Machine
  Learning}},\ }\href {https://doi.org/10.3847/1538-4357/ad2e8d} {\bibfield
  {journal} {\bibinfo  {journal} {\apj}\ }\textbf {\bibinfo {volume} {965}},\
  \bibinfo {eid} {47} (\bibinfo {year} {2024})},\ \Eprint
  {https://arxiv.org/abs/2309.11227} {arXiv:2309.11227 [nucl-th]} \BibitemShut
  {NoStop}%
\bibitem [{\citenamefont {Ferreira}\ and\ \citenamefont
  {Bejger}(2025)}]{PhysRevD.111.023035}%
  \BibitemOpen
  \bibfield  {author} {\bibinfo {author} {\bibfnamefont {M.}~\bibnamefont
  {Ferreira}}\ and\ \bibinfo {author} {\bibfnamefont {M.}~\bibnamefont
  {Bejger}},\ }\bibfield  {title} {\bibinfo {title} {Conditional variational
  autoencoder inference of neutron star equation of state from astrophysical
  observations},\ }\href {https://doi.org/10.1103/PhysRevD.111.023035}
  {\bibfield  {journal} {\bibinfo  {journal} {Phys. Rev. D}\ }\textbf {\bibinfo
  {volume} {111}},\ \bibinfo {pages} {023035} (\bibinfo {year}
  {2025})}\BibitemShut {NoStop}%
\bibitem [{\citenamefont {{Cuoco}}\ \emph {et~al.}(2021)\citenamefont
  {{Cuoco}}, \citenamefont {{Powell}}, \citenamefont {{Cavagli{\`a}}},
  \citenamefont {{Ackley}}, \citenamefont {{Bejger}}, \citenamefont
  {{Chatterjee}}, \citenamefont {{Coughlin}}, \citenamefont {{Coughlin}},
  \citenamefont {{Easter}}, \citenamefont {{Essick}}, \citenamefont
  {{Gabbard}}, \citenamefont {{Gebhard}}, \citenamefont {{Ghosh}},
  \citenamefont {{Haegel}}, \citenamefont {{Iess}}, \citenamefont {{Keitel}},
  \citenamefont {{Marka}}, \citenamefont {{Marka}}, \citenamefont {{Morawski}},
  \citenamefont {{Nguyen}}, \citenamefont {{Ormiston}}, \citenamefont
  {{Puerrer}}, \citenamefont {{Razzano}}, \citenamefont {{Staats}},
  \citenamefont {{Vajente}},\ and\ \citenamefont
  {{Williams}}}]{2021MLS&T...2a1002C}%
  \BibitemOpen
  \bibfield  {author} {\bibinfo {author} {\bibfnamefont {E.}~\bibnamefont
  {{Cuoco}}}, \bibinfo {author} {\bibfnamefont {J.}~\bibnamefont {{Powell}}},
  \bibinfo {author} {\bibfnamefont {M.}~\bibnamefont {{Cavagli{\`a}}}},
  \bibinfo {author} {\bibfnamefont {K.}~\bibnamefont {{Ackley}}}, \bibinfo
  {author} {\bibfnamefont {M.}~\bibnamefont {{Bejger}}}, \bibinfo {author}
  {\bibfnamefont {C.}~\bibnamefont {{Chatterjee}}}, \bibinfo {author}
  {\bibfnamefont {M.}~\bibnamefont {{Coughlin}}}, \bibinfo {author}
  {\bibfnamefont {S.}~\bibnamefont {{Coughlin}}}, \bibinfo {author}
  {\bibfnamefont {P.}~\bibnamefont {{Easter}}}, \bibinfo {author}
  {\bibfnamefont {R.}~\bibnamefont {{Essick}}}, \bibinfo {author}
  {\bibfnamefont {H.}~\bibnamefont {{Gabbard}}}, \bibinfo {author}
  {\bibfnamefont {T.}~\bibnamefont {{Gebhard}}}, \bibinfo {author}
  {\bibfnamefont {S.}~\bibnamefont {{Ghosh}}}, \bibinfo {author} {\bibfnamefont
  {L.}~\bibnamefont {{Haegel}}}, \bibinfo {author} {\bibfnamefont
  {A.}~\bibnamefont {{Iess}}}, \bibinfo {author} {\bibfnamefont
  {D.}~\bibnamefont {{Keitel}}}, \bibinfo {author} {\bibfnamefont
  {Z.}~\bibnamefont {{Marka}}}, \bibinfo {author} {\bibfnamefont
  {S.}~\bibnamefont {{Marka}}}, \bibinfo {author} {\bibfnamefont
  {F.}~\bibnamefont {{Morawski}}}, \bibinfo {author} {\bibfnamefont
  {T.}~\bibnamefont {{Nguyen}}}, \bibinfo {author} {\bibfnamefont
  {R.}~\bibnamefont {{Ormiston}}}, \bibinfo {author} {\bibfnamefont
  {M.}~\bibnamefont {{Puerrer}}}, \bibinfo {author} {\bibfnamefont
  {M.}~\bibnamefont {{Razzano}}}, \bibinfo {author} {\bibfnamefont
  {K.}~\bibnamefont {{Staats}}}, \bibinfo {author} {\bibfnamefont
  {G.}~\bibnamefont {{Vajente}}},\ and\ \bibinfo {author} {\bibfnamefont
  {D.}~\bibnamefont {{Williams}}},\ }\bibfield  {title} {\bibinfo {title}
  {{Enhancing gravitational-wave science with machine learning}},\ }\href
  {https://doi.org/10.1088/2632-2153/abb93a} {\bibfield  {journal} {\bibinfo
  {journal} {Machine Learning: Science and Technology}\ }\textbf {\bibinfo
  {volume} {2}},\ \bibinfo {eid} {011002} (\bibinfo {year} {2021})},\ \Eprint
  {https://arxiv.org/abs/2005.03745} {arXiv:2005.03745 [astro-ph.HE]}
  \BibitemShut {NoStop}%
\bibitem [{\citenamefont {{Stergioulas}}(2024)}]{2024arXiv240107406S}%
  \BibitemOpen
  \bibfield  {author} {\bibinfo {author} {\bibfnamefont {N.}~\bibnamefont
  {{Stergioulas}}},\ }\bibfield  {title} {\bibinfo {title} {{Machine Learning
  Applications in Gravitational Wave Astronomy}},\ }\href
  {https://doi.org/10.48550/arXiv.2401.07406} {\bibfield  {journal} {\bibinfo
  {journal} {arXiv e-prints}\ ,\ \bibinfo {eid} {arXiv:2401.07406}} (\bibinfo
  {year} {2024})},\ \Eprint {https://arxiv.org/abs/2401.07406}
  {arXiv:2401.07406 [gr-qc]} \BibitemShut {NoStop}%
\bibitem [{\citenamefont {{Cuoco}}\ \emph {et~al.}(2025)\citenamefont
  {{Cuoco}}, \citenamefont {{Cavagli{\`a}}}, \citenamefont {{Heng}},
  \citenamefont {{Keitel}},\ and\ \citenamefont
  {{Messenger}}}]{2025LRR....28....2C}%
  \BibitemOpen
  \bibfield  {author} {\bibinfo {author} {\bibfnamefont {E.}~\bibnamefont
  {{Cuoco}}}, \bibinfo {author} {\bibfnamefont {M.}~\bibnamefont
  {{Cavagli{\`a}}}}, \bibinfo {author} {\bibfnamefont {I.~S.}\ \bibnamefont
  {{Heng}}}, \bibinfo {author} {\bibfnamefont {D.}~\bibnamefont {{Keitel}}},\
  and\ \bibinfo {author} {\bibfnamefont {C.}~\bibnamefont {{Messenger}}},\
  }\bibfield  {title} {\bibinfo {title} {{Applications of machine learning in
  gravitational-wave research with current interferometric detectors:
  Applications of machine learning in gravitational-wave...}},\ }\href
  {https://doi.org/10.1007/s41114-024-00055-8} {\bibfield  {journal} {\bibinfo
  {journal} {Living Reviews in Relativity}\ }\textbf {\bibinfo {volume} {28}},\
  \bibinfo {eid} {2} (\bibinfo {year} {2025})},\ \Eprint
  {https://arxiv.org/abs/2412.15046} {arXiv:2412.15046 [gr-qc]} \BibitemShut
  {NoStop}%
\bibitem [{\citenamefont {Angelis}\ \emph {et~al.}(2023)\citenamefont
  {Angelis}, \citenamefont {Sofos},\ and\ \citenamefont
  {Karakasidis}}]{Angelis2023}%
  \BibitemOpen
  \bibfield  {author} {\bibinfo {author} {\bibfnamefont {D.}~\bibnamefont
  {Angelis}}, \bibinfo {author} {\bibfnamefont {F.}~\bibnamefont {Sofos}},\
  and\ \bibinfo {author} {\bibfnamefont {T.~E.}\ \bibnamefont {Karakasidis}},\
  }\bibfield  {title} {\bibinfo {title} {Artificial intelligence in physical
  sciences: Symbolic regression trends and perspectives},\ }\href
  {https://doi.org/10.1007/s11831-023-09922-z} {\bibfield  {journal} {\bibinfo
  {journal} {Archives of Computational Methods in Engineering}\ }\textbf
  {\bibinfo {volume} {30}},\ \bibinfo {pages} {3845} (\bibinfo {year}
  {2023})}\BibitemShut {NoStop}%
\bibitem [{\citenamefont {Tenachi}\ \emph {et~al.}(2023)\citenamefont
  {Tenachi}, \citenamefont {Ibata},\ and\ \citenamefont
  {Diakogiannis}}]{Tenachi_2023}%
  \BibitemOpen
  \bibfield  {author} {\bibinfo {author} {\bibfnamefont {W.}~\bibnamefont
  {Tenachi}}, \bibinfo {author} {\bibfnamefont {R.}~\bibnamefont {Ibata}},\
  and\ \bibinfo {author} {\bibfnamefont {F.~I.}\ \bibnamefont {Diakogiannis}},\
  }\bibfield  {title} {\bibinfo {title} {Deep symbolic regression for physics
  guided by units constraints: Toward the automated discovery of physical
  laws},\ }\href {https://doi.org/10.3847/1538-4357/ad014c} {\bibfield
  {journal} {\bibinfo  {journal} {The Astrophysical Journal}\ }\textbf
  {\bibinfo {volume} {959}},\ \bibinfo {pages} {99} (\bibinfo {year}
  {2023})}\BibitemShut {NoStop}%
\bibitem [{\citenamefont {Saeed}\ and\ \citenamefont
  {Omlin}(2023)}]{SAEED2023110273}%
  \BibitemOpen
  \bibfield  {author} {\bibinfo {author} {\bibfnamefont {W.}~\bibnamefont
  {Saeed}}\ and\ \bibinfo {author} {\bibfnamefont {C.}~\bibnamefont {Omlin}},\
  }\bibfield  {title} {\bibinfo {title} {Explainable ai (xai): A systematic
  meta-survey of current challenges and future opportunities},\ }\href
  {https://doi.org/https://doi.org/10.1016/j.knosys.2023.110273} {\bibfield
  {journal} {\bibinfo  {journal} {Knowledge-Based Systems}\ }\textbf {\bibinfo
  {volume} {263}},\ \bibinfo {pages} {110273} (\bibinfo {year}
  {2023})}\BibitemShut {NoStop}%
\bibitem [{\citenamefont {{Wong}}\ and\ \citenamefont
  {{Cranmer}}(2022)}]{2022mla..confE..25W}%
  \BibitemOpen
  \bibfield  {author} {\bibinfo {author} {\bibfnamefont {K.}~\bibnamefont
  {{Wong}}}\ and\ \bibinfo {author} {\bibfnamefont {M.}~\bibnamefont
  {{Cranmer}}},\ }\bibfield  {title} {\bibinfo {title} {{Automated discovery of
  interpretable gravitational-wave population models}},\ }in\ \href
  {https://doi.org/10.48550/arXiv.2207.12409} {\emph {\bibinfo {booktitle}
  {Machine Learning for Astrophysics}}}\ (\bibinfo {year} {2022})\ p.~\bibinfo
  {pages} {25},\ \Eprint {https://arxiv.org/abs/2207.12409} {arXiv:2207.12409
  [astro-ph.IM]} \BibitemShut {NoStop}%
\bibitem [{\citenamefont {{Nava}}\ \emph {et~al.}(2025)\citenamefont {{Nava}},
  \citenamefont {{Chatziioannou}}, \citenamefont {{Johnson}}, \citenamefont
  {{Miller}},\ and\ \citenamefont {{Thomas}}}]{2025AAS...24545007N}%
  \BibitemOpen
  \bibfield  {author} {\bibinfo {author} {\bibfnamefont {A.}~\bibnamefont
  {{Nava}}}, \bibinfo {author} {\bibfnamefont {K.}~\bibnamefont
  {{Chatziioannou}}}, \bibinfo {author} {\bibfnamefont {A.}~\bibnamefont
  {{Johnson}}}, \bibinfo {author} {\bibfnamefont {S.}~\bibnamefont
  {{Miller}}},\ and\ \bibinfo {author} {\bibfnamefont {L.}~\bibnamefont
  {{Thomas}}},\ }\bibfield  {title} {\bibinfo {title} {{Using Symbolic
  Regression to Characterise Degeneracies in Binary Black Hole Parameter
  Space}},\ }in\ \href@noop {} {\emph {\bibinfo {booktitle} {American
  Astronomical Society Meeting Abstracts}}},\ \bibinfo {series} {American
  Astronomical Society Meeting Abstracts}, Vol.\ \bibinfo {volume} {245}\
  (\bibinfo {year} {2025})\ p.\ \bibinfo {pages} {450.07}\BibitemShut {NoStop}%
\bibitem [{\citenamefont {{Yuan}}\ \emph {et~al.}(2025)\citenamefont {{Yuan}},
  \citenamefont {{Calz{\`a}}},\ and\ \citenamefont
  {{Pedrotti}}}]{2025arXiv250418270Y}%
  \BibitemOpen
  \bibfield  {author} {\bibinfo {author} {\bibfnamefont {G.-W.}\ \bibnamefont
  {{Yuan}}}, \bibinfo {author} {\bibfnamefont {M.}~\bibnamefont
  {{Calz{\`a}}}},\ and\ \bibinfo {author} {\bibfnamefont {D.}~\bibnamefont
  {{Pedrotti}}},\ }\bibfield  {title} {\bibinfo {title} {{ReGrayssion: Machine
  Learning-Based Analytical Expressions for Gray-Body Factors and Application
  to Primordial Black Holes}},\ }\href
  {https://doi.org/10.48550/arXiv.2504.18270} {\bibfield  {journal} {\bibinfo
  {journal} {arXiv e-prints}\ ,\ \bibinfo {eid} {arXiv:2504.18270}} (\bibinfo
  {year} {2025})},\ \Eprint {https://arxiv.org/abs/2504.18270}
  {arXiv:2504.18270 [astro-ph.CO]} \BibitemShut {NoStop}%
\bibitem [{\citenamefont {{Patra}}\ \emph {et~al.}(2025)\citenamefont
  {{Patra}}, \citenamefont {{Malik}}, \citenamefont {{Pais}}, \citenamefont
  {{Zhou}}, \citenamefont {{Agrawal}},\ and\ \citenamefont
  {{Provid{\^e}ncia}}}]{2025PhLB..86539470P}%
  \BibitemOpen
  \bibfield  {author} {\bibinfo {author} {\bibfnamefont {N.~K.}\ \bibnamefont
  {{Patra}}}, \bibinfo {author} {\bibfnamefont {T.}~\bibnamefont {{Malik}}},
  \bibinfo {author} {\bibfnamefont {H.}~\bibnamefont {{Pais}}}, \bibinfo
  {author} {\bibfnamefont {K.}~\bibnamefont {{Zhou}}}, \bibinfo {author}
  {\bibfnamefont {B.~K.}\ \bibnamefont {{Agrawal}}},\ and\ \bibinfo {author}
  {\bibfnamefont {C.}~\bibnamefont {{Provid{\^e}ncia}}},\ }\bibfield  {title}
  {\bibinfo {title} {{Inferring the equation of state from neutron star
  observables via machine learning}},\ }\href
  {https://doi.org/10.1016/j.physletb.2025.139470} {\bibfield  {journal}
  {\bibinfo  {journal} {Physics Letters B}\ }\textbf {\bibinfo {volume}
  {865}},\ \bibinfo {eid} {139470} (\bibinfo {year} {2025})},\ \Eprint
  {https://arxiv.org/abs/2502.20226} {arXiv:2502.20226 [nucl-th]} \BibitemShut
  {NoStop}%
\bibitem [{\citenamefont {{Imam}}\ \emph {et~al.}(2024)\citenamefont {{Imam}},
  \citenamefont {{Saxena}}, \citenamefont {{Malik}}, \citenamefont {{Patra}},\
  and\ \citenamefont {{Agrawal}}}]{2024PhRvD.110j3042I}%
  \BibitemOpen
  \bibfield  {author} {\bibinfo {author} {\bibfnamefont {S.~M.~A.}\
  \bibnamefont {{Imam}}}, \bibinfo {author} {\bibfnamefont {P.}~\bibnamefont
  {{Saxena}}}, \bibinfo {author} {\bibfnamefont {T.}~\bibnamefont {{Malik}}},
  \bibinfo {author} {\bibfnamefont {N.~K.}\ \bibnamefont {{Patra}}},\ and\
  \bibinfo {author} {\bibfnamefont {B.~K.}\ \bibnamefont {{Agrawal}}},\
  }\bibfield  {title} {\bibinfo {title} {{Calibrating global behavior of
  equation of state by combining nuclear and astrophysics inputs in a machine
  learning approach}},\ }\href {https://doi.org/10.1103/PhysRevD.110.103042}
  {\bibfield  {journal} {\bibinfo  {journal} {\prd}\ }\textbf {\bibinfo
  {volume} {110}},\ \bibinfo {eid} {103042} (\bibinfo {year} {2024})},\ \Eprint
  {https://arxiv.org/abs/2407.08553} {arXiv:2407.08553 [nucl-th]} \BibitemShut
  {NoStop}%
\bibitem [{\citenamefont {{Tolman}}(1939)}]{1939PhRv...55..364T}%
  \BibitemOpen
  \bibfield  {author} {\bibinfo {author} {\bibfnamefont {R.~C.}\ \bibnamefont
  {{Tolman}}},\ }\bibfield  {title} {\bibinfo {title} {{Static Solutions of
  Einstein's Field Equations for Spheres of Fluid}},\ }\href
  {https://doi.org/10.1103/PhysRev.55.364} {\bibfield  {journal} {\bibinfo
  {journal} {Physical Review}\ }\textbf {\bibinfo {volume} {55}},\ \bibinfo
  {pages} {364} (\bibinfo {year} {1939})}\BibitemShut {NoStop}%
\bibitem [{\citenamefont {{Oppenheimer}}\ and\ \citenamefont
  {{Volkoff}}(1939)}]{1939PhRv...55..374O}%
  \BibitemOpen
  \bibfield  {author} {\bibinfo {author} {\bibfnamefont {J.~R.}\ \bibnamefont
  {{Oppenheimer}}}\ and\ \bibinfo {author} {\bibfnamefont {G.~M.}\ \bibnamefont
  {{Volkoff}}},\ }\bibfield  {title} {\bibinfo {title} {{On Massive Neutron
  Cores}},\ }\href {https://doi.org/10.1103/PhysRev.55.374} {\bibfield
  {journal} {\bibinfo  {journal} {Physical Review}\ }\textbf {\bibinfo {volume}
  {55}},\ \bibinfo {pages} {374} (\bibinfo {year} {1939})}\BibitemShut
  {NoStop}%
\bibitem [{\citenamefont {{Love}}(1911)}]{1911spge.book.....L}%
  \BibitemOpen
  \bibfield  {author} {\bibinfo {author} {\bibfnamefont {A.~E.~H.}\
  \bibnamefont {{Love}}},\ }\href@noop {} {\emph {\bibinfo {title} {{Some
  Problems of Geodynamics}}}}\ (\bibinfo  {publisher} {Cambridge University
  Press},\ \bibinfo {year} {1911})\BibitemShut {NoStop}%
\bibitem [{\citenamefont {{Flanagan}}\ and\ \citenamefont
  {{Hinderer}}(2008)}]{2008PhRvD..77b1502F}%
  \BibitemOpen
  \bibfield  {author} {\bibinfo {author} {\bibfnamefont {{\'E}.~{\'E}.}\
  \bibnamefont {{Flanagan}}}\ and\ \bibinfo {author} {\bibfnamefont
  {T.}~\bibnamefont {{Hinderer}}},\ }\bibfield  {title} {\bibinfo {title}
  {{Constraining neutron-star tidal Love numbers with gravitational-wave
  detectors}},\ }\href {https://doi.org/10.1103/PhysRevD.77.021502} {\bibfield
  {journal} {\bibinfo  {journal} {\prd}\ }\textbf {\bibinfo {volume} {77}},\
  \bibinfo {eid} {021502} (\bibinfo {year} {2008})},\ \Eprint
  {https://arxiv.org/abs/0709.1915} {arXiv:0709.1915 [astro-ph]} \BibitemShut
  {NoStop}%
\bibitem [{\citenamefont {{Van Oeveren}}\ and\ \citenamefont
  {{Friedman}}(2017)}]{2017PhRvD..95h3014V}%
  \BibitemOpen
  \bibfield  {author} {\bibinfo {author} {\bibfnamefont {E.~D.}\ \bibnamefont
  {{Van Oeveren}}}\ and\ \bibinfo {author} {\bibfnamefont {J.~L.}\ \bibnamefont
  {{Friedman}}},\ }\bibfield  {title} {\bibinfo {title} {{Upper limit set by
  causality on the tidal deformability of a neutron star}},\ }\href
  {https://doi.org/10.1103/PhysRevD.95.083014} {\bibfield  {journal} {\bibinfo
  {journal} {\prd}\ }\textbf {\bibinfo {volume} {95}},\ \bibinfo {eid} {083014}
  (\bibinfo {year} {2017})},\ \Eprint {https://arxiv.org/abs/1701.03797}
  {arXiv:1701.03797 [gr-qc]} \BibitemShut {NoStop}%
\bibitem [{\citenamefont {{Cranmer}}(2023)}]{2023arXiv230501582C}%
  \BibitemOpen
  \bibfield  {author} {\bibinfo {author} {\bibfnamefont {M.}~\bibnamefont
  {{Cranmer}}},\ }\bibfield  {title} {\bibinfo {title} {{Interpretable Machine
  Learning for Science with PySR and SymbolicRegression.jl}},\ }\href
  {https://doi.org/10.48550/arXiv.2305.01582} {\bibfield  {journal} {\bibinfo
  {journal} {arXiv e-prints}\ ,\ \bibinfo {eid} {arXiv:2305.01582}} (\bibinfo
  {year} {2023})},\ \Eprint {https://arxiv.org/abs/2305.01582}
  {arXiv:2305.01582 [astro-ph.IM]} \BibitemShut {NoStop}%
\bibitem [{\citenamefont {{Python Software Foundation}}(2025)}]{python}%
  \BibitemOpen
  \bibfield  {author} {\bibinfo {author} {\bibnamefont {{Python Software
  Foundation}}},\ }\href@noop {} {\emph {\bibinfo {title} {Python Language
  Reference, version 3.13.3}}} (\bibinfo {year} {2025}),\ \bibinfo {note}
  {\url{https://docs.python.org/3/reference/}}\BibitemShut {NoStop}%
\bibitem [{\citenamefont {Bezanson}\ \emph {et~al.}(2017)\citenamefont
  {Bezanson}, \citenamefont {Edelman}, \citenamefont {Karpinski},\ and\
  \citenamefont {Shah}}]{doi:10.1137/141000671}%
  \BibitemOpen
  \bibfield  {author} {\bibinfo {author} {\bibfnamefont {J.}~\bibnamefont
  {Bezanson}}, \bibinfo {author} {\bibfnamefont {A.}~\bibnamefont {Edelman}},
  \bibinfo {author} {\bibfnamefont {S.}~\bibnamefont {Karpinski}},\ and\
  \bibinfo {author} {\bibfnamefont {V.~B.}\ \bibnamefont {Shah}},\ }\bibfield
  {title} {\bibinfo {title} {Julia: A fresh approach to numerical computing},\
  }\href {https://doi.org/10.1137/141000671} {\bibfield  {journal} {\bibinfo
  {journal} {SIAM Review}\ }\textbf {\bibinfo {volume} {59}},\ \bibinfo {pages}
  {65} (\bibinfo {year} {2017})},\ \Eprint
  {https://arxiv.org/abs/https://doi.org/10.1137/141000671}
  {https://doi.org/10.1137/141000671} \BibitemShut {NoStop}%
\bibitem [{\citenamefont {Smits}\ and\ \citenamefont
  {Kotanchek}(2005)}]{Smits2005}%
  \BibitemOpen
  \bibfield  {author} {\bibinfo {author} {\bibfnamefont {G.~F.}\ \bibnamefont
  {Smits}}\ and\ \bibinfo {author} {\bibfnamefont {M.}~\bibnamefont
  {Kotanchek}},\ }\bibinfo {title} {Pareto-front exploitation in symbolic
  regression},\ in\ \href {https://doi.org/10.1007/0-387-23254-0_17} {\emph
  {\bibinfo {booktitle} {Genetic Programming Theory and Practice II}}}\
  (\bibinfo  {publisher} {Springer US},\ \bibinfo {address} {Boston, MA},\
  \bibinfo {year} {2005})\ pp.\ \bibinfo {pages} {283--299}\BibitemShut
  {NoStop}%
\bibitem [{\citenamefont {{Tooper}}(1965)}]{1965ApJ...142.1541T}%
  \BibitemOpen
  \bibfield  {author} {\bibinfo {author} {\bibfnamefont {R.~F.}\ \bibnamefont
  {{Tooper}}},\ }\bibfield  {title} {\bibinfo {title} {{Adiabatic Fluid Spheres
  in General Relativity.}},\ }\href {https://doi.org/10.1086/148435} {\bibfield
   {journal} {\bibinfo  {journal} {\apj}\ }\textbf {\bibinfo {volume} {142}},\
  \bibinfo {pages} {1541} (\bibinfo {year} {1965})}\BibitemShut {NoStop}%
\bibitem [{\citenamefont {{Douchin}}\ and\ \citenamefont
  {{Haensel}}(2001)}]{DouchinH2001}%
  \BibitemOpen
  \bibfield  {author} {\bibinfo {author} {\bibfnamefont {F.}~\bibnamefont
  {{Douchin}}}\ and\ \bibinfo {author} {\bibfnamefont {P.}~\bibnamefont
  {{Haensel}}},\ }\bibfield  {title} {\bibinfo {title} {{A unified equation of
  state of dense matter and neutron star structure}},\ }\href
  {https://doi.org/10.1051/0004-6361:20011402} {\bibfield  {journal} {\bibinfo
  {journal} {\aap}\ }\textbf {\bibinfo {volume} {380}},\ \bibinfo {pages} {151}
  (\bibinfo {year} {2001})},\ \Eprint {https://arxiv.org/abs/astro-ph/0111092}
  {arXiv:astro-ph/0111092 [astro-ph]} \BibitemShut {NoStop}%
\bibitem [{\citenamefont {{Haensel}}\ and\ \citenamefont
  {{Pichon}}(1994)}]{1994A&A...283..313H}%
  \BibitemOpen
  \bibfield  {author} {\bibinfo {author} {\bibfnamefont {P.}~\bibnamefont
  {{Haensel}}}\ and\ \bibinfo {author} {\bibfnamefont {B.}~\bibnamefont
  {{Pichon}}},\ }\bibfield  {title} {\bibinfo {title} {{Experimental nuclear
  masses and the ground state of cold dense matter}},\ }\href
  {https://doi.org/10.48550/arXiv.nucl-th/9310003} {\bibfield  {journal}
  {\bibinfo  {journal} {\aap}\ }\textbf {\bibinfo {volume} {283}},\ \bibinfo
  {pages} {313} (\bibinfo {year} {1994})},\ \Eprint
  {https://arxiv.org/abs/nucl-th/9310003} {arXiv:nucl-th/9310003 [nucl-th]}
  \BibitemShut {NoStop}%
\bibitem [{\citenamefont {{Wiringa}}\ \emph {et~al.}(1988)\citenamefont
  {{Wiringa}}, \citenamefont {{Fiks}},\ and\ \citenamefont
  {{Fabrocini}}}]{1988PhRvC..38.1010W}%
  \BibitemOpen
  \bibfield  {author} {\bibinfo {author} {\bibfnamefont {R.~B.}\ \bibnamefont
  {{Wiringa}}}, \bibinfo {author} {\bibfnamefont {V.}~\bibnamefont {{Fiks}}},\
  and\ \bibinfo {author} {\bibfnamefont {A.}~\bibnamefont {{Fabrocini}}},\
  }\bibfield  {title} {\bibinfo {title} {{Equation of state for dense nucleon
  matter}},\ }\href {https://doi.org/10.1103/PhysRevC.38.1010} {\bibfield
  {journal} {\bibinfo  {journal} {\prc}\ }\textbf {\bibinfo {volume} {38}},\
  \bibinfo {pages} {1010} (\bibinfo {year} {1988})}\BibitemShut {NoStop}%
\bibitem [{\citenamefont {{Akmal}}\ \emph {et~al.}(1998)\citenamefont
  {{Akmal}}, \citenamefont {{Pandharipande}},\ and\ \citenamefont
  {{Ravenhall}}}]{1998PhRvC..58.1804A}%
  \BibitemOpen
  \bibfield  {author} {\bibinfo {author} {\bibfnamefont {A.}~\bibnamefont
  {{Akmal}}}, \bibinfo {author} {\bibfnamefont {V.~R.}\ \bibnamefont
  {{Pandharipande}}},\ and\ \bibinfo {author} {\bibfnamefont {D.~G.}\
  \bibnamefont {{Ravenhall}}},\ }\bibfield  {title} {\bibinfo {title}
  {{Equation of state of nucleon matter and neutron star structure}},\ }\href
  {https://doi.org/10.1103/PhysRevC.58.1804} {\bibfield  {journal} {\bibinfo
  {journal} {\prc}\ }\textbf {\bibinfo {volume} {58}},\ \bibinfo {pages} {1804}
  (\bibinfo {year} {1998})},\ \Eprint {https://arxiv.org/abs/nucl-th/9804027}
  {arXiv:nucl-th/9804027 [nucl-th]} \BibitemShut {NoStop}%
\bibitem [{\citenamefont {{Goriely}}\ \emph {et~al.}(2010)\citenamefont
  {{Goriely}}, \citenamefont {{Chamel}},\ and\ \citenamefont
  {{Pearson}}}]{2010PhRvC..82c5804G}%
  \BibitemOpen
  \bibfield  {author} {\bibinfo {author} {\bibfnamefont {S.}~\bibnamefont
  {{Goriely}}}, \bibinfo {author} {\bibfnamefont {N.}~\bibnamefont
  {{Chamel}}},\ and\ \bibinfo {author} {\bibfnamefont {J.~M.}\ \bibnamefont
  {{Pearson}}},\ }\bibfield  {title} {\bibinfo {title} {{Further explorations
  of Skyrme-Hartree-Fock-Bogoliubov mass formulas. XII. Stiffness and stability
  of neutron-star matter}},\ }\href
  {https://doi.org/10.1103/PhysRevC.82.035804} {\bibfield  {journal} {\bibinfo
  {journal} {\prc}\ }\textbf {\bibinfo {volume} {82}},\ \bibinfo {eid} {035804}
  (\bibinfo {year} {2010})},\ \Eprint {https://arxiv.org/abs/1009.3840}
  {arXiv:1009.3840 [nucl-th]} \BibitemShut {NoStop}%
\bibitem [{\citenamefont {{Bednarek}}\ \emph {et~al.}(2012)\citenamefont
  {{Bednarek}}, \citenamefont {{Haensel}}, \citenamefont {{Zdunik}},
  \citenamefont {{Bejger}},\ and\ \citenamefont
  {{Ma{\'n}ka}}}]{2012A&A...543A.157B}%
  \BibitemOpen
  \bibfield  {author} {\bibinfo {author} {\bibfnamefont {I.}~\bibnamefont
  {{Bednarek}}}, \bibinfo {author} {\bibfnamefont {P.}~\bibnamefont
  {{Haensel}}}, \bibinfo {author} {\bibfnamefont {J.~L.}\ \bibnamefont
  {{Zdunik}}}, \bibinfo {author} {\bibfnamefont {M.}~\bibnamefont {{Bejger}}},\
  and\ \bibinfo {author} {\bibfnamefont {R.}~\bibnamefont {{Ma{\'n}ka}}},\
  }\bibfield  {title} {\bibinfo {title} {{Hyperons in neutron-star cores and a
  2 M$_{{\ensuremath{\odot}}}$ pulsar}},\ }\href
  {https://doi.org/10.1051/0004-6361/201118560} {\bibfield  {journal} {\bibinfo
   {journal} {\aap}\ }\textbf {\bibinfo {volume} {543}},\ \bibinfo {eid} {A157}
  (\bibinfo {year} {2012})},\ \Eprint {https://arxiv.org/abs/1111.6942}
  {arXiv:1111.6942 [astro-ph.SR]} \BibitemShut {NoStop}%
\bibitem [{\citenamefont {Lalazissis}\ \emph {et~al.}(1997)\citenamefont
  {Lalazissis}, \citenamefont {K\"onig},\ and\ \citenamefont
  {Ring}}]{PhysRevC.55.540}%
  \BibitemOpen
  \bibfield  {author} {\bibinfo {author} {\bibfnamefont {G.~A.}\ \bibnamefont
  {Lalazissis}}, \bibinfo {author} {\bibfnamefont {J.}~\bibnamefont
  {K\"onig}},\ and\ \bibinfo {author} {\bibfnamefont {P.}~\bibnamefont
  {Ring}},\ }\bibfield  {title} {\bibinfo {title} {New parametrization for the
  lagrangian density of relativistic mean field theory},\ }\href
  {https://doi.org/10.1103/PhysRevC.55.540} {\bibfield  {journal} {\bibinfo
  {journal} {Phys. Rev. C}\ }\textbf {\bibinfo {volume} {55}},\ \bibinfo
  {pages} {540} (\bibinfo {year} {1997})}\BibitemShut {NoStop}%
\bibitem [{\citenamefont {{Sulaksono}}\ and\ \citenamefont
  {{Agrawal}}(2012)}]{2012NuPhA.895...44S}%
  \BibitemOpen
  \bibfield  {author} {\bibinfo {author} {\bibfnamefont {A.}~\bibnamefont
  {{Sulaksono}}}\ and\ \bibinfo {author} {\bibfnamefont {B.~K.}\ \bibnamefont
  {{Agrawal}}},\ }\bibfield  {title} {\bibinfo {title} {{Existence of hyperons
  in the pulsar PSRJ1614-2230}},\ }\href
  {https://doi.org/10.1016/j.nuclphysa.2012.09.006} {\bibfield  {journal}
  {\bibinfo  {journal} {\nphysa}\ }\textbf {\bibinfo {volume} {895}},\ \bibinfo
  {pages} {44} (\bibinfo {year} {2012})},\ \Eprint
  {https://arxiv.org/abs/1209.6160} {arXiv:1209.6160 [nucl-th]} \BibitemShut
  {NoStop}%
\bibitem [{\citenamefont {{Seidov}}(1971)}]{1971SvA....15..347S}%
  \BibitemOpen
  \bibfield  {author} {\bibinfo {author} {\bibfnamefont {Z.~F.}\ \bibnamefont
  {{Seidov}}},\ }\bibfield  {title} {\bibinfo {title} {{The Stability of a Star
  with a Phase Change in General Relativity Theory}},\ }\href@noop {}
  {\bibfield  {journal} {\bibinfo  {journal} {\sovast}\ }\textbf {\bibinfo
  {volume} {15}},\ \bibinfo {pages} {347} (\bibinfo {year} {1971})}\BibitemShut
  {NoStop}%
\bibitem [{\citenamefont {{Abbott}}\ \emph {et~al.}(2018)\citenamefont
  {{Abbott}}, \citenamefont {{Abbott}}, \citenamefont {{Abbott}}, \citenamefont
  {{Acernese}}, \citenamefont {{Ackley}}, \citenamefont {{Adams}},
  \citenamefont {{Adams}}, \citenamefont {{Addesso}}, \citenamefont
  {{Adhikari}}, \citenamefont {{Adya}}, \citenamefont {{Affeldt}},
  \citenamefont {{Agarwal}}, \citenamefont {{Agathos}}, \citenamefont
  {{Agatsuma}}, \citenamefont {{Aggarwal}}, \citenamefont {{Aguiar}},
  \citenamefont {{Aiello}}, \citenamefont {{Ain}}, \citenamefont {{Ajith}},
  \citenamefont {{Allen}}, \citenamefont {{Allen}}, \citenamefont {{Allocca}},
  \citenamefont {{Aloy}}, \citenamefont {{Altin}}, \citenamefont {{Amato}},
  \citenamefont {{Ananyeva}}, \citenamefont {{Anderson}}, \citenamefont
  {{Anderson}}, \citenamefont {{Angelova}}, \citenamefont {{Antier}},
  \citenamefont {{Appert}}, \citenamefont {{Arai}}, \citenamefont {{Araya}},
  \citenamefont {{Areeda}}, \citenamefont {{Ar{\`e}ne}}, \citenamefont
  {{Arnaud}}, \citenamefont {{Arun}}, \citenamefont {{Ascenzi}}, \citenamefont
  {{Ashton}}, \citenamefont {{Ast}}, \citenamefont {{Aston}}, \citenamefont
  {{Astone}}, \citenamefont {{Atallah}}, \citenamefont {{Aubin}}, \citenamefont
  {{Aufmuth}}, \citenamefont {{Aulbert}}, \citenamefont {{AultONeal}},
  \citenamefont {{Austin}}, \citenamefont {{Avila-Alvarez}}, \citenamefont
  {{Babak}}, \citenamefont {{Bacon}}, \citenamefont {{Badaracco}},
  \citenamefont {{Bader}}, \citenamefont {{Bae}}, \citenamefont {{Baker}},
  \citenamefont {{Baldaccini}}, \citenamefont {{Ballardin}}, \citenamefont
  {{Ballmer}}, \citenamefont {{Banagiri}}, \citenamefont {{Barayoga}},
  \citenamefont {{Barclay}}, \citenamefont {{Barish}}, \citenamefont
  {{Barker}}, \citenamefont {{Barkett}}, \citenamefont {{Barnum}},
  \citenamefont {{Barone}}, \citenamefont {{Barr}}, \citenamefont {{Barsotti}},
  \citenamefont {{Barsuglia}}, \citenamefont {{Barta}}, \citenamefont
  {{Bartlett}}, \citenamefont {{Bartos}}, \citenamefont {{Bassiri}},
  \citenamefont {{Basti}}, \citenamefont {{Batch}}, \citenamefont {{Bawaj}},
  \citenamefont {{Bayley}}, \citenamefont {{Bazzan}}, \citenamefont
  {{B{\'e}csy}}, \citenamefont {{Beer}}, \citenamefont {{Bejger}},
  \citenamefont {{Belahcene}}, \citenamefont {{Bell}}, \citenamefont
  {{Beniwal}}, \citenamefont {{Bensch}}, \citenamefont {{Berger}},
  \citenamefont {{Bergmann}}, \citenamefont {{Bernuzzi}}, \citenamefont
  {{Bero}}, \citenamefont {{Berry}}, \citenamefont {{Bersanetti}},
  \citenamefont {{Bertolini}}, \citenamefont {{Betzwieser}}, \citenamefont
  {{Bhandare}}, \citenamefont {{Bilenko}}, \citenamefont {{Bilgili}},
  \citenamefont {{Billingsley}}, \citenamefont {{Billman}}, \citenamefont
  {{Birch}}, \citenamefont {{Birney}}, \citenamefont {{Birnholtz}},
  \citenamefont {{Biscans}}, \citenamefont {{Biscoveanu}}, \citenamefont
  {{Bisht}}, \citenamefont {{Bitossi}}, \citenamefont {{Bizouard}},
  \citenamefont {{Blackburn}}, \citenamefont {{Blackman}}, \citenamefont
  {{Blair}}, \citenamefont {{Blair}}, \citenamefont {{Blair}}, \citenamefont
  {{Bloemen}}, \citenamefont {{Bock}}, \citenamefont {{Bode}}, \citenamefont
  {{Boer}}, \citenamefont {{Boetzel}}, \citenamefont {{Bogaert}}, \citenamefont
  {{Bohe}}, \citenamefont {{Bondu}}, \citenamefont {{Bonilla}}, \citenamefont
  {{Bonnand}}, \citenamefont {{Booker}}, \citenamefont {{Boom}}, \citenamefont
  {{Booth}}, \citenamefont {{Bork}}, \citenamefont {{Boschi}}, \citenamefont
  {{Bose}}, \citenamefont {{Bossie}}, \citenamefont {{Bossilkov}},
  \citenamefont {{Bosveld}}, \citenamefont {{Bouffanais}}, \citenamefont
  {{Bozzi}}, \citenamefont {{Bradaschia}}, \citenamefont {{Brady}},
  \citenamefont {{Bramley}}, \citenamefont {{Branchesi}}, \citenamefont
  {{Brau}}, \citenamefont {{Briant}}, \citenamefont {{Brighenti}},
  \citenamefont {{Brillet}}, \citenamefont {{Brinkmann}}, \citenamefont
  {{Brisson}}, \citenamefont {{Brockill}}, \citenamefont {{Brooks}},
  \citenamefont {{Brown}}, \citenamefont {{Brunett}}, \citenamefont
  {{Buchanan}}, \citenamefont {{Buikema}}, \citenamefont {{Bulik}},
  \citenamefont {{Bulten}}, \citenamefont {{Buonanno}}, \citenamefont
  {{Buskulic}}, \citenamefont {{Buy}}, \citenamefont {{Byer}}, \citenamefont
  {{Cabero}}, \citenamefont {{Cadonati}}, \citenamefont {{Cagnoli}},
  \citenamefont {{Cahillane}}, \citenamefont {{Calder{\'o}n Bustillo}},
  \citenamefont {{Callister}}, \citenamefont {{Calloni}}, \citenamefont
  {{Camp}}, \citenamefont {{Canepa}}, \citenamefont {{Canizares}},
  \citenamefont {{Cannon}}, \citenamefont {{Cao}}, \citenamefont {{Cao}},
  \citenamefont {{Capano}}, \citenamefont {{Capocasa}}, \citenamefont
  {{Carbognani}}, \citenamefont {{Caride}}, \citenamefont {{Carney}},
  \citenamefont {{Carullo}}, \citenamefont {{Casanueva Diaz}}, \citenamefont
  {{Casentini}}, \citenamefont {{Caudill}}, \citenamefont {{Cavagli{\`a}}},
  \citenamefont {{Cavalier}}, \citenamefont {{Cavalieri}}, \citenamefont
  {{Cella}}, \citenamefont {{Cepeda}}, \citenamefont {{Cerd{\'a}-Dur{\'a}n}},
  \citenamefont {{Cerretani}}, \citenamefont {{Cesarini}}, \citenamefont
  {{Chaibi}}, \citenamefont {{Chamberlin}}, \citenamefont {{Chan}},
  \citenamefont {{Chao}}, \citenamefont {{Charlton}}, \citenamefont {{Chase}},
  \citenamefont {{Chassande-Mottin}}, \citenamefont {{Chatterjee}},
  \citenamefont {{Chatziioannou}}, \citenamefont {{Cheeseboro}}, \citenamefont
  {{Chen}}, \citenamefont {{Chen}}, \citenamefont {{Chen}}, \citenamefont
  {{Cheng}}, \citenamefont {{Chia}},\ and\ \citenamefont
  {{Chincarini}}}]{2018PhRvL.121p1101A}%
  \BibitemOpen
  \bibfield  {author} {\bibinfo {author} {\bibfnamefont {B.~P.}\ \bibnamefont
  {{Abbott}}}, \bibinfo {author} {\bibfnamefont {R.}~\bibnamefont {{Abbott}}},
  \bibinfo {author} {\bibfnamefont {T.~D.}\ \bibnamefont {{Abbott}}}, \bibinfo
  {author} {\bibfnamefont {F.}~\bibnamefont {{Acernese}}}, \bibinfo {author}
  {\bibfnamefont {K.}~\bibnamefont {{Ackley}}}, \bibinfo {author}
  {\bibfnamefont {C.}~\bibnamefont {{Adams}}}, \bibinfo {author} {\bibfnamefont
  {T.}~\bibnamefont {{Adams}}}, \bibinfo {author} {\bibfnamefont
  {P.}~\bibnamefont {{Addesso}}}, \bibinfo {author} {\bibfnamefont {R.~X.}\
  \bibnamefont {{Adhikari}}}, \bibinfo {author} {\bibfnamefont {V.~B.}\
  \bibnamefont {{Adya}}}, \bibinfo {author} {\bibfnamefont {C.}~\bibnamefont
  {{Affeldt}}}, \bibinfo {author} {\bibfnamefont {B.}~\bibnamefont
  {{Agarwal}}}, \bibinfo {author} {\bibfnamefont {M.}~\bibnamefont
  {{Agathos}}}, \bibinfo {author} {\bibfnamefont {K.}~\bibnamefont
  {{Agatsuma}}}, \bibinfo {author} {\bibfnamefont {N.}~\bibnamefont
  {{Aggarwal}}}, \bibinfo {author} {\bibfnamefont {O.~D.}\ \bibnamefont
  {{Aguiar}}}, \bibinfo {author} {\bibfnamefont {L.}~\bibnamefont {{Aiello}}},
  \bibinfo {author} {\bibfnamefont {A.}~\bibnamefont {{Ain}}}, \bibinfo
  {author} {\bibfnamefont {P.}~\bibnamefont {{Ajith}}}, \bibinfo {author}
  {\bibfnamefont {B.}~\bibnamefont {{Allen}}}, \bibinfo {author} {\bibfnamefont
  {G.}~\bibnamefont {{Allen}}}, \bibinfo {author} {\bibfnamefont
  {A.}~\bibnamefont {{Allocca}}}, \bibinfo {author} {\bibfnamefont {M.~A.}\
  \bibnamefont {{Aloy}}}, \bibinfo {author} {\bibfnamefont {P.~A.}\
  \bibnamefont {{Altin}}}, \bibinfo {author} {\bibfnamefont {A.}~\bibnamefont
  {{Amato}}}, \bibinfo {author} {\bibfnamefont {A.}~\bibnamefont {{Ananyeva}}},
  \bibinfo {author} {\bibfnamefont {S.~B.}\ \bibnamefont {{Anderson}}},
  \bibinfo {author} {\bibfnamefont {W.~G.}\ \bibnamefont {{Anderson}}},
  \bibinfo {author} {\bibfnamefont {S.~V.}\ \bibnamefont {{Angelova}}},
  \bibinfo {author} {\bibfnamefont {S.}~\bibnamefont {{Antier}}}, \bibinfo
  {author} {\bibfnamefont {S.}~\bibnamefont {{Appert}}}, \bibinfo {author}
  {\bibfnamefont {K.}~\bibnamefont {{Arai}}}, \bibinfo {author} {\bibfnamefont
  {M.~C.}\ \bibnamefont {{Araya}}}, \bibinfo {author} {\bibfnamefont {J.~S.}\
  \bibnamefont {{Areeda}}}, \bibinfo {author} {\bibfnamefont {M.}~\bibnamefont
  {{Ar{\`e}ne}}}, \bibinfo {author} {\bibfnamefont {N.}~\bibnamefont
  {{Arnaud}}}, \bibinfo {author} {\bibfnamefont {K.~G.}\ \bibnamefont
  {{Arun}}}, \bibinfo {author} {\bibfnamefont {S.}~\bibnamefont {{Ascenzi}}},
  \bibinfo {author} {\bibfnamefont {G.}~\bibnamefont {{Ashton}}}, \bibinfo
  {author} {\bibfnamefont {M.}~\bibnamefont {{Ast}}}, \bibinfo {author}
  {\bibfnamefont {S.~M.}\ \bibnamefont {{Aston}}}, \bibinfo {author}
  {\bibfnamefont {P.}~\bibnamefont {{Astone}}}, \bibinfo {author}
  {\bibfnamefont {D.~V.}\ \bibnamefont {{Atallah}}}, \bibinfo {author}
  {\bibfnamefont {F.}~\bibnamefont {{Aubin}}}, \bibinfo {author} {\bibfnamefont
  {P.}~\bibnamefont {{Aufmuth}}}, \bibinfo {author} {\bibfnamefont
  {C.}~\bibnamefont {{Aulbert}}}, \bibinfo {author} {\bibfnamefont
  {K.}~\bibnamefont {{AultONeal}}}, \bibinfo {author} {\bibfnamefont
  {C.}~\bibnamefont {{Austin}}}, \bibinfo {author} {\bibfnamefont
  {A.}~\bibnamefont {{Avila-Alvarez}}}, \bibinfo {author} {\bibfnamefont
  {S.}~\bibnamefont {{Babak}}}, \bibinfo {author} {\bibfnamefont
  {P.}~\bibnamefont {{Bacon}}}, \bibinfo {author} {\bibfnamefont
  {F.}~\bibnamefont {{Badaracco}}}, \bibinfo {author} {\bibfnamefont
  {M.~K.~M.}\ \bibnamefont {{Bader}}}, \bibinfo {author} {\bibfnamefont
  {S.}~\bibnamefont {{Bae}}}, \bibinfo {author} {\bibfnamefont {P.~T.}\
  \bibnamefont {{Baker}}}, \bibinfo {author} {\bibfnamefont {F.}~\bibnamefont
  {{Baldaccini}}}, \bibinfo {author} {\bibfnamefont {G.}~\bibnamefont
  {{Ballardin}}}, \bibinfo {author} {\bibfnamefont {S.~W.}\ \bibnamefont
  {{Ballmer}}}, \bibinfo {author} {\bibfnamefont {S.}~\bibnamefont
  {{Banagiri}}}, \bibinfo {author} {\bibfnamefont {J.~C.}\ \bibnamefont
  {{Barayoga}}}, \bibinfo {author} {\bibfnamefont {S.~E.}\ \bibnamefont
  {{Barclay}}}, \bibinfo {author} {\bibfnamefont {B.~C.}\ \bibnamefont
  {{Barish}}}, \bibinfo {author} {\bibfnamefont {D.}~\bibnamefont {{Barker}}},
  \bibinfo {author} {\bibfnamefont {K.}~\bibnamefont {{Barkett}}}, \bibinfo
  {author} {\bibfnamefont {S.}~\bibnamefont {{Barnum}}}, \bibinfo {author}
  {\bibfnamefont {F.}~\bibnamefont {{Barone}}}, \bibinfo {author}
  {\bibfnamefont {B.}~\bibnamefont {{Barr}}}, \bibinfo {author} {\bibfnamefont
  {L.}~\bibnamefont {{Barsotti}}}, \bibinfo {author} {\bibfnamefont
  {M.}~\bibnamefont {{Barsuglia}}}, \bibinfo {author} {\bibfnamefont
  {D.}~\bibnamefont {{Barta}}}, \bibinfo {author} {\bibfnamefont
  {J.}~\bibnamefont {{Bartlett}}}, \bibinfo {author} {\bibfnamefont
  {I.}~\bibnamefont {{Bartos}}}, \bibinfo {author} {\bibfnamefont
  {R.}~\bibnamefont {{Bassiri}}}, \bibinfo {author} {\bibfnamefont
  {A.}~\bibnamefont {{Basti}}}, \bibinfo {author} {\bibfnamefont {J.~C.}\
  \bibnamefont {{Batch}}}, \bibinfo {author} {\bibfnamefont {M.}~\bibnamefont
  {{Bawaj}}}, \bibinfo {author} {\bibfnamefont {J.~C.}\ \bibnamefont
  {{Bayley}}}, \bibinfo {author} {\bibfnamefont {M.}~\bibnamefont {{Bazzan}}},
  \bibinfo {author} {\bibfnamefont {B.}~\bibnamefont {{B{\'e}csy}}}, \bibinfo
  {author} {\bibfnamefont {C.}~\bibnamefont {{Beer}}}, \bibinfo {author}
  {\bibfnamefont {M.}~\bibnamefont {{Bejger}}}, \bibinfo {author}
  {\bibfnamefont {I.}~\bibnamefont {{Belahcene}}}, \bibinfo {author}
  {\bibfnamefont {A.~S.}\ \bibnamefont {{Bell}}}, \bibinfo {author}
  {\bibfnamefont {D.}~\bibnamefont {{Beniwal}}}, \bibinfo {author}
  {\bibfnamefont {M.}~\bibnamefont {{Bensch}}}, \bibinfo {author}
  {\bibfnamefont {B.~K.}\ \bibnamefont {{Berger}}}, \bibinfo {author}
  {\bibfnamefont {G.}~\bibnamefont {{Bergmann}}}, \bibinfo {author}
  {\bibfnamefont {S.}~\bibnamefont {{Bernuzzi}}}, \bibinfo {author}
  {\bibfnamefont {J.~J.}\ \bibnamefont {{Bero}}}, \bibinfo {author}
  {\bibfnamefont {C.~P.~L.}\ \bibnamefont {{Berry}}}, \bibinfo {author}
  {\bibfnamefont {D.}~\bibnamefont {{Bersanetti}}}, \bibinfo {author}
  {\bibfnamefont {A.}~\bibnamefont {{Bertolini}}}, \bibinfo {author}
  {\bibfnamefont {J.}~\bibnamefont {{Betzwieser}}}, \bibinfo {author}
  {\bibfnamefont {R.}~\bibnamefont {{Bhandare}}}, \bibinfo {author}
  {\bibfnamefont {I.~A.}\ \bibnamefont {{Bilenko}}}, \bibinfo {author}
  {\bibfnamefont {S.~A.}\ \bibnamefont {{Bilgili}}}, \bibinfo {author}
  {\bibfnamefont {G.}~\bibnamefont {{Billingsley}}}, \bibinfo {author}
  {\bibfnamefont {C.~R.}\ \bibnamefont {{Billman}}}, \bibinfo {author}
  {\bibfnamefont {J.}~\bibnamefont {{Birch}}}, \bibinfo {author} {\bibfnamefont
  {R.}~\bibnamefont {{Birney}}}, \bibinfo {author} {\bibfnamefont
  {O.}~\bibnamefont {{Birnholtz}}}, \bibinfo {author} {\bibfnamefont
  {S.}~\bibnamefont {{Biscans}}}, \bibinfo {author} {\bibfnamefont
  {S.}~\bibnamefont {{Biscoveanu}}}, \bibinfo {author} {\bibfnamefont
  {A.}~\bibnamefont {{Bisht}}}, \bibinfo {author} {\bibfnamefont
  {M.}~\bibnamefont {{Bitossi}}}, \bibinfo {author} {\bibfnamefont {M.~A.}\
  \bibnamefont {{Bizouard}}}, \bibinfo {author} {\bibfnamefont {J.~K.}\
  \bibnamefont {{Blackburn}}}, \bibinfo {author} {\bibfnamefont
  {J.}~\bibnamefont {{Blackman}}}, \bibinfo {author} {\bibfnamefont {C.~D.}\
  \bibnamefont {{Blair}}}, \bibinfo {author} {\bibfnamefont {D.~G.}\
  \bibnamefont {{Blair}}}, \bibinfo {author} {\bibfnamefont {R.~M.}\
  \bibnamefont {{Blair}}}, \bibinfo {author} {\bibfnamefont {S.}~\bibnamefont
  {{Bloemen}}}, \bibinfo {author} {\bibfnamefont {O.}~\bibnamefont {{Bock}}},
  \bibinfo {author} {\bibfnamefont {N.}~\bibnamefont {{Bode}}}, \bibinfo
  {author} {\bibfnamefont {M.}~\bibnamefont {{Boer}}}, \bibinfo {author}
  {\bibfnamefont {Y.}~\bibnamefont {{Boetzel}}}, \bibinfo {author}
  {\bibfnamefont {G.}~\bibnamefont {{Bogaert}}}, \bibinfo {author}
  {\bibfnamefont {A.}~\bibnamefont {{Bohe}}}, \bibinfo {author} {\bibfnamefont
  {F.}~\bibnamefont {{Bondu}}}, \bibinfo {author} {\bibfnamefont
  {E.}~\bibnamefont {{Bonilla}}}, \bibinfo {author} {\bibfnamefont
  {R.}~\bibnamefont {{Bonnand}}}, \bibinfo {author} {\bibfnamefont
  {P.}~\bibnamefont {{Booker}}}, \bibinfo {author} {\bibfnamefont {B.~A.}\
  \bibnamefont {{Boom}}}, \bibinfo {author} {\bibfnamefont {C.~D.}\
  \bibnamefont {{Booth}}}, \bibinfo {author} {\bibfnamefont {R.}~\bibnamefont
  {{Bork}}}, \bibinfo {author} {\bibfnamefont {V.}~\bibnamefont {{Boschi}}},
  \bibinfo {author} {\bibfnamefont {S.}~\bibnamefont {{Bose}}}, \bibinfo
  {author} {\bibfnamefont {K.}~\bibnamefont {{Bossie}}}, \bibinfo {author}
  {\bibfnamefont {V.}~\bibnamefont {{Bossilkov}}}, \bibinfo {author}
  {\bibfnamefont {J.}~\bibnamefont {{Bosveld}}}, \bibinfo {author}
  {\bibfnamefont {Y.}~\bibnamefont {{Bouffanais}}}, \bibinfo {author}
  {\bibfnamefont {A.}~\bibnamefont {{Bozzi}}}, \bibinfo {author} {\bibfnamefont
  {C.}~\bibnamefont {{Bradaschia}}}, \bibinfo {author} {\bibfnamefont {P.~R.}\
  \bibnamefont {{Brady}}}, \bibinfo {author} {\bibfnamefont {A.}~\bibnamefont
  {{Bramley}}}, \bibinfo {author} {\bibfnamefont {M.}~\bibnamefont
  {{Branchesi}}}, \bibinfo {author} {\bibfnamefont {J.~E.}\ \bibnamefont
  {{Brau}}}, \bibinfo {author} {\bibfnamefont {T.}~\bibnamefont {{Briant}}},
  \bibinfo {author} {\bibfnamefont {F.}~\bibnamefont {{Brighenti}}}, \bibinfo
  {author} {\bibfnamefont {A.}~\bibnamefont {{Brillet}}}, \bibinfo {author}
  {\bibfnamefont {M.}~\bibnamefont {{Brinkmann}}}, \bibinfo {author}
  {\bibfnamefont {V.}~\bibnamefont {{Brisson}}}, \bibinfo {author}
  {\bibfnamefont {P.}~\bibnamefont {{Brockill}}}, \bibinfo {author}
  {\bibfnamefont {A.~F.}\ \bibnamefont {{Brooks}}}, \bibinfo {author}
  {\bibfnamefont {D.~D.}\ \bibnamefont {{Brown}}}, \bibinfo {author}
  {\bibfnamefont {S.}~\bibnamefont {{Brunett}}}, \bibinfo {author}
  {\bibfnamefont {C.~C.}\ \bibnamefont {{Buchanan}}}, \bibinfo {author}
  {\bibfnamefont {A.}~\bibnamefont {{Buikema}}}, \bibinfo {author}
  {\bibfnamefont {T.}~\bibnamefont {{Bulik}}}, \bibinfo {author} {\bibfnamefont
  {H.~J.}\ \bibnamefont {{Bulten}}}, \bibinfo {author} {\bibfnamefont
  {A.}~\bibnamefont {{Buonanno}}}, \bibinfo {author} {\bibfnamefont
  {D.}~\bibnamefont {{Buskulic}}}, \bibinfo {author} {\bibfnamefont
  {C.}~\bibnamefont {{Buy}}}, \bibinfo {author} {\bibfnamefont {R.~L.}\
  \bibnamefont {{Byer}}}, \bibinfo {author} {\bibfnamefont {M.}~\bibnamefont
  {{Cabero}}}, \bibinfo {author} {\bibfnamefont {L.}~\bibnamefont
  {{Cadonati}}}, \bibinfo {author} {\bibfnamefont {G.}~\bibnamefont
  {{Cagnoli}}}, \bibinfo {author} {\bibfnamefont {C.}~\bibnamefont
  {{Cahillane}}}, \bibinfo {author} {\bibfnamefont {J.}~\bibnamefont
  {{Calder{\'o}n Bustillo}}}, \bibinfo {author} {\bibfnamefont {T.~A.}\
  \bibnamefont {{Callister}}}, \bibinfo {author} {\bibfnamefont
  {E.}~\bibnamefont {{Calloni}}}, \bibinfo {author} {\bibfnamefont {J.~B.}\
  \bibnamefont {{Camp}}}, \bibinfo {author} {\bibfnamefont {M.}~\bibnamefont
  {{Canepa}}}, \bibinfo {author} {\bibfnamefont {P.}~\bibnamefont
  {{Canizares}}}, \bibinfo {author} {\bibfnamefont {K.~C.}\ \bibnamefont
  {{Cannon}}}, \bibinfo {author} {\bibfnamefont {H.}~\bibnamefont {{Cao}}},
  \bibinfo {author} {\bibfnamefont {J.}~\bibnamefont {{Cao}}}, \bibinfo
  {author} {\bibfnamefont {C.~D.}\ \bibnamefont {{Capano}}}, \bibinfo {author}
  {\bibfnamefont {E.}~\bibnamefont {{Capocasa}}}, \bibinfo {author}
  {\bibfnamefont {F.}~\bibnamefont {{Carbognani}}}, \bibinfo {author}
  {\bibfnamefont {S.}~\bibnamefont {{Caride}}}, \bibinfo {author}
  {\bibfnamefont {M.~F.}\ \bibnamefont {{Carney}}}, \bibinfo {author}
  {\bibfnamefont {G.}~\bibnamefont {{Carullo}}}, \bibinfo {author}
  {\bibfnamefont {J.}~\bibnamefont {{Casanueva Diaz}}}, \bibinfo {author}
  {\bibfnamefont {C.}~\bibnamefont {{Casentini}}}, \bibinfo {author}
  {\bibfnamefont {S.}~\bibnamefont {{Caudill}}}, \bibinfo {author}
  {\bibfnamefont {M.}~\bibnamefont {{Cavagli{\`a}}}}, \bibinfo {author}
  {\bibfnamefont {F.}~\bibnamefont {{Cavalier}}}, \bibinfo {author}
  {\bibfnamefont {R.}~\bibnamefont {{Cavalieri}}}, \bibinfo {author}
  {\bibfnamefont {G.}~\bibnamefont {{Cella}}}, \bibinfo {author} {\bibfnamefont
  {C.~B.}\ \bibnamefont {{Cepeda}}}, \bibinfo {author} {\bibfnamefont
  {P.}~\bibnamefont {{Cerd{\'a}-Dur{\'a}n}}}, \bibinfo {author} {\bibfnamefont
  {G.}~\bibnamefont {{Cerretani}}}, \bibinfo {author} {\bibfnamefont
  {E.}~\bibnamefont {{Cesarini}}}, \bibinfo {author} {\bibfnamefont
  {O.}~\bibnamefont {{Chaibi}}}, \bibinfo {author} {\bibfnamefont {S.~J.}\
  \bibnamefont {{Chamberlin}}}, \bibinfo {author} {\bibfnamefont
  {M.}~\bibnamefont {{Chan}}}, \bibinfo {author} {\bibfnamefont
  {S.}~\bibnamefont {{Chao}}}, \bibinfo {author} {\bibfnamefont
  {P.}~\bibnamefont {{Charlton}}}, \bibinfo {author} {\bibfnamefont
  {E.}~\bibnamefont {{Chase}}}, \bibinfo {author} {\bibfnamefont
  {E.}~\bibnamefont {{Chassande-Mottin}}}, \bibinfo {author} {\bibfnamefont
  {D.}~\bibnamefont {{Chatterjee}}}, \bibinfo {author} {\bibfnamefont
  {K.}~\bibnamefont {{Chatziioannou}}}, \bibinfo {author} {\bibfnamefont
  {B.~D.}\ \bibnamefont {{Cheeseboro}}}, \bibinfo {author} {\bibfnamefont
  {H.~Y.}\ \bibnamefont {{Chen}}}, \bibinfo {author} {\bibfnamefont
  {X.}~\bibnamefont {{Chen}}}, \bibinfo {author} {\bibfnamefont
  {Y.}~\bibnamefont {{Chen}}}, \bibinfo {author} {\bibfnamefont {H.~P.}\
  \bibnamefont {{Cheng}}}, \bibinfo {author} {\bibfnamefont {H.~Y.}\
  \bibnamefont {{Chia}}},\ and\ \bibinfo {author} {\bibfnamefont
  {A.}~\bibnamefont {{Chincarini}}},\ }\bibfield  {title} {\bibinfo {title}
  {{GW170817: Measurements of Neutron Star Radii and Equation of State}},\
  }\href {https://doi.org/10.1103/PhysRevLett.121.161101} {\bibfield  {journal}
  {\bibinfo  {journal} {\prl}\ }\textbf {\bibinfo {volume} {121}},\ \bibinfo
  {eid} {161101} (\bibinfo {year} {2018})},\ \Eprint
  {https://arxiv.org/abs/1805.11581} {arXiv:1805.11581 [gr-qc]} \BibitemShut
  {NoStop}%
\bibitem [{\citenamefont {{LIGO-P1800115}}(2018)}]{P1800115}%
  \BibitemOpen
  \bibfield  {author} {\bibinfo {author} {\bibnamefont {{LIGO-P1800115}}},\
  }\href {https://dcc.ligo.org/LIGO-P1800115/public} {\bibinfo {title} {Ligo
  document p1800115-v12 (lvc)}} (\bibinfo {year} {2018}),\ \bibinfo {note}
  {accessed: 2025-04-22}\BibitemShut {NoStop}%
\bibitem [{\citenamefont {Virtanen}\ \emph {et~al.}(2020)\citenamefont
  {Virtanen}, \citenamefont {Gommers}, \citenamefont {Oliphant}, \citenamefont
  {Haberland}, \citenamefont {Reddy}, \citenamefont {Cournapeau}, \citenamefont
  {Burovski}, \citenamefont {Peterson}, \citenamefont {Weckesser},
  \citenamefont {Bright}, \citenamefont {{van der Walt}}, \citenamefont
  {Brett}, \citenamefont {Wilson}, \citenamefont {Millman}, \citenamefont
  {Mayorov}, \citenamefont {Nelson}, \citenamefont {Jones}, \citenamefont
  {Kern}, \citenamefont {Larson}, \citenamefont {Carey}, \citenamefont {Polat},
  \citenamefont {Feng}, \citenamefont {Moore}, \citenamefont {{VanderPlas}},
  \citenamefont {Laxalde}, \citenamefont {Perktold}, \citenamefont {Cimrman},
  \citenamefont {Henriksen}, \citenamefont {Quintero}, \citenamefont {Harris},
  \citenamefont {Archibald}, \citenamefont {Ribeiro}, \citenamefont
  {Pedregosa}, \citenamefont {{van Mulbregt}},\ and\ \citenamefont {{SciPy 1.0
  Contributors}}}]{2020SciPy-NMeth}%
  \BibitemOpen
  \bibfield  {author} {\bibinfo {author} {\bibfnamefont {P.}~\bibnamefont
  {Virtanen}}, \bibinfo {author} {\bibfnamefont {R.}~\bibnamefont {Gommers}},
  \bibinfo {author} {\bibfnamefont {T.~E.}\ \bibnamefont {Oliphant}}, \bibinfo
  {author} {\bibfnamefont {M.}~\bibnamefont {Haberland}}, \bibinfo {author}
  {\bibfnamefont {T.}~\bibnamefont {Reddy}}, \bibinfo {author} {\bibfnamefont
  {D.}~\bibnamefont {Cournapeau}}, \bibinfo {author} {\bibfnamefont
  {E.}~\bibnamefont {Burovski}}, \bibinfo {author} {\bibfnamefont
  {P.}~\bibnamefont {Peterson}}, \bibinfo {author} {\bibfnamefont
  {W.}~\bibnamefont {Weckesser}}, \bibinfo {author} {\bibfnamefont
  {J.}~\bibnamefont {Bright}}, \bibinfo {author} {\bibfnamefont {S.~J.}\
  \bibnamefont {{van der Walt}}}, \bibinfo {author} {\bibfnamefont
  {M.}~\bibnamefont {Brett}}, \bibinfo {author} {\bibfnamefont
  {J.}~\bibnamefont {Wilson}}, \bibinfo {author} {\bibfnamefont {K.~J.}\
  \bibnamefont {Millman}}, \bibinfo {author} {\bibfnamefont {N.}~\bibnamefont
  {Mayorov}}, \bibinfo {author} {\bibfnamefont {A.~R.~J.}\ \bibnamefont
  {Nelson}}, \bibinfo {author} {\bibfnamefont {E.}~\bibnamefont {Jones}},
  \bibinfo {author} {\bibfnamefont {R.}~\bibnamefont {Kern}}, \bibinfo {author}
  {\bibfnamefont {E.}~\bibnamefont {Larson}}, \bibinfo {author} {\bibfnamefont
  {C.~J.}\ \bibnamefont {Carey}}, \bibinfo {author} {\bibfnamefont
  {{\.I}.}~\bibnamefont {Polat}}, \bibinfo {author} {\bibfnamefont
  {Y.}~\bibnamefont {Feng}}, \bibinfo {author} {\bibfnamefont {E.~W.}\
  \bibnamefont {Moore}}, \bibinfo {author} {\bibfnamefont {J.}~\bibnamefont
  {{VanderPlas}}}, \bibinfo {author} {\bibfnamefont {D.}~\bibnamefont
  {Laxalde}}, \bibinfo {author} {\bibfnamefont {J.}~\bibnamefont {Perktold}},
  \bibinfo {author} {\bibfnamefont {R.}~\bibnamefont {Cimrman}}, \bibinfo
  {author} {\bibfnamefont {I.}~\bibnamefont {Henriksen}}, \bibinfo {author}
  {\bibfnamefont {E.~A.}\ \bibnamefont {Quintero}}, \bibinfo {author}
  {\bibfnamefont {C.~R.}\ \bibnamefont {Harris}}, \bibinfo {author}
  {\bibfnamefont {A.~M.}\ \bibnamefont {Archibald}}, \bibinfo {author}
  {\bibfnamefont {A.~H.}\ \bibnamefont {Ribeiro}}, \bibinfo {author}
  {\bibfnamefont {F.}~\bibnamefont {Pedregosa}}, \bibinfo {author}
  {\bibfnamefont {P.}~\bibnamefont {{van Mulbregt}}},\ and\ \bibinfo {author}
  {\bibnamefont {{SciPy 1.0 Contributors}}},\ }\bibfield  {title} {\bibinfo
  {title} {{{SciPy} 1.0: Fundamental Algorithms for Scientific Computing in
  Python}},\ }\href {https://doi.org/10.1038/s41592-019-0686-2} {\bibfield
  {journal} {\bibinfo  {journal} {Nature Methods}\ }\textbf {\bibinfo {volume}
  {17}},\ \bibinfo {pages} {261} (\bibinfo {year} {2020})}\BibitemShut
  {NoStop}%
\bibitem [{Note1()}]{Note1}%
  \BibitemOpen
  \bibinfo {note} {How to constrain all the constants to be positive: \protect
  \url {https://github.com/MilesCranmer/PySR/discussions/449}}\BibitemShut
  {NoStop}%
\bibitem [{\citenamefont {Akiba}\ \emph {et~al.}(2019)\citenamefont {Akiba},
  \citenamefont {Sano}, \citenamefont {Yanase}, \citenamefont {Ohta},\ and\
  \citenamefont {Koyama}}]{akiba2019optuna}%
  \BibitemOpen
  \bibfield  {author} {\bibinfo {author} {\bibfnamefont {T.}~\bibnamefont
  {Akiba}}, \bibinfo {author} {\bibfnamefont {S.}~\bibnamefont {Sano}},
  \bibinfo {author} {\bibfnamefont {T.}~\bibnamefont {Yanase}}, \bibinfo
  {author} {\bibfnamefont {T.}~\bibnamefont {Ohta}},\ and\ \bibinfo {author}
  {\bibfnamefont {M.}~\bibnamefont {Koyama}},\ }\bibfield  {title} {\bibinfo
  {title} {{O}ptuna: A next-generation hyperparameter optimization framework},\
  }in\ \href@noop {} {\emph {\bibinfo {booktitle} {The 25th ACM SIGKDD
  International Conference on Knowledge Discovery \& Data Mining}}}\ (\bibinfo
  {year} {2019})\ pp.\ \bibinfo {pages} {2623--2631}\BibitemShut {NoStop}%
\bibitem [{\citenamefont {{PySR Application Programming Interface (API)
  documentation}}(2025)}]{pysr2025}%
  \BibitemOpen
  \bibfield  {author} {\bibinfo {author} {\bibnamefont {{PySR Application
  Programming Interface (API) documentation}}},\ }\href
  {https://ai.damtp.cam.ac.uk/pysr/api/} {\bibinfo {title} {Pysr: Symbolic
  regression in python and julia}} (\bibinfo {year} {2025}),\ \bibinfo {note}
  {accessed: 2025-04-21}\BibitemShut {NoStop}%
\end{thebibliography}%

\end{document}